\documentclass[a4paper,11p]{article}
\pdfoutput=1
\usepackage[T1]{fontenc} 

\usepackage{
graphicx,
hyperref,
amsmath,
amssymb,
charter,
xcolor,
ifluatex,
longtable,
booktabs,
multirow,
bbold,cancel}

\usepackage{comment}

\usepackage{tikz}
\usepackage{tikz-feynman}
\tikzfeynmanset{compat=1.1.0}

\usepackage{multirow}
\newcommand*{\myalign}[2]{\multicolumn{1}{#1}{#2}}

\usepackage{float}
\usepackage{colortbl}

\definecolor{Gray}{gray}{0.95}
\definecolor{RGray}{gray}{0.85}
\definecolor{CGray}{gray}{0.92}

\usepackage{multicol}
\definecolor{tit}{rgb}{0.1,0.2,0.4}
\definecolor{blus}{cmyk}{1,1,0,0.6}
\definecolor{verde}{cmyk}{0.92,0,0.59,0.25}

\usetikzlibrary{matrix}

\usepackage{tipa}
\usepackage{amsmath,amssymb,amsfonts,bm}
\usepackage{dsfont}
\usepackage{epsfig}
\usepackage{graphicx}
\usepackage{slashed}
\usepackage{color}

\usepackage{caption}
\usepackage{subcaption}
\usepackage{slashed} 

\usepackage{adjustbox}
\usepackage{commath}
\usepackage{calc}

\usepackage{sidecap} 

\newcommand{\be}{\begin{equation}}
\newcommand{\ee}{\end{equation}}

\newcommand{\bea}{\begin{eqnarray}}
\newcommand{\eea}{\end{eqnarray}}

\newcommand{\bfig}{\begin{figure}}
\newcommand{\efig}{\end{figure}}

\usepackage{slashed}

\usepackage{commath}
\usepackage{calc}

\newcommand{\D}{{\cal D}}
\newcommand{\U}{{\cal U}}

\newcommand{\N}{{\cal N}}
\newcommand{\M}{{\cal M}}

\makeatletter
\newcommand*{\rom}[1]{\expandafter\@slowromancap\romannumeral #1@}
\makeatother

\usepackage[english]{babel}

\usepackage{amsmath}

\usepackage{booktabs}

\newcommand{\iab}{{\ensuremath\rm ab^{-1}}}

\newcommand{\e}[1]{\cdot 10^{#1}}

\usepackage{mathtools}

\usepackage{color,hyperref}
\definecolor{darkblue}{rgb}{0.0,0.0,0.3}
\hypersetup{colorlinks,breaklinks,
            linkcolor=darkblue,urlcolor=darkblue,
            anchorcolor=darkblue,citecolor=darkblue}

\def\1by3{\ensuremath{\frac{1}{3}}}
\def\4by3{\ensuremath{\frac{4}{3}}}
\def\2by3{\ensuremath{\frac{2}{3}}}

\providecommand*\url[1]{\href{#1}{#1}}
\renewcommand*\url[1]{\href{#1}{\texttt{#1}}}

\usepackage{amsfonts}

\usepackage[T1]{fontenc}
\usepackage{tikz}
\usetikzlibrary{arrows,calc,positioning}

\tikzstyle{intt}=[draw,text centered,minimum size=6em,text width=5.25cm,text height=0.34cm]
\tikzstyle{intl}=[draw,text centered,minimum size=2em,text width=2.75cm,text height=0.34cm]
\tikzstyle{int}=[draw,minimum size=2.5em,text centered,text width=3.5cm]
\tikzstyle{intg}=[draw,minimum size=3em,text centered,text width=6.cm]
\tikzstyle{sum}=[draw,shape=circle,inner sep=2pt,text centered,node distance=3.5cm]
\tikzstyle{summ}=[drawshape=circle,inner sep=4pt,text centered,node distance=3.cm]

\begin{document} 

\allowdisplaybreaks
\vspace*{-2.5cm}
\begin{flushright}
{\small
IIT-BHU
}
\end{flushright}

\vspace{2cm}

\begin{center}
{\LARGE \bf \color{tit}  Dark-technicolor  at colliders }\\[1cm]

{\large\bf Gauhar Abbas$^{a}$\footnote{email: gauhar.phy@iitbhu.ac.in}},~
{\large\bf Vartika Singh$^{a}$\footnote{email: vartikasingh.rs.phy19@itbhu.ac.in }},~ 
{\large\bf Neelam Singh$^{a}$\footnote{email: neelamsingh.rs.phy19@itbhu.ac.in }   }  
\\[7mm]
{\it $^a$ } {\em Department of Physics, Indian Institute of Technology (BHU), Varanasi 221005, India}\\[3mm]

\vspace{1cm}
{\large\bf\color{blus} Abstract}
\begin{quote}
We demonstrate that QCD-like gauge dynamics can be consistently embedded within the Dark Technicolor paradigm by invoking the extended Most Attractive Channel hypothesis, thereby revitalizing conventional technicolor scenarios. In this framework, the Higgs mass is generated dynamically while remaining consistent with electroweak precision tests, including constraints from the $S$ parameter. The flavor problem is resolved by incorporating the Standard Hierarchical VEVs Model, whereas a simple Froggatt--Nielsen construction is shown to be incompatible. Couplings of techni-hadrons such as $\rho_{\rm TC}$ and $\eta_{\rm TC}^\prime$ to Standard Model fermions are highly suppressed, leading to negligible direct fermionic signatures. Nevertheless, DTC mesons remain testable at the HL-LHC, HE-LHC, and future 100~TeV collider, with promising discovery channels including $\bar{b}b$, $\tau^+\tau^-$, $t\bar{t}$, and $\gamma\gamma$. 
\end{quote}

\thispagestyle{empty}
\end{center}

\begin{quote}
{\large\noindent\color{blus} 
}

\end{quote}

\newpage
\setcounter{footnote}{0}

\section{Introduction}
\label{intro}
The origin of mass remains one of the most fundamental open questions in particle physics. Within the Standard Model (SM), this is addressed by the Higgs mechanism, an elementary scalar doublet acquires a vacuum expectation value (VEV), spontaneously breaking electroweak symmetry and generating particle masses. However, the dynamical origin of this VEV lies beyond the SM.

Analogous mechanisms appear in other physical systems. In the Ginzburg–Landau theory of superconductivity~\cite{Ginzburg:1950sr}, the phenomenological order parameter was later understood in the Bardeen–Cooper Schrieffer (BCS) theory~\cite{Bardeen:1957kj,Bardeen:1957mv} as a condensate of Cooper pairs bound by short-range interactions. In hadron physics, low-energy dynamics are described by the Gell-Mann–Lévy (GML) $\sigma$ model~\cite{Gell-Mann:1960mvl}, with a microscopic foundation in the Nambu–Jona-Lasinio (NJL) model~\cite{Nambu:1961tp,Nambu:1961fr}, which parallels BCS theory. At the QCD level, the GML order parameter $\langle \sigma \rangle = f_\pi = 95~\text{MeV}$ corresponds to the quark condensate $\langle \bar{q}q \rangle = \mathcal{O}(f_\pi^3)$, dynamically generated via the short-range color force. This is a classic example of dynamical symmetry breaking with a composite local order parameter.

Nature therefore seems to favor symmetry breaking via composite order parameters across diverse systems, from superconductivity to hadron physics, suggesting that a similar mechanism could underlie electroweak symmetry breaking at higher scales. This motivates scenarios beyond the SM in which the Higgs emerges dynamically from strong interactions.

Technicolor (TC) models provide such a framework~\cite{Weinberg:1975gm,Susskind:1978ms}; for reviews and phenomenology, see~\cite{Kaul:1981uk,Chivukula:1995hr,Chivukula:1995dt,Chivukula:1998if,Chivukula:2000mb,Chivukula:2011ue,Chivukula:2011jrk,Hill:2002ap,Sannino:2008ha,Sannino:2010ca,DiChiara:2010xb,Dietrich:2005jn,Belyaev:2008yj,Foadi:2008xj,Andersen:2011yj,Cacciapaglia:2020kgq,Eichten:1986eq,Doff:2019vav,Alanne:2016rpe}. Besides offering a dynamical origin for electroweak symmetry breaking, TC models also address the naturalness problem: composite scalars receive additive radiative corrections only of order $\Lambda_{\rm TC}$, the TC scale~\cite{Hill:2002ap}. However, QCD-like TC models face severe challenges. In particular, large flavor-changing neutral currents (FCNC) push the extended technicolor (ETC)  \cite{Dimopoulos:1979es, Eichten:1979ah} scale to $\Lambda_{\rm ETC} \gtrsim 10^6$ GeV~\cite{Hill:2002ap}, leading to fermion masses
\begin{align}
m_f \sim \frac{\Lambda_{\rm TC}^3}{\Lambda_{\rm ETC}^2},
\end{align}
which are too small for realistic spectra. Moreover, such models conflict with electroweak precision observables.

There are alternatives models, such as walking dynamics~\cite{Holdom:1981rm,Holdom:1984sk,Yamawaki:1985zg,Appelquist:1986an,Appelquist:1986tr,Appelquist:1987fc}.  Even with walking technicolor, obtaining the physical top-quark mass solely from ETC interactions 
remains highly problematic, since such a mechanism inevitably induces unacceptably large violations 
of weak isospin~\cite{Chivukula:1988qr}. The most compelling and widely studied resolution is to 
supplement technicolor with a new strong interaction, ``topcolor''~\cite{Hill:1991(a)t}, which 
dynamically generates the dominant part of the top-quark mass. In this framework the ETC sector 
contributes only subdominantly, thereby avoiding large isospin-breaking effects~\cite{Hill:1994hp}.  For other alternative approches, see Refs.~\cite{Cacciapaglia:2020kgq,Sannino:2009aw,Mojaza:2012zd,Dietrich:2006cm,Sannino:2012wy,Hong:2004td}.

We notice that the long-standing flavor problem of the SM, its unexplained hierarchies of masses, mixing angles, and neutrino properties, remains deeply connected to the origin of mass itself~\cite{Abbas:2023ivi,Abbas:2024wzp}. A compelling framework to address both issues is the \emph{dark technicolor} (DTC) paradigm~\cite{Abbas:2020frs}, defined by
\begin{align} 
\mathcal{G}_{\rm DTC} \equiv \rm SU(N_{\rm TC}) \times SU(N_{\rm DTC}) \times SU(N_{\rm D}), \end{align}
where $\rm TC$ denotes technicolor, $\rm DTC$ represents dark technicolor, and $\rm D$ corresponds to dark QCD (DQCD). In contrast to conventional technicolor, the DTC framework remains QCD-like but naturally disentangles fermion mass generation from the FCNC problem. The extended technicolor (ETC) scale is kept high ($\sim 10^6$ GeV), while fermion masses and mixings, including those of leptons emerge dynamically from DTC interactions~\cite{Abbas:2020frs}. 

In this work, we investigate a novel fermionic mass-generation mechanism for technicolor-type theories which, within the DTC paradigm, simultaneously achieves electroweak symmetry breaking, reproduces the observed Higgs mass, and explains the SM flavor structure through the Extended Most Attractive Channel (EMAC) hypothesis~\cite{Aoki:1983ae,Aoki:1983za,Aoki:1983yy}. This fermionic mechanism within the DTC framework thereby overcomes the longstanding problems of conventional technicolor and walking dynamics, such as large isospin-breaking effects~\cite{Hill:1994hp}.

One effective low-energy realization of DTC is the \emph{Hierarchical VEVs Model} (HVM)~\cite{Abbas:2017vws}, in which multi-fermion condensates give rise to hierarchical VEVs of six gauge-singlet scalar fields~\cite{Abbas:2020frs}. A standard realization, the \emph{Standard HVM} (SHVM), has been shown to predict precise leptonic mixing observables~\cite{Abbas:2023dpf}. Another viable limit is the Froggatt–Nielsen (FN) mechanism~\cite{Froggatt:1978nt}, which can be embedded in DTC using discrete flavor symmetries such as $\mathcal{Z}_N \times \mathcal{Z}_M$~\cite{Abbas:2018lga}. We show explicitly that both SHVM and the FN mechanism can be naturally embedded within the DTC framework, thereby providing a unified resolution of the SM flavor problem. In the present work, however, only the SHVM limit reproduces the full SM flavor structure.

We also study collider signatures of the DTC paradigm, with emphasis on the SHVM limit, at the High-Luminosity LHC (HL-LHC), High-Energy LHC (HE-LHC)~\cite{FCC:2018bvk}, and future 100~TeV colliders such as FCC-hh~\cite{FCC:2018byv}. 

\medskip

This paper is organized as follows: In Sec.~\ref{sec2}, we review the DTC paradigm, followed by the EMAC hypothesis in Sec.~\ref{emac}. Experimental constraints are discussed in Sec.~\ref{exp_dtc}. Effective low-energy limits of the paradigm are presented in Sec.~\ref{sec4}, and the minimal version of DTC is discussed in Sec.~\ref{min_dtc}. The mass spectrum is analyzed via scaling relations in Sec.~\ref{scaling}.  Collider physics is investigated in section \ref{col_sig}. Conclusions are summarized in Sec.~\ref{sum}.

\section{Dark-technicolor paradigm}
\label{sec2}
We now discuss the DTC paradigm, which can give rise to the SHVM and the FN mechanism at low energies \cite{Abbas:2020frs}.  The DTC-paradigm was first proposed in reference \cite{Abbas:2020frs}.  The DTC paradigm is based on symmetry $\mathcal{G} \equiv \rm SU(\rm N_{\rm TC}) \times SU(\rm N_{\rm DTC}) \times SU(\rm{N}_{\rm D})$.  The TC dynamics is defined by the TC fermionic doublet obeying the following transformations under the $\rm SU(3)_c \times \rm SU(2)_L \times \rm U(1)_Y \times \mathcal{G}$ as \cite{Abbas:2020frs},
\begin{eqnarray}
T  &\equiv&  \begin{pmatrix}
T  \\
B
\end{pmatrix}_L :(1,2,0,\rm{N}_{\rm TC},1,1), ~
T_{R} : (1,1,1,\text{N}_{\rm{TC}},1,1), B_{R} : (1,1,-1,\rm{N}_{\rm TC},1,1), 
\end{eqnarray}
where   electric charges  are  $Q_{T} = +\frac{1}{2}$ and $Q_{B} = -\frac{1}{2}$.  

The fermions of the  DTC symmetry $\rm SU(\rm N_{\rm DTC}) $   transform under  $\rm SU(3)_c \times SU(2)_L \times U(1)_Y \times \mathcal{G}$ as,
\begin{eqnarray}
 D^i &\equiv& C_{L,R}^i  : (1,1, \rm Y,1,\rm N_{\rm DTC},1),~S_{L,R}^i  : (1,1,-Y,1,\rm N_{\rm DTC},1), 
\end{eqnarray}
where  $i=1,2,3 \cdots $,  and  electric charge of $C$   is  $+\frac{1}{2}$  and that of $ S$ is  $-\frac{1}{2}$ for $Y=1$.  

The DQCD symmetry $\rm SU(N_{\rm D})$ contains fermions  transforming under   $\rm SU(3)_c \times SU(2)_L \times U(1)_Y \times \mathcal{G}$ as,
\begin{eqnarray}
F_{L,R} &\equiv &U_{L,R}^i :  (3,1,\dfrac{4}{3},1,1,\rm N_{\rm D}),
D_{L,R}^{i} :   (3,1,-\dfrac{2}{3},1,1,\rm N_{\rm D}),  \\ \nonumber 
&& N_{L,R}^i :   (1,1,0,1,1,\rm N_{\rm D}), ~E_{L,R}^{i} :   (1,1,-2,1,1,\rm N_{\rm D}),
\end{eqnarray}
where  $i=1,2,3 \cdots $. 

In the DTC paradigm,  the TC symmetry  $ \rm SU(\rm N_{\rm TC})$ contains the left-handed TC doublet fermions with zero hypercharge.  On the other side, the right handed TC fermions $T_R$ and $B_R$ form a vector-like pair.  Thus, the gauge anomaly vanishes for the TC dynamics based on the $ \rm SU(\rm N_{\rm TC})$ symmetry.  The Witten global  anomaly \cite{Witten:1982fp} remains absent for even $\rm N_{\rm TC}$ for any number of doublets  \cite{Hill:2002ap}.  The gauge dynamics of the symmetries $\rm SU(\rm N_{\rm DTC}) $ and $\rm SU(\rm{N}_{\rm D})$ only have vector-like fermions resulting in the absence of the gauge anomalies  for these symmetries.
We further assume that the TC fermions,   the left-handed SM fermions, and the $F_R$ fermions are accommodated in an ETC symmetry.  On the other hand, there exists an extended DTC (EDTC) symmetry containing the  DTC fermions,  the right-handed SM fermions, and the $F_L$ fermions. It is important to note that the  $\rm SU(\rm N_{\rm D})$ symmetry  is a connecting bridge between the TC and DTC dynamics.  This results in a suppression of the mixing between the dynamics of the TC and the DTC by the factor $1 / \Lambda$, where $\Lambda$  is the scale of the DQCD.

We notice that the symmetry $\mathcal{G}$ gives rise to three global anomalous  $\rm U(1)_{\rm A}$  symmetries denoted by   $\rm U(1)_{\rm A}^{\rm TC }$,   $\rm U(1)_{\rm A}^{\rm DTC }$  and $\rm U(1)_{\rm A}^{\rm D }$ in the DTC paradigm.  In general, a global axial symmetry $\rm U(1)_A$ can be broken by instantons.   This breaking provides a $2\rm K$-fermion operator with a non-vanishing VEV, and $2\rm K$ conserved quantum numbers \cite{Harari:1981bs}. This can be written as,
\begin{align}
\mathrm{U(1)}_{X_{\mathrm{TC,DTC,DQCD}}} \longrightarrow  \mathbb{Z}_{2K}.
\end{align}
where $\rm K$ denotes the  massless flavors of the $\rm SU(\rm N)$  gauge dynamics in the $\rm N$-dimensional representation.   

Thus, the DTC paradigm creates a generic residual   $\mathcal{Z}_{\rm N} \times \mathcal{Z}_{\rm M} \times \mathcal{Z}_{\rm P}$ flavor symmetry, where $\rm N= 2 \rm K_{\rm TC}$, $\rm M= 2 \rm K_{\rm DTC}$, and $\rm P= 2 \rm K_{\rm D}$. This results in certain conserved axial charges modulo $2\rm K$ \cite{Harari:1981bs}.

\section{The extended most attractive channel hypothesis}
\label{emac}
In a series of papers by Aoki and Bando (AB) \cite{Aoki:1983ae,Aoki:1983za,Aoki:1983yy}, it was shown that a $2n$-body multi-fermion state $(\bar{\psi}_L \psi_R)^n$ becomes more attractive as $n$ increases.  This fact can be parametrized in terms of the spin and chiral structure of multi-fermion systems \cite{Aoki:1983yy}.  We briefly review the details before addressing the experimental constraints on the DTC paradigm.

For the sake of demonstration \cite{Aoki:1983ae,Aoki:1983za,Aoki:1983yy}, we first consider a two-fermion system in a non-Abelian color gauge theory.  The potential between fermions, assuming one-gauge boson exchange, can be written as,
\begin{align}
\label{pot}
V = g^2 F(i_1 i_2 i_1^\prime i_2^\prime) \langle \lambda^a (1) \lambda^a (2) \rangle,
\end{align}
where $\lambda^a (n) $ show the generators of the gauge group $\rm SU(\rm N)$ for fermion $n$, $g$ denotes the gauge coupling constant, and $i_n^{(\prime)}$ stands for all degrees of freedom such as momentum, spin, and chirality, excluding color.

In the hypothesis of ``most attractive channel" (MAC) \cite{Raby:1979my}, the factor $F$ is common among fermions that live in different representations of the gauge group $\rm SU(N)$, and condensation can be realized only in $\psi_L \psi_L$ or $\bar{\psi}_L \psi_R$ channels.  Thus, the factor $F$  behaves trivially within the MAC framework.

In the EMAC hypothesis \cite{Aoki:1983ae,Aoki:1983za,Aoki:1983yy}, the factor $F$ exhibits a non-trivially chiral dependence through the number of fermions present in the chiral condensate.  The argument of AB goes as follows:  since the potential given in equation  \ref{pot} is attractive, any chiral condensate occurs with the scale $\mu$ satisfying the equation 
\begin{align}
- V (g^2 (\mu^2)) \sim 1,
\end{align}
where the coupling constant $g^2 (\mu^2)$ runs as,
\begin{align}
g^2 (\mu^2) = \frac{1}{\beta_0 \log \mu^2/\Lambda^2}.
\end{align}
The solution of the running coupling can be written as,
\begin{align}
\mu^2 \sim \Lambda^2 \exp (-F \langle \lambda \lambda \rangle/\beta_0).
\end{align}
We observe from this solution that the chiral difference among various channels is exponentialized due to the chiral degrees of freedom, which are parametrized inside $F$. However, the full structure of $F$ is not known.  

AB used $j=1/2$  modes as a trial wave function for all left- and right-handed fermions to determine the dependence of $F$ on the fermionic body number.  This is achieved by defining an effective Hamiltonian of a multi-fermion system, which is obtained by summing up all combinations of two-fermions.  The effective Hamiltonian can be written as the sum of the electric and magnetic parts.  For the electric part of the interaction energy, all color singlet states of the $n$-body system are degenerate. However,  it turns out that the magnetic interaction energy is attractive for  a color singlet and spin-zero system with maximum chirality  $(\bar{\psi}_R \psi_L)^{n/2}$ for even $n$.  For the case of a two-body system, the most attractive channels are $\bar{\psi}_R \psi_L$ and $\psi_L \psi_L$ for spin-zero states, which are in agreement with the MAC hypothesis.  In the case of a color singlet and spin one, the most attractive channel with maximum chirality is $\bar{\psi}_L \psi_L$ for a two-body system.  For more details, see \cite{Aoki:1983ae,Aoki:1983za,Aoki:1983yy}. 

In general, a $n$-body  color singlet and spin-zero   multi-fermion condensate  $(\bar{\psi}_R \psi_L)^{n/2}$ with maximum chirality  for even $n$ can be defined in terms of its energy as \cite{Aoki:1983ae,Aoki:1983za,Aoki:1983yy},
\bea
\bar{E} (n) = \dfrac{1}{n} E (\bar{\psi}_R^{n/2} \psi_L^{n/2}) \lesssim V_{E}^{LL} \dfrac{N^2-1}{N} - V_M^{LL} \dfrac{N-1}{N} (n+3 N+1),
\eea
where $V_E^{LL}$ and $V_M^{LL}$ show  the electric and magnetic part of the Hamiltonian of two fermions.

We observe from the above result that $\bar{E} (n)$ has a linear decrease in $n$, resulting in more attractive multi-fermion systems.  Thus,  the  larger values of $n$ result in  the hierarchical structure for the multi-fermion chiral condensations in the pattern
\be
\label{mult_cond}
\langle \bar{\psi}_R \psi_L \rangle << \langle \bar{\psi}_R  \bar{\psi}_R \psi_L  \psi_L \rangle <<  \langle \bar{\psi}_R \bar{\psi}_R  \bar{\psi}_R \psi_L  \psi_L \psi_L \rangle << \cdots.
\ee
The above series is terminated by $n_{\rm max}$, which is equal to or less than the types of fermions present in theory \cite{Aoki:1983ae,Aoki:1983za,Aoki:1983yy}.  

Thus, the dependence of $F$ on the fermionic body number $n$,  turns  out to be,
\be
F \propto \Delta \chi,
\ee
where $\Delta \chi$ is  the chirality of a multi-fermion  operator.  The hierarchy of  a chiral  multi-fermion condensate can be parametrized as  \cite{Aoki:1983yy}, 
\be 
\label{VEV_h}
\langle  ( \bar{\psi}_R \psi_L )^n \rangle \sim \left(  \Lambda \exp(k \Delta \chi) \right)^{3n},
\ee
where $k$ stands for a constant, and $\Lambda$ denotes the scale of the underlying gauge dynamics.   


\section{Experimental constraints on the DTC paradigm} 
\label{exp_dtc}
In this section, we discusse different experimental constraints on the DTC dynamics.

\subsection{Higgs mass constraint}
\label{higgs_cons}
In QCD, the lightest scalar resonance, the $\sigma$ meson, can be estimated as~\cite{delbourgo1982}
\be
\label{sigma_mass}
m_{\sigma} \approx 2 m_{\rm dyn} ,
\ee
where $m_{\rm dyn}$ is the nonperturbatively generated dynamical fermion mass. Taking $m_{\rm dyn} \approx \Lambda_{\rm QCD} \approx 250~\text{MeV}$ yields $m_\sigma \approx 500~\text{MeV}$, in good agreement with experimental determinations~\cite{pdg24}.

By analogy, a composite Higgs boson in a QCD-like TC theory is predicted as~\cite{Elias:1984zh}
\be
m_{\rm H} \approx 2 M_{\rm dyn,\,TC}.
\ee

The non-perturbative mass $M_{\rm dyn}^{\rm TC}$ is related to the technifermion condensate via~\cite{Elias:1984zh}
\be 
\label{mdyc}
- \langle \bar{T}T \rangle  = \frac{\rm N_{\rm TC}}{4\pi^2} M_{\rm dyn,\,TC}^{3} \, \alpha_{\rm TC}(\mu^2).
\ee
Using Eqs.~\eqref{chi_con} and \eqref{mdyc}, the Higgs mass can be expressed as
\be
\label{higgs_mass_general}
m_{\rm H} \approx 2 \Lambda_{\rm TC} \, e^{\,k_{\rm TC} \, \Delta \chi_{\rm TC}} .
\ee
For $k_{\rm TC} = 0$, the exponential factor reduces to unity, giving
\be
\label{higgs_mass}
m_{\rm H} \approx 2 \Lambda_{\rm TC}.
\ee
This case corresponds to $F=0$, which means that the TC dynamics is governed solely by the MAC hypothesis, with the EMAC hypothesis playing no role in the mass spectrum. Under the MAC hypothesis, the most attractive channel for the lowest-lying scalars (S-wave) is either $\psi_L \psi_L$ or $\bar{\psi}_L \psi_R$. Therefore, we may identify the SM Higgs as a composite state of the form $\bar{\psi}_L \psi_R$.

By scaling up two-flavor QCD, the mass of the lightest scalar singlet in the TC type theories is estimated to lie in the 
range $1.0~{\rm TeV}\lesssim M_{\rm dyn,\,TC} \lesssim 1.4~{\rm TeV}$, which is higher than the value 
suggested by experiment~\cite{Foadi:2012bb}. However, Foadi, Frandsen, and Sannino (FFS) showed 
that technicolor (TC) dynamics can still accommodate a TC Higgs with a physical mass of 
$125~{\rm GeV}$, either with or without walking effects~\cite{Foadi:2012bb}. Their analysis 
demonstrated that SM top-quark radiative corrections naturally lower the TC Higgs 
mass toward the observed value~\cite{atlas,cms}.

After including the SM top-quark radiative corrections, the physical Higgs mass is given by\cite{Foadi:2012bb},

\begin{align}
    m_{\rm H}^2 = M_{\rm dyn,\,TC}^2 - 12 \kappa^2 r_t^2 m_t^2,
\end{align}
where $r_t =1$ will provide the SM-like Yukawa coupling of the top quark, and $\kappa$ is a number of order one.  For more details, see Ref. \cite{Foadi:2012bb}.

FFS showed that for fermions in the fundamental representation of $\rm SU(\rm N_{\rm TC})$, the scale $1.0~{\rm TeV}\lesssim M_{\rm dyn,\,TC} \lesssim 1.4~{\rm TeV}$ can easily recovered for one technic-doublet.  Therefore, we use $\Lambda_{\rm TC} = M_{\rm dyn,\,TC} = 1$ TeV in this work.

\subsection{S-parameter}
The dynamics of TC theories is tightly constrained by the electroweak oblique parameters~\cite{Peskin:1990zt,Peskin:1991sw}, with the $S$ parameter being particularly sensitive to the underlying strong dynamics. Current experimental determinations of $S$ and $T$ read~\cite{pdg24}:
\begin{equation}  
S = -0.04 \pm 0.10, 
\qquad 
T = 0.01 \pm 0.12.
\end{equation}

For strongly-coupled scenarios, next-to-leading order (NLO) expressions for these parameters have been obtained in Refs.~\cite{Pich:2025ywr,Pich:2013fea}. The $S$ parameter at NLO can be written as
\begin{align}
S_{\mathrm{NLO}}  
&> \frac{4 \pi v^{2}}{M_{V}^{r}} 
+ \left.\Delta S_{\mathrm{NLO}}\right|_{\varphi \varphi, h \varphi} 
+ \left.\Delta S_{\mathrm{NLO}}\right|_{\psi \bar{\psi}},
\label{S_par}
\end{align}
with
\begin{align}
\left.\Delta S_{\mathrm{NLO}}\right|_{\varphi \varphi, h \varphi}
&= \frac{1}{12 \pi} \bigg[
\left(1-\kappa_{W}^{2}\right)\left(\log \frac{M_{V}^{2}}{m_{h}^{2}} - \frac{11}{6}\right)
- \kappa_{W}^{2}\left(\log \frac{M_{A}^{2}}{M_{V}^{2}} - 1 + \frac{M_{A}^{2}}{M_{V}^{2}}\right)
\bigg],
\label{delS_par}
\end{align}
\begin{align}
\left.\Delta S_{\mathrm{NLO}}\right|_{\psi \bar{\psi}} 
&= -\frac{F_{V}^{2}\left(C_{0}^{V_{3}^{1}}\right)^{2}}{3 \pi M_{V}^{2}}
\left(1 - \frac{M_{V}^{2}}{M_{A}^{2}} + \log \frac{M_{A}^{2}}{M_{V}^{2}}\right),
\end{align}
and
\begin{equation*}
\left.T\right|_{\varphi \varphi, h \varphi} 
= \frac{3}{16 \pi \cos^{2} \theta_{W}}
\left[
(1-\kappa_{W}^{2})\left(1-\log \frac{M_{V}^{2}}{m_{h}^{2}}\right) 
+ \kappa_{W}^{2} \log \frac{M_{A}^{2}}{M_{V}^{2}}
\right].
\end{equation*}
Here, only the first Weinberg sum rule is assumed, and $\kappa_W = M_V^2 / M_A^2$ parametrizes the coupling of the lightest scalar (the Higgs boson) to two electroweak gauge bosons ($W^+W^-$ or $ZZ$).  

For $\kappa_W = 1.023 \pm 0.026$, the bounds on vector resonances in a QCD-like framework are~\cite{Pich:2025ywr}:
\begin{equation}
\label{s_bound}
M_A \geq M_V \geq 2~\text{TeV} 
\qquad (95\%~\text{C.L.}).
\end{equation}

\subsubsection{Phenomenological interpretation in the DTC framework}  
\label{s-pheno}
A natural question arises: why must the lightest vector TC meson satisfy $m_{\rho_{\mathrm{TC}}} \gtrsim 2~\mathrm{TeV}$, despite the presence of a comparatively light Higgs boson at $125~\mathrm{GeV}$?   

In our framework, the TC spectrum is governed by the MAC hypothesis ($k_{\rm TC}=0$), whereas the EMAC ($k_{\rm TC}\neq 0$) plays no role. Under this assumption, the constant $F$ is universal across all TC fermions, with condensation occurring in $\psi_L \psi_L$ or $\bar{\psi}_L \psi_R$ channels. Condensates in alternative channels, particularly vector ones such as $\bar{\psi}_L \psi_R$ and $\bar{\psi}_L \psi_L$, are disfavored. The only constraint on $F$ is that it remains positive.  

Although the TC sector exhibits QCD-like gauge dynamics, its spectrum need not be a simple rescaling of QCD. Since QCD with $N_c=3$ is not strictly in the large-$ N_c$ limit, it is crucial to clarify what we mean by ``large-$ N_c$." A more robust definition is provided by large-$ N_c$ scaling: for fixed fermion flavor number $N_f$, the ratio of the lightest resonance mass $M$ to the pion decay constant $f_\pi$ scales as  
\begin{equation}
\frac{M}{f_\pi} \propto \frac{1}{\sqrt{\rm N_{\mathrm{TC}}}}, 
\label{eq:MFpi-scaling}
\end{equation}
as $\rm N_{\mathrm{TC}}\to\infty$~\cite{tHooft:1973alw,Witten:1979kh}. Consequently, decreasing $\rm N_{\mathrm{TC}}$ typically increases $M/f_\pi$, a trend also supported by lattice studies~\cite{Lucini:2012gg}. We therefore define large-$\rm N_{\mathrm{TC}}$ as the regime in which this $1/\sqrt{\rm N_{\mathrm{TC}}}$ scaling is manifested.  

As emphasized in Ref.~\cite{Chivukula:1992nw}, the ratio $M/f_\pi$ is constrained  by,  
\begin{equation}
\frac{M}{f_\pi} \leq \frac{4\pi}{\sqrt{N_f}}.
\label{eq:NDA-bound}
\end{equation}
Once this upper bound is saturated, further decreasing $\rm N_{\mathrm{TC}}$ no longer increases $M/f_\pi$.  

\begin{figure}[H]
\centering
\includegraphics[width=0.5\linewidth]{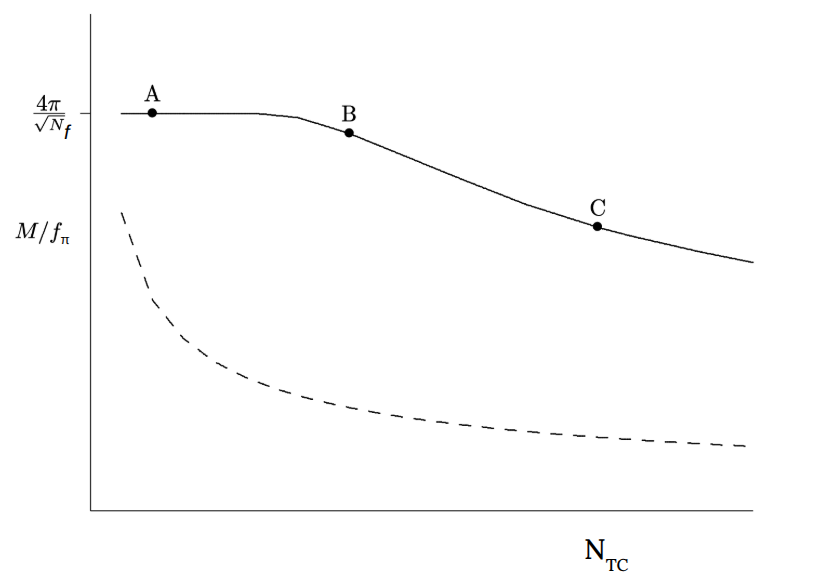}
\caption{Possible behaviors of $M/f_\pi$, with $f_\pi=F_{\Pi_{\rm TC}}$, in an $\rm SU(N_{\rm TC})$ gauge theory at fixed fermion flavor number $N_f$. The solid line shows $M/f_\pi \propto 1/\sqrt{\rm N_{\rm TC}}$ at large $\rm N_{\rm TC}$, saturating the bound in Eq.~\eqref{eq:NDA-bound} at intermediate $\rm N_{\rm TC}$, and eventually losing asymptotic freedom at very small $\rm N_{\rm TC}$. The dashed line represents a monotonic decrease without saturation. Adapted from Ref.~\cite{Chivukula:1992nw}.}
\label{fig_M_NTC}
\end{figure}

Two scenarios are possible which are illustrated in Fig.~\ref{fig_M_NTC}. The solid line corresponds to $M/f_\pi \propto 1/\sqrt{\rm N_{\mathrm{TC}}}$ at large $\rm N_{\mathrm{TC}}$, followed by a plateau where Eq.~\eqref{eq:NDA-bound} is saturated, and finally terminating when asymptotic freedom is lost at very small $\rm N_{\mathrm{TC}}$. The dashed line, in contrast, depicts a monotonic decrease without saturating the bound.  

In QCD, with two light flavors and $N_c=3$, the scale  
\begin{equation}
\frac{4\pi f_\pi}{\sqrt{N_f}} \simeq 825~\mathrm{MeV},
\end{equation}
is numerically close to the $\rho(770)$ mass, suggesting saturation of the bound. Thus, QCD lies on the solid curve, with $N_c=3$ near point A or B rather than C. If QCD corresponds to A, the $1/\sqrt{N_c}$ scaling emerges only for $N_c\gg3$.  

For our TC model, we find $\Lambda_{\rm TC}=10^3~\mathrm{GeV}$ and $f_\pi=F_{\Pi_{\rm TC}}=246~\mathrm{GeV}$ (for two flavors). The corresponding bound is  
\begin{equation}
\frac{4\pi F_{\Pi_{\rm TC}}}{\sqrt{N_f}} \simeq 2186~\mathrm{GeV}.
\end{equation}
We conjecture that in this model the $1/\sqrt{\rm N_{\rm TC}}$ scaling becomes effective only for $\rm N_{\rm TC} > 3$, such that the $\rho_{\rm TC}$ mass already saturates the upper bound~\eqref{eq:NDA-bound}. Accordingly, our TC dynamics follow the solid-curve scenario, with $\rm N_{\rm TC} =  3$ placing the theory near A or B, while remaining consistent with the $S$-parameter bound of Eq.~\eqref{s_bound}. This conjecture is further supported by recent lattice computations.  

Indeed, Ref.~\cite{Nogradi:2019iek} finds that for $N_c=3$ the ratio $M_\rho/F_\Pi$ in the chiral limit is essentially independent of $N_f$:  
\begin{equation}
\frac{M_\rho}{F_\Pi} \Bigg \rvert^{N_f=2-6}_{N_c=3} = 7.95(15).
\end{equation}

At large $N_c$, quenched lattice studies instead find~\cite{Bali:2013kia}  
\begin{equation}
\sqrt{\frac{N_c}{3}} \,\frac{M_\rho}{F_\Pi} \Bigg \rvert_{N_c\to \infty} = 7.08(10).
\end{equation}

For $\rm N_{\rm TC} = 3$~\cite{Bali:2013kia},  
\begin{equation}
\frac{M_{\rho_{\rm TC}}}{\sqrt{\sigma}}  = 1.749(26),
\end{equation}
where the string tension $\sqrt{\sigma}$ is related to the decay constant via~\cite{Bali:2013kia}  
\begin{equation}
\sqrt{\frac{3}{\rm N_{\rm TC}}} \,\frac{F_{\Pi}}{\sqrt{\sigma}} = 0.2174 (30).
\end{equation}

Applying these relations to our model with $\rm N_{\rm TC} = 3$ and $F_{\Pi_{\rm TC}} = 246~\mathrm{GeV}$ yields  
\begin{equation}\label{eq:mass_rho}
M_{\rho_{\rm TC}} = 1980~\mathrm{GeV},
\end{equation} 
which is reasonably consistent with the bound in Eq.~\eqref{s_bound}. Using the simple scaling relation \cite{Tandean:1995ci},
\begin{equation}
   M_{\rho_{\rm TC}} = \frac{F_{\Pi_{\rm TC}}}{f_{\pi}} \sqrt{\frac{3}{\rm N_{\rm TC}}} m_{\rho}=2007 \mathrm{GeV},
\end{equation}
  which is in agreement with Eq. \eqref{eq:mass_rho}. We have used $m_\rho=775.26$ MeV \cite{pdg24}.
  
In addition, these vector mesons can naturally be made even heavier by gauging them under a local symmetry which may be broken at a high scale, such as through the dark-QCD dynamics. For instance, we can gauge the three techni-rho $\rho_{\rm TC}$ vector mesons under a local $\rm SU(2)_{R}$ symmetry which is broken by the  dark-QCD dynamics through the condensate formed by the $\rm SU(2)_{R}$ doublets transforming under $\rm SU(3)_c \times \rm SU(2)_L \times \rm SU(2)_R \times \rm U(1)_Y \times \mathcal{G}$ as
\begin{eqnarray}
F_R&\equiv&  
\begin{pmatrix} U \\ D \end{pmatrix}_R 
:(3,1, 2, \dfrac{1}{3},1,1,\rm N_{\rm D}),
\quad
U_{L}:  (3,1, 1, \dfrac{4}{3},1,1,\rm N_{\rm D}),
\quad
D_{L} : (3,1,1, -\dfrac{2}{3},1,1,\rm N_{\rm D}), \\ \nonumber 
F_L&\equiv&  
\begin{pmatrix} U \\ D \end{pmatrix}_L 
:(3,1, 2, \dfrac{1}{3},1,1,\rm N_{\rm D}),
\quad
U_{R} :  (3,1, 1, \dfrac{4}{3},1,1,\rm N_{\rm D}),
\quad
D_{R} : (3,1,1, -\dfrac{2}{3},1,1,\rm N_{\rm D}).
\end{eqnarray} 

\begin{figure}[H]
	\centering
 \includegraphics[width=0.45\linewidth]{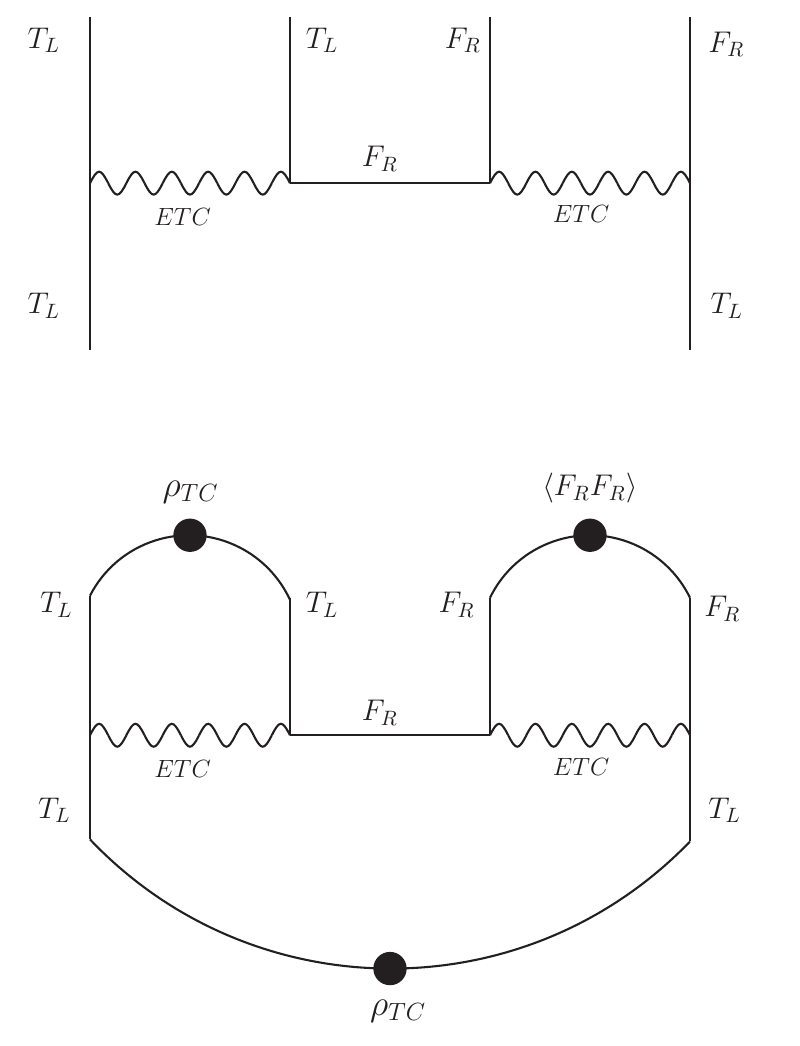}
    \caption{The mass generation of the techni-rho meson from the DQCD dynamics.  The blob denotes either the formation of a meson or a condensate. }
 \label{rhotc_mass}	
 \end{figure}

The mass generation of the $\rho_{\rm TC}$ meson is  schematically depicted in figure \ref{rhotc_mass}.  This setup is similar to the left-right symmetric models~\cite{Mohapatra:1974hk,Senjanovic:1975rk}, and identical to models discussed in references \cite{Abbas:2017vle,Abbas:2017hzw}. The  contribution to the mass of  $\rho_{\rm TC}$ is of the order,
\begin{align}
    M_{\rho_{\rm TC}} & \approx g_R \sqrt{N_d} F_{\Pi_{\rm DQCD}},
\end{align}
where $g_R$ is the coupling constant corresponding to the symmetry $\rm SU(2)_{R}$, and $N_d$ are the number of $\rm SU(2)_{R}$ doublets.  Thus, if $F_{\Pi_{\rm DQCD}}$ is sufficiently large, the masses of the $\rho_{\rm TC}$ vector mesons become naturally large.  The interactions of $\rho_{\rm TC}$ to the SM fermions will be discussed in \ref{scaling}.

We notice that for  satisfying  the bound of Eq.~\eqref{s_bound} requires $\rm N_{\rm TC}\leq 3$.  
In this work we choose $\rm N_{\rm TC}=3$.  
This choice also avoids the additional Goldstone bosons present for $\rm N_{\rm TC}=2$,\footnote{
$\rm N_{\rm TC}=2$ corresponds to a pseudoreal representation.}
but it does introduce a Witten global anomaly. Avoiding this anomaly requires a minimal and
well-motivated modification of the DTC paradigm. Therefore, we extend the TC fermionic content by adding a leptonic doublet which is a singlet under the TC symmetry,
\begin{eqnarray}
T_L &\equiv&  
\begin{pmatrix} T \\ B \end{pmatrix}_L 
:(1,2,Y,\rm N_{\rm TC},1,1),
\quad
T_{R} : (1,1,Y+1,\rm N_{\rm TC},1,1),
\quad
B_{R} : (1,1,Y-1,\rm N_{\rm TC},1,1), \\ \nonumber
L_L &\equiv&  
\begin{pmatrix} N \\ E \end{pmatrix}_L 
:(1,2,Y,1,1,1),
\quad
N_{R} : (1,1,Y+1,1,1,1),
\quad
E_{R} : (1,1,Y-1,1,1,1).
\end{eqnarray}

For creating mass of the  heavy leptons $L$, we assume that its mass is generated similar to top quark through the operator of the form given in Eq. \ref{mass2} in the next section,
\bea
{\mathcal{L}} &=& \dfrac{1}{\Lambda }\Bigl[  y^L  \bar{L}_{L}  \tilde{\varphi} L_{R}   \chi_r \Bigr]  
+  {\rm H.c.},
\eea
where $\chi_r$ provides a heavy  multi-fermion-condensate in the form of a large VEV.

\section{Effective low energy limits of the DTC paradigm }
\label{sec4}
\begin{figure}[H]
\centering
 \includegraphics[width=0.66\linewidth]{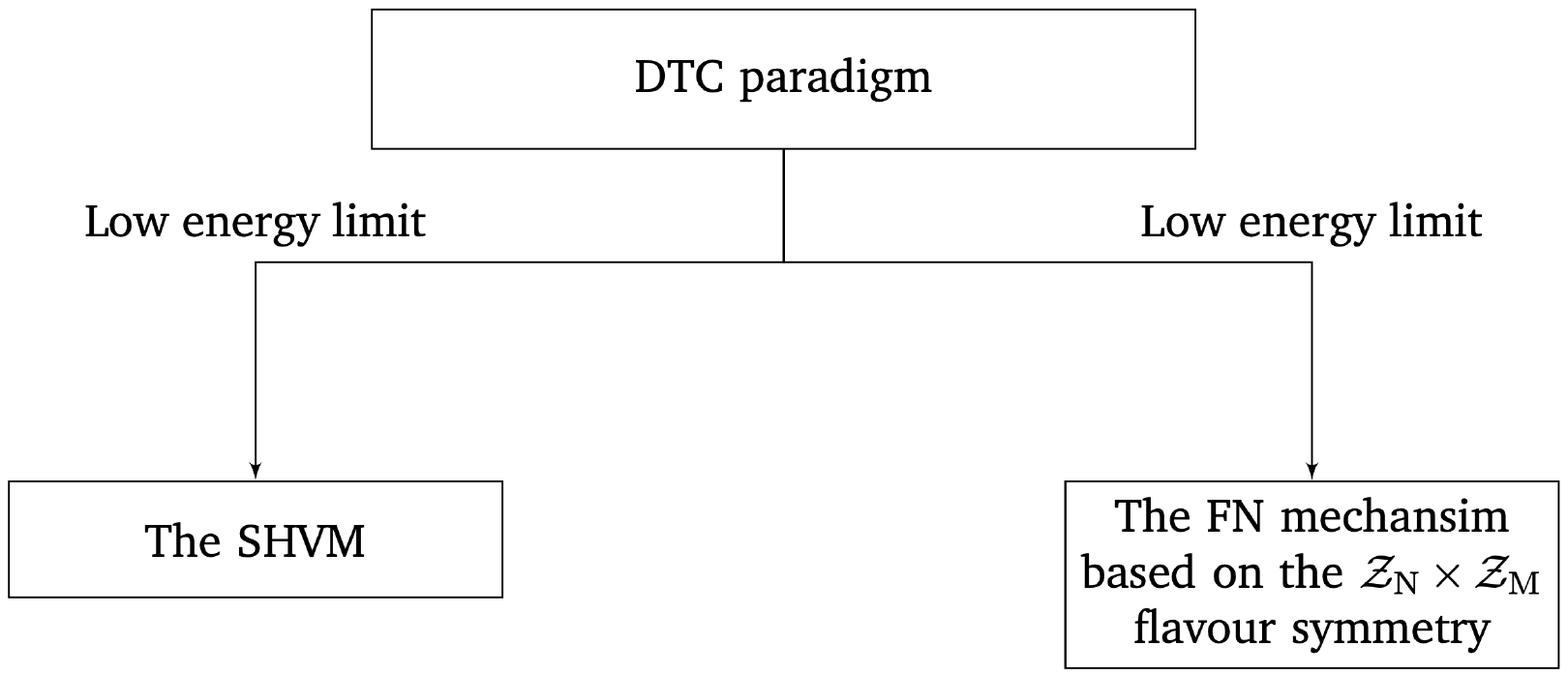}
    \caption{ At low energies, the DTC paradigm may effectively reduce to either the SHVM or the FN mechanism.}
 \label{fig:effective_limits}	
 \end{figure}

The DTC  paradigm  can be mapped onto the SHVM and the FN mechanism based on the  $\mathcal{Z}_{\rm N} \times \mathcal{Z}_{\rm M}$ symmetry at low energies, as shown in figure \ref{fig:effective_limits}.  In this section, we discuss how to achieve their effective low-energy manifestations.

\subsection{Standard HVM }
One possible low-energy realization of the DTC paradigm is the SHVM, 
in which the flavor problem is addressed by introducing six gauge-singlet scalar fields, 
$\chi_r$ ($r=1,\dots,6$), coupled to the charged fermions, and a seventh singlet scalar field, $\chi_7$, 
which couples exclusively to the neutrino sector~\cite{Abbas:2017vws,Abbas:2020frs,Abbas:2023dpf,Abbas:2024jut}. 
In the SHVM framework, the scalar fields $\chi_r$ are interpreted as multi-fermion chiral condensates 
arising from the underlying strong dynamics of the DTC paradigm. 
In this section, we briefly review the details of a specific SHVM realization that is free from 
large FCNC effects, as discussed in Ref.~\cite{Abbas:2024jut}.

The gauge singlet scalar  fields  $\chi_r $  transform under the SM symmetry  $\mathcal{G}_{\rm SM} \equiv \rm SU(3)_c \times SU(2)_L \times U(1)_Y$ as,
\begin{eqnarray}
\chi_r :(1,1,0),
 \end{eqnarray} 
where $r=1-6$.

The masses of the charged fermions originate from the following  dimension-5 Lagrangian,
\bea
\label{mass2}
{\mathcal{L}} &=& \dfrac{1}{\Lambda }\Bigl[  y_{ij}^u  \bar{\psi}_{L_i}^{q}  \tilde{\varphi} \psi_{R_i}^{u}   \chi _r +     
   y_{ij}^d  \bar{\psi}_{L_i}^{q}   \varphi \psi_{R_i}^{d}  \chi _{r}   +   y_{ij}^\ell  \bar{\psi}_{L_i}^{\ell}   \varphi \psi_{R_i}^{\ell}  \chi _{r} \Bigr]  
+  {\rm H.c.},
\eea
$i$ and $j$   stand for family indices,  $ \psi_{L}^q,  \psi_{L}^\ell  $ denote  the  quark and leptonic doublets,  $ \psi_{R}^u,  \psi_{R}^d, \psi_{R}^\ell$ are the right-handed up,  down-type  quarks and  leptons,  $\varphi$ and $ \tilde{\varphi}= -i \sigma_2 \varphi^* $  show the SM Higgs field, and its conjugate, where  $\sigma_2$ is  the second Pauli matrix. 

The charged fermion mass pattern and quark mixing is obtained  by assigning the following generic charges  under the   $\mathcal{Z}_{\rm N} \times \mathcal{Z}_{\rm M} \times \mathcal{Z}_{\rm P}$ flavor symmetry,
\begin{align}
\psi_{L_1}^{q} &: (+, 1, \omega^{\rm P-3
}_{14}),~ \psi_{L_2}^{q}: (+, 1, \omega^{9}_{14}),~  \psi_{L_3}^{q}: (-, 1, \omega^{ 8}_{14}), \\ \nonumber
u_{R} &: (-, \omega_4, \omega^{\rm P-5}_{14}),~ c_{R}: (+, 1, \omega^4_{14}),~  t_{R}: (+, 1, 1), \\ \nonumber
d_{R} &: (-, \omega_4, \omega^{\rm 9}_{14}),~ s_{R}: (+, 1, \omega^{12}_{14}),~  b_{R}: (-, 1, \omega_{14}^2), \\ \nonumber
\psi_{L_1}^{\ell} &: (+, \omega^{3}_{4}, \omega^{12}_{14}),~ \psi_{L_2}^{\ell}: (+, \omega^{3}_4, \omega^{10}_{14}),~  \psi_{L_3}^{\ell}: (+, \omega^{3}_4, \omega^{6}_{14}), \\ \nonumber
e_{R} &: (-, 1, \omega^{10}_{14}),~ \mu_{R}: (+, \omega^{3}_4, \omega^{13}_{14}),~  \tau_{R}: (+, \omega^{3}_4, \omega_{14}), \\ \nonumber
\nu_{e_{R}} &: (+, 1, \omega^{8}_{14}),~ \nu_{\mu_{R}}: (-, \omega_4, \omega^{3}_{14}),~  \nu_{\tau_{R}}: (-, \omega_4, \omega_{14}^3), \\ \nonumber
\chi_1 &: (-, \omega^3_4, \omega^{2}_{14}),~ \chi_2: (+, 1, \omega^{5}_{14}),~  \chi_3: (-, 1, \omega^8_{14}), \\ \nonumber
\chi_4 &: (+, 1, \omega^{13}_{14}),~ \chi_5: (+, 1, \omega^{11}_{14}),~  \chi_6: (+, 1, \omega^{6}_{14}), 
\end{align}
where $\omega_4$ denotes the fourth  and  $\omega_{14}$ is the fourteenth root of unity corresponding to the symmetries $\mathcal{Z}_4$  and $\mathcal{Z}_{14}$,  respectively.  Moreover, we need  $\rm N= 2$, $\rm M\geq 4$,  and $\rm P \geq 14$ for producing the charged flavor pattern. We observe that  the desired flavor structure of the charged fermions requires $\rm N= 2$.

We recover the  neutrino masses  by adding   three right-handed  neutrinos $\nu_{eR}$, $\nu_{\mu R}$, $\nu_{\tau R}$  and the singlet scalar field $\chi_7$ to the SM, and by writing the dimension-6 operators as,
\begin{eqnarray}
\label{mass_N}
-{\mathcal{L}}_{\rm Yukawa}^{\nu} &=&      y_{ij}^\nu \bar{ \psi}_{L_i}^\ell   \tilde{\varphi}  \nu_{f_R} \left[  \dfrac{ \chi_r \chi_7 (\text{or}~ \chi_r  \chi_7^\dagger)}{\Lambda^2} \right] +  {\rm H.c.}. 
\end{eqnarray}

It is remarkable that to produce the normal ordered neutrino masses and the observables of leptonic mixing, we must have $\rm P = 14$ for the symmetry $\mathcal{Z}_{\rm P}$, which makes $\rm P = 14$ a magic number.  For instance, we assign charges to different fermionic and scalar fields under the $\mathcal{Z}_2 \times \mathcal{Z}_4 \times \mathcal{Z}_{14} $ flavor symmetry as shown in table \ref{tab1}. 

 \begin{table}[ht]
\begin{center}
\begin{tabular}{|c|c|c|c||c|c|c|c||c|c|c|c||c|c|c|c|}
  \hline
  Fields                               &   $\mathcal{Z}_2$  &  $\mathcal{Z}_4$   &  $\mathcal{Z}_{14}$   & Fields   &  $\mathcal{Z}_2$   &  $\mathcal{Z}_4$ &  $\mathcal{Z}_{14}$ & Fields   & $\mathcal{Z}_2$  & $\mathcal{Z}_4$  &  $\mathcal{Z}_{14}$  & Fields  &  $\mathcal{Z}_2$   &  $\mathcal{Z}_4$  &  $\mathcal{Z}_{14}$ \\
  \hline
 $u_{R}$                        &     -   &     $\omega_4$          &    $\omega^{9}_{14}$ & $d_{R}$    &     -    & $\omega_4$        &    $\omega^{9}_{14}$  & $ \psi_{L_3}^{q} $       &    -     &  $1$    &   $\omega^8_{14}$   &  $\tau_R$      &   +  &  $\omega^3_{4}$      &     $\omega_{14}$             \\
  $c_{R}$                       &     +   &    $1$          &    $\omega^4_{14}$   &  $s_R$                         &      +  &  $1$      &  $\omega^{12}_{14}$  &  $ \psi_{L_1}^\ell $                          &     +   & $\omega^3_4$     &  $\omega^{12}_{14}$                          &   $\nu_{e_R}$   &    +   &   $1$     &      $\omega^{8}_{14}$        \\
   $t_{R}$                        &     +   &    $1$         &    $1$   & $b_R$                         &      -  &  $1$    &  $\omega^{2}_{14}$  &  $ \psi_{L_2}^{\ell} $     &      +  & $\omega^3_4$      &  $\omega^{10}_{14}$     & $\nu_{\mu_R}$                   &     -  &   $\omega_4$    &   $\omega^3_{14}$                    \\
  $\chi _1$                        &      -  &   $\omega^3_4$      &    $\omega^2_{14}$    &   $\chi _4$                          &      +  &  $1$      &   $ \omega^{13}_{14}$    &   $ \psi_{L_3}^{ \ell} $       &    +     &  $\omega^3_4$    &   $\omega^6_{14}$                                 & $\nu_{\tau_R}$                    &     -   &  $\omega_4$     &   $\omega^3_{14}$          \\
  $\chi _2$                   & +     &       $1$      &  $\omega^5_{14}$   & $ \psi_{L_1}^q $                          &      +  &  $1$      &  $\omega^{11}_{14}$  & $e_R$    &      -   &   $1$       &    $\omega^{10}_{14}$      &  $\chi_7 $                          &      -   &  $\omega^2_4$   &     $\omega^8_{14}$                                               \\
   $\chi _3$                  &    -   &       $1$    & $ \omega^8_{14}$        & $ \psi_{L_2}^{q} $     &      +  & $1 $      &  $\omega^{9}_{14}$  &   $ \mu_R$     &   +  & $\omega^3_4$       &     $\omega^{13}_{14}$      &  $ \varphi $                           &      +  &1     &   1                                            \\
    $\chi _5$                  &    +  &       $1$    & $ \omega^{11}_{14}$        &  $\chi _6$         &      +  & $1 $      &  $\omega^{6}_{14}$  &       &     &       &         &                           &       &     &                                             \\
  \hline
\end{tabular}
\end{center}
\caption{The transformation charges of left- and right-handed fermions, as well as scalar fields, under the $\mathcal{Z}_2$, $\mathcal{Z}_4$, and $\mathcal{Z}_{14}$ symmetries for the normal mass ordering are presented. Here, $\omega_4$ and $\omega_{14}$ represent the fourth and fourteenth roots of unity associated with the $\mathcal{Z}_4$ and $\mathcal{Z}_{14}$ symmetries, respectively.}
 \label{tab1}
\end{table} 

The masses of charged fermions are now produced by the Lagrangian,
\begin{align}
\label{mass22}
{\mathcal{L}_{f}}  =& \dfrac{1}{\Lambda }\Bigl[  y_{11}^u  \bar{\psi}_{L_1}^{q}  \tilde{\varphi} \psi_{R_1}^{u}   \chi _1 +  y_{13}^u  \bar{\psi}_{L_1}^{q}  \tilde{\varphi} \psi_{R_3}^{u}   \chi _5  +  y_{22}^u  \bar{\psi}_{L_2}^{q}  \tilde{\varphi} \psi_{R_2}^{u}   \chi_2  +  y_{23}^u  \bar{\psi}_{L_2}^{q}  \tilde{\varphi} \psi_{R_3}^{u}   \chi_2^\dagger  
  +  y_{33}^u  \bar{\psi}_{L_3}^{q}  \tilde{\varphi} \psi_{R_3}^{u}   \chi_3 \\ \nonumber
  & +   y_{11}^d  \bar{\psi}_{L_1}^{q}   \varphi \psi_{R_1}^{d}  \chi_{1} +     
   y_{12}^d  \bar{\psi}_{L_1}^{q}   \varphi \psi_{R_2}^{d}  \chi_{4}           
    + y_{22}^d  \bar{\psi}_{L_2}^{q}   \varphi \psi_{R_2}^{d}  \chi_{5}  + y_{33}^d  \bar{\psi}_{L_3}^{q}   \varphi \psi_{R_3}^{d}  \chi_{6}  \\ \nonumber
   & +   y_{11}^\ell  \bar{\psi}_{L_1}^{\ell}   \varphi \psi_{R_1}^{\ell}  \chi _{1}  +   y_{12}^\ell  \bar{\psi}_{L_1}^{\ell}   \varphi \psi_{R_2}^{\ell}  \chi _{4}  +   y_{13}^\ell  \bar{\psi}_{L_1}^{\ell}   \varphi \psi_{R_3}^{\ell}  \chi _{5}  +   y_{22}^\ell  \bar{\psi}_{L_2}^{\ell}   \varphi \psi_{R_2}^{\ell}  \chi _{5}  +   y_{23}^\ell  \bar{\psi}_{L_2}^{\ell}   \varphi \psi_{R_3}^{\ell}  \chi _{2}^\dagger  \\ \nonumber
& +   y_{33}^\ell  \bar{\psi}_{L_3}^{\ell}   \varphi \psi_{R_3}^{\ell}  \chi _{2} +  {\rm H.c.} \Bigr].
\end{align}

The  fermionic mass pattern can be explained  in terms of  the   VEVs pattern    $ \langle \chi _4 \rangle > \langle \chi _1 \rangle $, $ \langle \chi _2 \rangle >> \langle \chi _5 \rangle $, $ \langle \chi _3 \rangle >> \langle \chi _6 \rangle $, $ \langle \chi _{3} \rangle >> \langle \chi _{2} \rangle >> \langle \chi _{1} \rangle $, and  $ \langle \chi _6 \rangle >> \langle \chi _5 \rangle >> \langle \chi _4 \rangle $.

The mass matrices of up, down-type quarks and leptons read as,
\begin{align}
\label{mUD}
\M_\U & =   \dfrac{ v }{\sqrt{2}} 
\begin{pmatrix}
y_{11}^u  \epsilon_1 &  0  & y_{13}^u  \epsilon_{5}    \\
0    & y_{22}^u \epsilon_{2}  &  y_{23}^u  \epsilon_{2}   \\
0   &  0    &  y_{33}^u  \epsilon_{3} 
\end{pmatrix},  
\M_\D = \dfrac{ v }{\sqrt{2}} 
 \begin{pmatrix}
  y_{11}^d \epsilon_{1} &    y_{12}^d \epsilon_{4} &  0 \\
0 &     y_{22}^d \epsilon_{5} &  0\\
  0 &    0  &   y_{33}^d \epsilon_{6}\\
\end{pmatrix},  
\M_\ell  =\dfrac{ v }{\sqrt{2}} 
  \begin{pmatrix}
  y_{11}^\ell \epsilon_1 &    y_{12}^\ell \epsilon_4  &   y_{13}^\ell \epsilon_5 \\
 0 &    y_{22}^\ell \epsilon_5 &   y_{23}^\ell \epsilon_2\\
   0  &    0  &   y_{33}^\ell \epsilon_2 \\
\end{pmatrix},
\end{align} 
where $\epsilon_r = \dfrac{\langle \chi_r \rangle }{\Lambda}$ and  $\epsilon_r<1$.  

The masses of charged fermions can be written as,
\begin{eqnarray}
\label{mass1a}
m_t  &=& \ \left|y^u_{33} \right| \epsilon_{3} v/\sqrt{2}, ~
m_c  = \   \left|y^u_{22} \epsilon_{2} \right|  v /\sqrt{2} ,~
m_u  =  |y_{11}^u  |\,  \epsilon_1 v /\sqrt{2},\nonumber \\
m_b  &\approx& \ |y^d_{33}| \epsilon_{6} v/\sqrt{2}, 
m_s  \approx \   \left|y^d_{22}  \right| \epsilon_{5} v /\sqrt{2},
m_d  \approx  \left|y_{11}^d    \right|\,  \epsilon_{1} v /\sqrt{2},\nonumber \\
m_\tau  &\approx& \ |y^\ell_{33}| \epsilon_{2} v/\sqrt{2}, ~
m_\mu  \approx \   |y^\ell_{22} | \epsilon_{5} v /\sqrt{2} ,~
m_e  =  |y_{11}^\ell   |\,  \epsilon_1 v /\sqrt{2}.
\end {eqnarray}

The quark mixing angles are given by,
\begin{eqnarray}
\sin \theta_{12}  & \simeq&   \left|\frac{ y_{12}^d}{ y_{22}^d} \right| \frac{ \epsilon_{4}}{ \epsilon_{5}}, ~ 
\sin \theta_{23}  \simeq   \left|\frac{ y_{23}^u}{ y_{33}^u} \right| \frac{ \epsilon_{2}}{ \epsilon_{3}}, 
\sin \theta_{13}  \simeq    \left|\frac{ y_{13}^u}{ y_{33}^u} \right| \frac{ \epsilon_{5}}{ \epsilon_{3}}.
\end{eqnarray}

In general, the $\epsilon_r$  parameters are \cite{Abbas:2023dpf},
\begin{equation}
\label{epsi}
\epsilon_1 = 3.16 \times 10^{-6},~ \epsilon_2 = 0.0031,~ \epsilon_3 = 0.87,~\epsilon_4 = 0.000061,~\epsilon_5 = 0.000270,~\epsilon_6 = 0.0054,~\epsilon_7 = 7.18 \times 10^{-10}.  
\end{equation}

The  SHVM allows only Dirac-type neutrinos.  The mass matrix for neutrinos is given by,
\begin{equation}
\label{NM1}
\M_{\N} = \dfrac{v}{\sqrt{2}}  
\begin{pmatrix}
y_{11}^\nu   \epsilon_1 \epsilon_7   &  y_{12}^\nu   \epsilon_4 \epsilon_7  & y_{13}^\nu  \epsilon_4  \epsilon_7 \\
0   & y_{22}^\nu  \epsilon_4  \epsilon_7 &  y_{23}^\nu  \epsilon_4  \epsilon_7 \\
0   &   y_{32}^\nu  \epsilon_5  \epsilon_7   &  y_{33}^\nu  \epsilon_5  \epsilon_7
\end{pmatrix}.
\end{equation}

The neutrino masses can be written as,
\begin{eqnarray}
\label{mass_neutrino}
m_3  &\approx&  |y^\nu_{33}|  \epsilon_5 \epsilon_7 v/\sqrt{2}, 
m_2  \approx     |y^\nu_{22} - \dfrac{y_{23}^\nu  y_{32}^\nu}{y_{33}^\nu} |  \epsilon_4 \epsilon_7 v /\sqrt{2},
m_1  \approx  |y_{11}^\nu  |\,  \epsilon_1 \epsilon_7 v /\sqrt{2}, \quad \quad.
\end {eqnarray}
The masses of neutrinos are of the order $\{m_3,m_2,m_1\} = \{5.05 \times 10^{-2}, 8.67 \times 10^{-3}, 2.67 \times 10^{-4} \}\, \text{eV}$ \cite{Abbas:2023dpf}.

The  neutrino mixing angles are the main predictions of the SHVM, and are given by,
\begin{eqnarray}
\sin \theta_{12}^\ell  &\simeq& \left|{y_{12}^\ell \epsilon_4  \over y_{22}^\ell  \epsilon_5}-{y_{12}^\nu  \over y_{22}^\nu }   +  {y_{23}^{\ell *} y_{13}^\nu \epsilon_4  \over y_{33}^\ell y_{33}^\nu  \epsilon_5}   \right| , 
\sin \theta_{23}^\ell  \simeq  \left|{y_{23}^\ell   \over y_{33}^\ell  }    -{y_{23}^\nu \epsilon_4  \over y_{33}^\nu  \epsilon_5}   \right| ,~
\sin \theta_{13}^\ell    \simeq \left|  {y_{13}^\ell   \epsilon_5  \over y_{33}^\ell  \epsilon_2 }   -{y_{13}^\nu \epsilon_4  \over y_{33}^\nu  \epsilon_5}   \right| .
\end{eqnarray}  
We assume all the couplings of the order one,  and write,
\begin{eqnarray}
\sin \theta_{12}^\ell  &\simeq&  \left|-{y_{12}^\nu  \over y_{22}^\nu } +{y_{12}^\ell \epsilon_4  \over y_{22}^\ell  \epsilon_5}+  {y_{23}^{\ell *} y_{13}^\nu \epsilon_4  \over y_{33}^\ell y_{33}^\nu  \epsilon_5}   \right| \geq   \left|-{y_{12}^\nu  \over y_{22}^\nu }   \right| -   \left| {y_{12}^\ell   \over y_{22}^\ell  } +  {y_{23}^{\ell *} y_{13}^\nu   \over y_{33}^\ell y_{33}^\nu  }  \right|  {\epsilon_4  \over   \epsilon_5} \approx  1 - 2 \sin \theta_{12}, \\ \nonumber
\sin \theta_{23}^\ell  &\simeq&  \left|{y_{23}^\ell  \over y_{33}^\ell } - {y_{23}^\nu \epsilon_4  \over y_{33}^\nu  \epsilon_5} \right| \geq   \left|{y_{23}^\ell  \over y_{33}^\ell }   \right| -   \left| {y_{23}^\nu   \over y_{33}^\nu  }  \right|   { \epsilon_4  \over  \epsilon_5}\approx  1 -  \sin \theta_{12}, \\ \nonumber
\sin \theta_{13}^\ell    &\simeq& \left|    -{y_{13}^\nu \epsilon_4  \over y_{33}^\nu  \epsilon_5} + {y_{13}^\ell   \epsilon_5  \over y_{33}^\ell  \epsilon_2 }    \right| \geq  \left|    - {y_{13}^\nu   \over y_{33}^\nu  }  \right|  { \epsilon_4  \over   \epsilon_5}-  \left| {y_{13}^\ell     \over y_{33}^\ell   }  \right| { \epsilon_5  \over   \epsilon_2}  \approx \sin \theta_{12} - \frac{m_s}{m_c},
\end{eqnarray}
where $ m_s /m_c =      \epsilon_5  /   \epsilon_2    $. This result shows that the leptonic mixing angles can be predicted in terms of the Cabibbo angle and masses of strange and charm quarks.  Moreover, notable result are the prediction of correct and precise order of mixing angles,  and the pattern   $\sin \theta_{23}^\ell > \sin \theta_{12}^\ell >> \sin \theta_{13}^\ell  $. flavor bounds and collider signatures of the SHVM are discussed in the reference \cite{Abbas:2024jut}.

\subsection{Scalar potential of the SHVM}
\label{scalar_potential}
To construct the scalar potential of the SHVM, we introduce an extra $\mathcal{Z}_2^\prime$ symmetry. Under this new  $\mathcal{Z}_2^\prime$ symmetry, the right-handed  fermions transform as  $\psi_{R_{u,d,\ell,\nu}} : -$, the singlet scalar fields  as $\chi_r: -$,  where $r=1-6$, and the field $\chi_7$ as $\chi_7:+$ . This assignment eliminates all cubic interactions among the scalar fields, leaving the flavor structure intact and significantly simplifying the resulting phenomenology.  With these transformation properties, the scalar potential takes the form 
\bea 
V &=& -\mu^2 \varphi^\dagger \varphi + \lambda (\varphi^\dagger \varphi)^2  -   \mu_{\chi_1}^2 |\chi_1|^2   -   \mu_{\chi_2}^2 |\chi_2|^2   -   \mu_{\chi_3}^2 |\chi_3|^2    -   \mu_{\chi_4}^2|\chi_4|^2    -   \mu_{\chi_5}^2 |\chi_5|^2    -   \mu_{\chi_6}^2 |\chi_6|^2    
 \\ \nonumber  
 &&-   \mu_{\chi_7}^2 |\chi_7|^2 
+ \lambda_{\chi_1}  |\chi_1|^4  + \lambda_{\chi_2}  |\chi_2|^4+ \lambda_{\chi_3}  |\chi_3|^4+ \lambda_{\chi_4}  |\chi_4|^4+ \lambda_{\chi_5}  |\chi_5|^4+ \lambda_{\chi_6}  |\chi_6|^4+ \lambda_{\chi_7}  |\chi_7|^4   \\ \nonumber 
&&+     \lambda_{\varphi \chi_{ij}}  \varphi^\dagger \varphi   \chi^\dagger_i \chi_j  + \lambda_{\chi_{12}}  |\chi_1|^2 |\chi_2|^2  + \lambda_{\chi_{13}}  |\chi_1|^2 |\chi_3|^2  + \lambda_{\chi_{14}}  |\chi_1|^2 |\chi_4|^2  + \lambda_{\chi_{15}}  |\chi_1|^2 |\chi_5|^2   \\ \nonumber
&&+ \lambda_{\chi_{16}}  |\chi_1|^2 |\chi_6|^2  + \lambda_{\chi_{17}}  |\chi_1|^2 |\chi_7|^2 
+ \lambda_{\chi_{23}}  |\chi_2|^2 |\chi_3|^2  + \lambda_{\chi_{24}}  |\chi_2|^2 |\chi_4|^2  + \lambda_{\chi_{25}}  |\chi_2|^2 |\chi_5|^2  + \lambda_{\chi_{26}}  |\chi_2|^2 |\chi_6|^2   \\ \nonumber
&&+ \lambda_{\chi_{27}}  |\chi_2|^2 |\chi_7|^2  
+ \lambda_{\chi_{23}}  |\chi_2|^3 |\chi_3|^4  + \lambda_{\chi_{25}}  |\chi_2|^3 |\chi_5|^2  + \lambda_{\chi_{36}}  |\chi_3|^2 |\chi_6|^2  + \lambda_{\chi_{37}}  |\chi_3|^2 |\chi_7|^2   \\ \nonumber
&&+ \lambda_{\chi_{45}}  |\chi_4|^2 |\chi_5|^2  + \lambda_{\chi_{46}}  |\chi_4|^2 |\chi_6|^2  + \lambda_{\chi_{47}}  |\chi_4|^2 |\chi_7|^2 + \lambda_{\chi_{56}}  |\chi_5|^2 |\chi_6|^2  + \lambda_{\chi_{57}}  |\chi_5|^2 |\chi_7|^2  \\ \nonumber
&&  + \lambda_{\chi_{67}}  |\chi_6|^2 |\chi_7|^2  
+ \rm H.c..
\label{SP} 
\eea

We can parametrize the scalar fields as,
\begin{align}
 \chi_r(x) &=\frac{v_r + s_r(x) +i\, a_r(x)}{\sqrt{2}}, ~ \varphi =\left( \begin{array}{c}
G^+ \\
\frac{v+h+i G^0}{\sqrt{2}} \\
\end{array} \right).
\end{align}
Throughout our analysis, we take the quartic couplings in the potential to be generically of order one. Moreover, we assume $ \lambda_{\varphi \chi_{ij}} =0$.  For providing masses to axial degrees of freedom, we extend the scalar potential by adding soft symmetry-breaking terms,
\begin{align}
V_{\rm soft}
= -\rho_r^2 \ \chi_r^2    \rm +  H.c. .
\label{soft_pot}
\end{align}

Using $v_r = \sqrt{2} \epsilon_r \Lambda$, we can write the masses of scalars  approximately as,
\begin{align}
\label{scal_mass}
m_{s_r}^2 \approx  & 16 \epsilon_r^2 \Lambda^2 .
\end{align}

The pseudoscalar mass-matrix is completely diagonal, and the masses of pseudoscalars are given by
\begin{align}
\label{ps_mass}
m_{a_r}^2 = & 4 \rho_r^2 .
\end{align}

\subsection{The SHVM within the DTC-paradigm}
\label{SHVM_UV}
\begin{figure}[H]
	\centering
 \includegraphics[width=\linewidth]{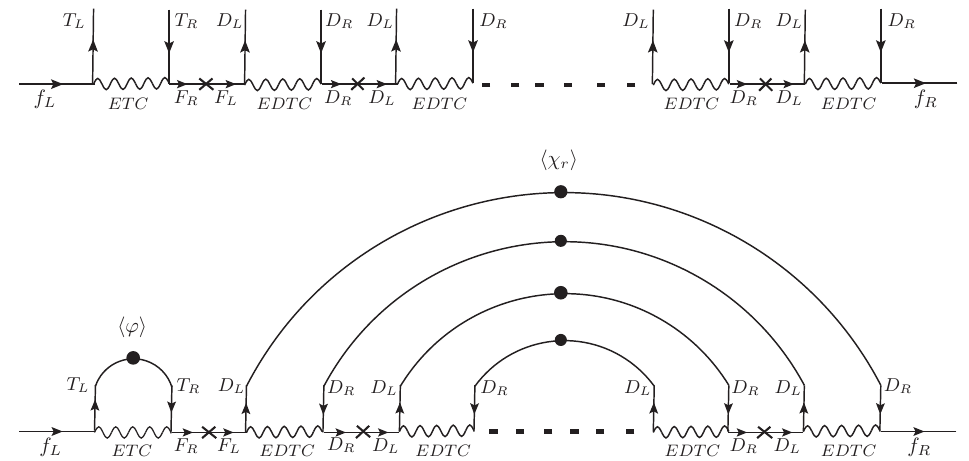}
    \caption{The Feynman diagrams for the masses of  charged fermions in the  DTC paradigm.  The top part shows the generic interactions of the SM, TC, DQCD and DTC fermions.  In the lower part of figure, the formations of the TC  chiral condensates, $\langle  \varphi \rangle$ (circular blob), a   generic   multi-fermion  chiral condensates  $\langle  \chi_r \rangle$ (collection of circular blobs),    and the resulting mass of the SM charged fermion is depicted. }
 \label{fig_shvm}	
 \end{figure}

The SHVM is one of the possible low-energy limits of the DTC paradigm. In this section, we show how to accommodate the SHVM within the framework of the DTC paradigm.  The DTC paradigm provides the interactions responsible for creating  the charged fermion mass matrix in the SHVM as  shown in the upper part of figure  \ref{fig_shvm}.  We show the formation of chiral TC condensates, which play the role of the Higgs VEV, and the DTC multi-fermion chiral condensates   denoted by $\langle  \chi_r \rangle$  in the lower part of figure  \ref{fig_shvm}.

In the TC sector, we assume only one doublet, that is 2 flavours.  Thus, as discussed earlier, the axial symmetry is broken to 
\begin{align}
\mathrm{U(1)}_{X_{\mathrm{TC}}} \xrightarrow[\text{}]{\text{anomaly and instantons}}  \mathbb{Z}_{4} \xrightarrow[\text{}]{ \langle \bar{\psi}_R \psi_L \rangle }  \mathbb{Z}_{2}.
\end{align}

However, we have assumed $k_{\rm TC} = 0$ \footnote{Numerical fits also prefer this value.}.  The symmetry $\mathbb{Z}_{2}$ enters into multifermion condensates through the equation
\begin{equation}
\langle (\bar{\psi}_R \psi_L)^n \rangle \sim \bigl( \Lambda\, e^{k_{\rm TC}\,\Delta \chi} \bigr)^{3n}.
\end{equation}

Thus, in the TC sector,  the factor $e^{k_{\rm TC}\,\Delta \chi}$ is trivial, thus, the factor $F$  does not play any role in the formation of the chiral condensate, and dynamics is identical to  the MAC framework, where only two fermions condensate occurs.  The series
\begin{align}
\langle \bar{\psi}_R \psi_L \rangle << \langle \bar{\psi}_R  \bar{\psi}_R \psi_L  \psi_L \rangle <<  \langle \bar{\psi}_R \bar{\psi}_R  \bar{\psi}_R \psi_L  \psi_L \psi_L \rangle << \cdots.
\end{align}
 is terminated with formation of only two fermion condensate~\cite{Aoki:1983ae,Aoki:1983za,Aoki:1983yy}.

The chiral condensate for a QCD-like theory, using equation \ref{VEV_h},  is given by \cite{Miransky:1994vk},
\begin{align}
\label{chi_con}
\langle \bar{T} T \rangle_{\rm \Lambda_{\rm ETC}}   \approx &  - \dfrac{\rm N_{\rm TC}}{4 \pi^2 } \left[ \Lambda_{\rm TC} \exp(k_{\rm TC} \Delta \chi_{\rm TC}) \right]^3, \\ \nonumber
\langle \bar{D} D \rangle_{\rm \Lambda_{\rm EDTC}}   \approx &  - \dfrac{\rm N_{\rm DTC}}{4 \pi^2 } \left[ \Lambda_{\rm DTC} \exp(k_{\rm DTC} \Delta \chi_{\rm DTC}) \right]^3, \\ \nonumber
\langle \bar{F} F \rangle_{\rm \Lambda_{\rm GUT}}   \approx & -  \dfrac{\rm N_{\rm D}}{4 \pi^2 } \left[ \Lambda \exp(k_{\rm D} \Delta \chi_{\rm D}) \right]^3.
\end{align}
The mass matrices of the charged fermions in equation \ref{mUD} are now approximately given by, 
\bea \label{TC_masses1} \M_{\U,\D,\ell} & \propto &  {\rm N}_{\rm D}^{n} \left[ \frac{g_{\rm ETC}^2}{\Lambda_{\rm ETC} 2} \langle \bar{T} T \rangle_{\rm \Lambda_{\rm ETC}} \right]     \dfrac{1}{\Lambda} \left[ \frac{g_{\rm EDTC}^{2n}}{\Lambda_{\rm EDTC}^{3n-1}} \left(\langle \bar{D} D \rangle_{\rm \Lambda_{\rm EDTC}} \right)^n \right].
\eea
where $n = 1,2,3 \cdots$ and $f=u,d,\ell$.

 Using equation \ref{chi_con},  the mass matrices can further be written   as,
\bea
\label{TC_masses2}
\M_{\U,\D,\ell} & = & y_{ij}^f {{\rm N}_{\rm D}^{n_i/2}  \dfrac{\rm N_{\rm TC}}{4 \pi^2 }  \frac{\Lambda_{\rm TC}^{3}}{\Lambda_{\rm ETC}^2}} \exp(6 k_{\rm TC} )   \dfrac{1}{\Lambda} \left[\dfrac{\rm {N_{DTC}}}{4 \pi^2 }\right]^{n_i/2} \frac{\Lambda_{\rm DTC}^{ n_i + 1}}{\Lambda_{\rm EDTC}^{n_i}} \left[\exp(3 n_i k_{\rm DTC}) \right]^{n_i/2},~
\eea
where we have assumed that the TC chiral condensate is of the type $\langle \bar{T}_R T_L \rangle $ and $\Delta \chi_{\rm TC} =2$.  Moreover, $g_{\rm ETC} = g_{\rm EDTC}=( 1-4\pi)$, and $k_{\rm TC}>0$ are assumed. Furthermore,   $ n_i = 2,4,6, \cdots 2 n $ are  the number of fermions in a multi-fermion chiral DTC condensate that plays the role of the VEV $ \langle \chi_r \rangle$ \cite{Abbas:2020frs}, and $\Lambda_{\rm TC}$, $\Lambda_{\rm DTC}$, and  $\Lambda $ denote   the scale of the TC, DTC, and DQCD dynamics respectively. 

 We conclude  from  equation \ref{TC_masses2} that 
\bea
\label{map1}
\epsilon_r \propto \dfrac{1}{\Lambda} \left[\dfrac{{\rm N_{\rm DTC}}}{4 \pi^2 }\right]^{n_i/2} \frac{\Lambda_{\rm DTC}^{n_i + 1}}{\Lambda_{\rm EDTC}^{n_i}} \left[\exp(3  n_i k_{\rm DTC}) \right]^{n_i/2}.
\eea
Thus, the masses of charged fermions given in equation \ref{mass1a} can be written  in terms of equation \ref{map1}, and at the leading order, the mass of a charged fermion is,
\bea
\label{TC_masses3}
m_{f} & \approx & |y_{11}^f| {{\rm N}_{\rm D}^{n_i/2} \dfrac{\rm N_{\rm TC}}{4 \pi^2 }  \frac{\Lambda_{\rm TC}^{3}}{\Lambda_{\rm ETC}^2} } \exp(6 k_{\rm TC} )  \dfrac{1}{\Lambda} \left[\dfrac{{\rm N_{DTC}}}{4 \pi^2 }\right]^{n_i/2} \frac{\Lambda_{\rm DTC}^{n_i + 1}}{\Lambda_{\rm EDTC}^{n_i}} \left[\exp(3 n_i k_{\rm DTC}) \right]^{n_i/2}.
\eea

\begin{figure}[h]
	\centering
 \includegraphics[width=\linewidth]{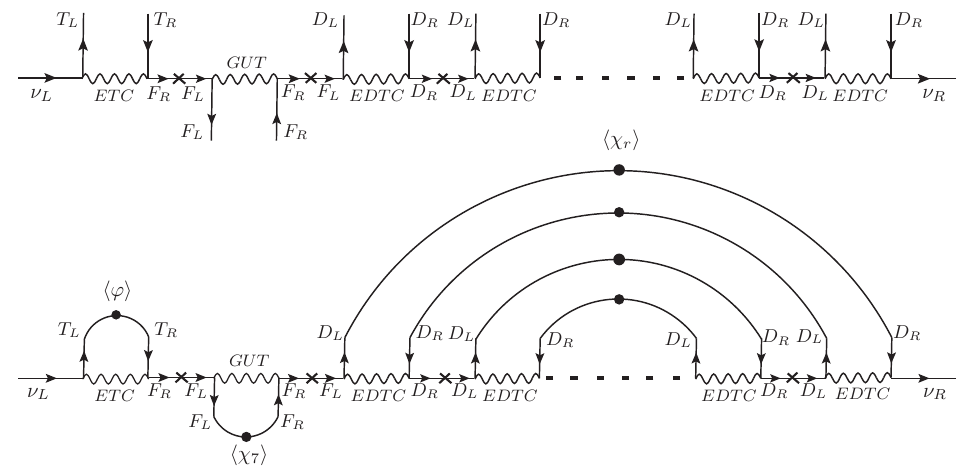}
    \caption{The Feynman diagrams for the masses of neutrinos in the DTC paradigm. On the top, there are generic interactions involving the SM, TC, DQCD and DTC gauge sectors mediated by ETC, EDTC and GUT gauge bosons.  In the bottom, we show the generic Feynman diagram after the formation of the fermionic condensates. }
 \label{fig_nu}	
 \end{figure}

For obtaining neutrino masses, we assume that the ETC and EDTC symmetries further unify in a GUT theory,  leading to dimension-6 operators given in equation \ref{mass_N} from which the neutrino masses originate.  The resulting interactions are shown in the upper part of Figure \ref{fig_nu}, which are mediated by the GUT gauge bosons among the $F_L$ and $F_R$ fermions.  The chiral condensate  $\langle \bar{F}_L F_R \rangle$  (circular blob) acts like the VEV $\langle \chi_7 \rangle$.  A generic formation of the neutrino mass term is shown in the lower part of the figure \ref{fig_nu}.

The neutrino mass matrix given in equation \ref{NM1} is  recovered as,
\bea
\label{TC_nmassesN}
\M_{\N}  =  y_{ij}^\nu {\rm N_{\rm D}}^{(n_i+2)/2} \dfrac{{\rm N_{\rm TC}}}{4 \pi^2 }  \frac{\Lambda_{\rm TC}^{3}}{\Lambda_{\rm ETC}^2} \exp(6 k_{\rm TC} )   \dfrac{1}{\Lambda} \left[\dfrac{{\rm N_{\rm DTC}}}{4 \pi^2 }\right]^{n_i/2} \frac{\Lambda_{\rm DTC}^{n_i + 1}}{\Lambda_{\rm EDTC}^{n_i}} \left[\exp(n_i k_{\rm DTC}) \right]^{n_i/2} \dfrac{1}{\Lambda} \dfrac{{\rm N_{D}}}{4 \pi^2 } \frac{\Lambda^{3}}{\Lambda_{\rm GUT}^{2}} \exp(6 k_{\rm D}),
\eea
where,  
\bea
\epsilon_7 \propto  \dfrac{1}{\Lambda} \dfrac{\rm N_{\rm D}}{4 \pi^2 } \frac{\Lambda^{3}}{\Lambda_{\rm GUT}^{2}} \exp(6 k_{\rm D}).
\eea

The  masses of neutrinos at the leading order turn out to be,
\bea
\label{neutrino_mass}
m_{\nu}  =  |y_{11}^\nu| {\rm N}_{\rm D}^{(n_i+2)/2} \dfrac{{\rm N}_{\rm TC}}{4 \pi^2 }  \frac{\Lambda_{\rm TC}^{3}}{\Lambda_{\rm ETC}^2} \exp(6 k_{\rm TC} )   \dfrac{1}{\Lambda} \left[\dfrac{{\rm N_{DTC}}}{4 \pi^2 }\right]^{n_i/2} \frac{\Lambda_{\rm DTC}^{n_i + 1}}{\Lambda_{\rm EDTC}^{n_i}} \left[\exp(3 n_i k_{\rm DTC}) \right]^{n_i/2} \dfrac{1}{\Lambda} \dfrac{{\rm N}_{\rm D}}{4 \pi^2 } \frac{\Lambda^{3}}{\Lambda_{\rm GUT}^{2}} \exp(6 k_{\rm D}).
\eea

We observe that in the DTC-paradigm, the underlying TC, DTC, and DQCD sectors are all QCD-like confining gauge theories. In such theories, the Higgs and the additional scalars arise as composite states generated by strong dynamics. Their masses and interactions are protected by approximate global symmetries, and the absence of fundamental scalars ensures that quadratic divergences do not arise. Thus, the framework inherits the notion of ``strong naturalness'' familiar from QCD~~\cite{Hill:2002ap}.

Each strong sector possesses an approximate custodial scale symmetry at high energies. The dynamical scales $\Lambda_{\rm TC}$, $\Lambda_{\rm DTC}$, and $\Lambda$ originate from the explicit breaking of scale invariance through the trace anomaly~~\cite{Hill:2002ap}. As the approximate scale invariance is restored in the ultraviolet (UV) limit, these scales become negligible. Consequently, the scalar potential is protected by an approximate global scale symmetry.

\begin{figure}[H]
	\centering
 \includegraphics[width=0.6\linewidth]{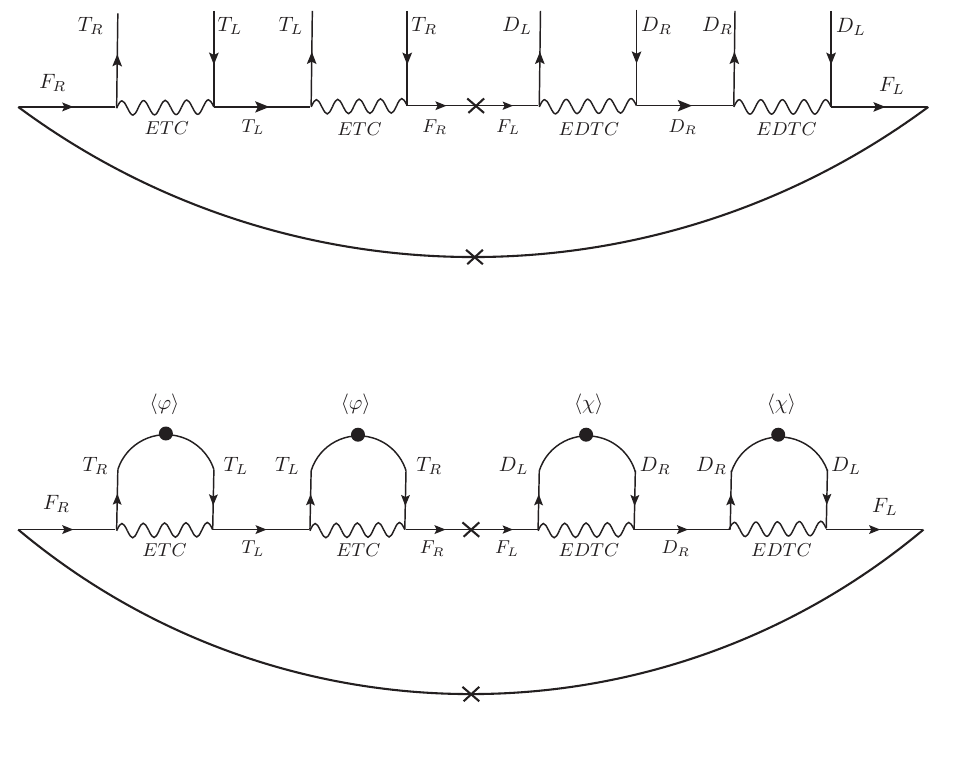}
    \caption{The Feynman diagrams  showing the mixing between the TC and the DTC dynamics. }
 \label{tc_dtc_mix}	
 \end{figure}

In the DTC paradigm, the mixing between the TC and DTC sectors is mediated by DQCD dynamics as shown schematically  in figure \ref{tc_dtc_mix}.  This mixing is suppressed by the factor, 
\begin{align}
\frac{\Lambda_{\rm TC}\,\Lambda_{\rm DTC}}{\Lambda^{2}}.    
\end{align}
Therefore, at the leading order, the SM Higgs potential effectively decouples from the DTC sector. Only the scalar potential involving the fields $\chi_{r}$ needs to be analyzed, and which is done in section \ref{scalar_potential}.

Finally, since the scalar states of TC, DTC, and DQCD are composite resonances formed at their respective strong scales, they receive only additive renormalizations of order $\Lambda_{\rm TC}$, $\Lambda_{\rm DTC}$, or $\Lambda$. UV sensitivity beyond these scales is therefore absent, just as in QCD. This is the sense in which the composite scalar sector of the DTC-paradigm enjoys strong naturalness.

For describing the symmetry structure of the mutli-fermion condensates, we adopt the strategy proposed in reference~\cite{Harari:1981bs}.  As discussed earlier, the framework contains three axial symmetries,
$\mathrm{U(1)}_{X_{\mathrm{TC}}}$, $\mathrm{U(1)}_{X_{\mathrm{DTC}}}$, and $\mathrm{U(1)}_{X_{\mathrm{DQCD}}}$.
We assign an axial charge $X = +1$ to all left-handed fermions and $X = -1$ to their
right-handed counterparts.

In general, we notice that the multi-fermion condensates (VEVs) may be parametrized as \cite{Aoki:1983yy},
\begin{equation}
\label{VEV_h}
\langle (\bar{\psi}_R \psi_L)^n \rangle \sim \bigl( \Lambda\, e^{k\,\Delta \chi} \bigr)^{3n}.
\end{equation}

This expression suggests a hierarchical breaking of the axial symmetries 
$\mathrm{U}(1)_{X_{\mathrm{TC}}}$, $\mathrm{U}(1)_{X_{\mathrm{DTC}}}$, and 
$\mathrm{U}(1)_{X_{\mathrm{DQCD}}}$  such that~\cite{Aoki:1983yy},
\begin{align}
\mathrm{U}(1)_{\mathrm{A}} &
\xrightarrow[\text{}]{\text{anomaly and instantons}}
\mathbb{Z}_{2K}
\cdots
\mathbb{Z}_{24}
\rightarrow 
\mathbb{Z}_{22}
\rightarrow 
\mathbb{Z}_{20}
\rightarrow 
\mathbb{Z}_{18}
\rightarrow 
\mathbb{Z}_{16}
\rightarrow 
\mathbb{Z}_{12}
\rightarrow 
\mathbb{Z}_{10}
\rightarrow 
\mathbb{Z}_{8}
\rightarrow 
\mathbb{Z}_{6}
\rightarrow 
\mathbb{Z}_{4}
\rightarrow 
\mathbb{Z}_{2},
\end{align}
where the first breaking step is generated by instanton effects of the strong dynamics.
Thus, each multi-fermion condensate is associated with a residual discrete
subgroup $\mathbb{Z}_{2K}$, which acts as a conserved quantum number.

The axial $\mathrm{U(1)}_{X_{\rm DTC}}$ symmetry is anomalous under the corresponding strong
dynamics. Instanton effects generate a $2K$-fermion operator carrying an axial charge
$X_{\mathrm{DTC}} = 2K$, where $K$ denotes the number of massless flavors in an
$N$-dimensional representation of the confining gauge group $\mathrm{SU}(N)$ \cite{Harari:1981bs}. The associated
operator acquires a nonvanishing VEV, thereby breaking the axial
symmetry according to \cite{Harari:1981bs},
\begin{align}
\mathrm{U(1)}_{X_{\mathrm{DTC}}} \xrightarrow[\text{}]{\text{anomaly and instantons}}  \mathbb{Z}_{2K}.
\end{align}
Thus, the axial quantum numbers $X_{\mathrm{DTC}}$ are conserved only modulo $2K$.

Suppose a first multi-fermion condensate is formed using $N_1$ massless flavors. A second,
distinct multi-fermion condensate may then be formed by adding an additional set of
$N_2$ massless flavors to the first structure. In order for this second condensate to be
distinct, the added flavors must transform differently under the global symmetry.
This distinction is naturally provided by the conserved axial charge
$X_{\mathrm{DTC}}$ modulo $2K$\cite{Harari:1981bs}. Consequently,   the residual discrete axial symmetry
serves as a robust label distinguishing in-equivalent multi-fermion condensate structures
within the theory. Thus, the multi-fermion structure of the theory is protected by a richer symmetry blue print consists of continuous chiral, and discrete residual symmetries.  This explanation cannot be achieved within the effective framework of the SHVM.  

The dynamics of the fields $\chi_r$ is inherently nonperturbative, and thus a standard loop expansion for quantum corrections is not applicable. However, as emphasized earlier, the theory contains no fundamental scalars, and therefore quadratic divergences are absent. Any perturbative corrections that arise are finite and at most of the order  of the dynamical scale.

For fitting fermion masses and mixings, we map the fermionic mass matrices given in Eqs.~\ref{mUD} and \ref{NM1} onto Eqs.~\ref{TC_masses2} and \ref{TC_nmassesN}, respectively. The numerical values of fermion masses at $1$~TeV are taken from Ref.~\cite{Xing:2007fb}:  
\begin{eqnarray}
\{m_t, m_c, m_u\} &\simeq& \{150.7 \pm 3.4,~ 0.532^{+0.074}_{-0.073},~ (1.10^{+0.43}_{-0.37}) \times 10^{-3}\}~{\rm GeV}, \nonumber \\
\{m_b, m_s, m_d\} &\simeq& \{2.43\pm 0.08,~ 4.7^{+1.4}_{-1.3} \times 10^{-2},~ 2.50^{+1.08}_{-1.03} \times 10^{-3}\}~{\rm GeV}, \nonumber \\
\{m_\tau, m_\mu, m_e\} &\simeq& \{1.78\pm 0.2,~ 0.105^{+9.4 \times 10^{-9}}_{-9.3 \times 10^{-9}},~ (4.96\pm 0.00000043) \times 10^{-4}\}~{\rm GeV}.
\end{eqnarray}

The magnitudes and phase of the CKM mixing matrix are taken from Ref.~\cite{Zyla:2021}:  
\bea
|V_{ud}| = 0.97370 \pm 0.00014, \quad |V_{cb}| = 0.0410 \pm 0.0014, \quad |V_{ub}| = 0.00382 \pm 0.00024, \quad \delta = 1.196^{+0.045}_{-0.043}.
\eea  

For the normal mass ordering, the neutrino mass-squared differences and leptonic mixing angles are adopted from the global fit in Ref.~\cite{deSalas:2020pgw}:  
\bea
\Delta m_{21}^2 &=& (7.50^{+0.64}_{-0.56}) \times 10^{-5}~{\rm eV}^2, \quad |\Delta m_{31}^2| = (2.55\pm 0.08) \times 10^{-3}~{\rm eV}^2, \nonumber \\
\sin \theta_{12}^\ell &=& 0.564^{+0.044}_{-0.043}, \quad \sin \theta_{23}^\ell = 0.758^{+0.023}_{-0.099}, \quad \sin \theta_{13}^\ell = 0.1483^{+0.0067}_{-0.0069},
\eea  
where the quoted uncertainties correspond to the $3\sigma$ ranges.  

To quantify the fit, we define the $\chi^2$ function as  
\begin{align}
\chi^2 &= 
 \frac{(m_q - m_q^{\rm model})^2}{\sigma_{m_q}^2} 
+ \frac{(\sin \theta_{ij} - \sin \theta_{ij}^{\rm model})^2}{\sigma_{\sin \theta_{ij}}^2} 
+ \frac{(m_\ell - m_\ell^{\rm model})^2}{\sigma_{m_\ell}^2} \nonumber \\
&\quad + \frac{(\Delta m_{21}^2 - \Delta m_{21}^{2,\,\rm model})^2}{\sigma_{\Delta m_{21}^2}^2} 
+ \frac{(\Delta m_{31}^2 - \Delta m_{31}^{2,\,\rm model})^2}{\sigma_{\Delta m_{31}^2}^2} \nonumber \\
&\quad + \frac{(\sin \theta_{ij}^\ell - \sin \theta_{ij}^{\ell,\,\rm model})^2}{\sigma_{\sin \theta_{ij}^\ell}^2}, 
\end{align}  
where $q=\{u,d,c,s,t,b\}$, $\ell=\{e,\mu,\tau\}$, and $i,j=1,2,3$.  

The dimensionless coefficients are parameterized as 
\[
y_{ij}^{u,d}= |y_{ij}^{u,d}| e^{i \phi_{ij}^{q}}, 
\]
with $|y_{ij}^{u,d}| \in [0.3, 4 \pi]$ and $ \phi_{ij}^{q} \in [0,2\pi]$. The fit is performed for three benchmark scenarios.  

The parameter space of the SHVM embedded in the DTC framework is found to be smooth, and the best-fit results are typically obtained for  
\begin{align}
 \{n_1, n_2, n_3, n_4, n_5, n_6, n_7\} &= \{8, 12, 14, 8, 10, 12, 2\}, \nonumber \\
 \rm N_{\rm TC} = 3, \quad \rm N_{\rm DTC} &= 20, \quad \rm N_{\rm D} = 12.
\end{align}  

\subsubsection{$\Lambda = \Lambda_{\rm DTC}$}

In this scenario the fit results are,
\begin{eqnarray}
 \label{fit_SHVM1}
 && 
     \Lambda_{\rm ETC}= 10^7 \text{ GeV}, \Lambda_{\rm DTC}= 500\text{ GeV},   
 \Lambda_{\rm EDTC}= 510 \text{ GeV}, \nonumber \\
&& \Lambda = 500 \text{ GeV}, \Lambda_{\rm GUT}= 3.2 \times 10^{7} \text{ GeV},  k_{\rm DTC}= 0.014,k_{\rm D} = 0.001.   
 \end{eqnarray}
The dimensionless couplings $y_{ij}^{u,d}$ are,
 \begin{equation}
y^u_{ij} = \begin{pmatrix}
 -0.47-0.16 i & 0 & -11.92-0.44 i \\
 0 & -0.72-0.25 i & -8.80+0.16 i \\
 0 & 0 & -11.35+3.90 i \\
\end{pmatrix},  
\end{equation}
\begin{equation}
y^d_{ij} = \begin{pmatrix}
 -0.68-0.16 i & -2.94-0.54 i & 0 \\
 0 & -1.02+0.18 i & 0 \\
 0 & 0 & -3.48+0.46 i \\
\end{pmatrix},  
\end{equation}
The dimensionless couplings $y_{ij}^{\ell,\nu}$ are,
\begin{equation}
y^\ell_{ij} = \begin{pmatrix}
 0.97\, +0.22 i & 1 & 1 \\
 0 & 0.75\, -0.81 i & 1 \\
 0 & 0 & -1.10+0.06 i \\
\end{pmatrix},
\end{equation}
\begin{equation}
y^\nu_{ij} = \begin{pmatrix}
 0.91\, -0.54 i & 1 & 1 \\
 0 & 1.49\, +2.66 i & -1.41-1.01 i \\
 0 & 1.66\, -0.89 i & 0.47\, +1.12 i \\
\end{pmatrix}, 
\end{equation}
These results are obtained for $\chi^2_{\rm min} = 4.01$.

\subsubsection{$\Lambda > \Lambda_{\rm DTC}$} 

 In this case, we have set the values of scales as follows,
\begin{equation}
\Lambda_{\rm ETC}= 10^7 \text{ GeV}, \Lambda_{\rm DTC}= 500\text{ GeV},   
 \Lambda_{\rm EDTC}= 510\text{ GeV}, \Lambda =1 \text{ TeV}.
\end{equation}
The fit results are,
\begin{eqnarray}
 \label{fit_SHVM2}
 \Lambda_{\rm GUT}= 6.3 \times 10^{7} \text{ GeV},  k_{\rm DTC}= 0.019,  k_{\rm D} = 0.029.  
 \end{eqnarray}
The dimensionless couplings $y_{ij}^{u,d}$ are,
 \begin{equation}
y^u_{ij} = \begin{pmatrix}
 -0.40-0.30 i & 0 & -2.81-11.67 i \\
 0 & -0.38-0.42 i & -5.82+3.02 i \\
 0 & 0 & -1.65-6.03 i \\
\end{pmatrix},  
\end{equation}
\begin{equation}
y^d_{ij} = \begin{pmatrix}
 -0.58+0.69 i & -3.43+1.75 i & 0 \\
 0 & -0.69-0.78 i & 0 \\
 0 & 0 & -2.57-0.45 i \\
\end{pmatrix},  
\end{equation}

The dimensionless couplings $y_{ij}^{\ell,\nu}$ are,
\begin{equation}
y^\ell_{ij} = \begin{pmatrix}
 -0.19-0.98 i & 1 & 1 \\
 0 & 1.1\, +0. i & 1 \\
 0 & 0 & 0.74\, -0.81 i \\
\end{pmatrix},
\end{equation}
\begin{equation}
y^\nu_{ij} = \begin{pmatrix}
 -0.01+1.13 i & 1 & 1 \\
 0 & 3.15\, +2.46 i & 1.05\, +0.24 i \\
 0 & 1.29\, -0.17 i & 0.30\, -0.98 i \\
\end{pmatrix}, 
\end{equation}
and  $\chi^2_{\rm min } =1.51$.

 \subsubsection{$\Lambda < \Lambda_{\rm DTC}$}

For this case we have set the values of scales as follows,
\begin{equation}
\Lambda_{\rm ETC}= 10^7 \text{ GeV}, \Lambda_{\rm DTC}= 1\text{ TeV},   
 \Lambda_{\rm EDTC}= 1.15\text{ TeV}, \Lambda = 500 \text{ GeV},
\end{equation}
and the fit results are,
\begin{eqnarray}
 \label{fit_SHVM3}
    \Lambda_{\rm GUT}= 4 \times 10^{7} \text{ GeV}, k_{\rm DTC}= 0.018,  k_{\rm D} = 0.006.  
 \end{eqnarray}

The dimensionless couplings $y_{ij}^{u,d}$ are,
 \begin{equation}
y^u_{ij} = \begin{pmatrix}
 0.25\, +0.43 i & 0 & 5.91\, +10.41 i \\
 0 & -0.48-0.62 i & -0.76+8.97 i \\
 0 & 0 & -1.96+11.84 i \\
\end{pmatrix},  
\end{equation}
\begin{equation}
y^d_{ij} = \begin{pmatrix}
 -0.55+0.37 i & -2.43-1.44 i & 0 \\
 0 & 0.51\, -0.90 i & 0 \\
 0 & 0 & -1.87+3.06 i \\
\end{pmatrix}.
\end{equation}

The dimensionless couplings $y_{ij}^{\ell,\nu}$ are,
\begin{equation}
y^\ell_{ij} = \begin{pmatrix}
 0.97\, +0.22 i & 1 & 1 \\
 0 & 0.42\, -1.02 i & 1 \\
 0 & 0 & -1.10+0.06 i \\
\end{pmatrix},
\end{equation}
\begin{equation}
y^\nu_{ij} = \begin{pmatrix}
 0.93\, +0.75 i & 1 & 1 \\
 0 & -0.38+3.37 i & -0.89+0.99 i \\
 0 & 0.73\, -0.58 i & -1.85+0.32 i \\
\end{pmatrix}.
\end{equation}
These results are obtained for $\chi^2_{min} =4.97$.

\subsection{Masses of scalars corresponding to $\chi_r$ fields}
The scalar degrees of freedom associated with the fields 
$\chi_r$ ($r=1,\dots,7$) are first treated within the effective SHVM framework as elementary degrees of freedom. In this effective description, their masses are estimated by assuming
order-one quartic couplings $\lambda_{\chi_r}$.

However, in the UV-complete theory their masses originate from the underlying
DTC and EDTC strong interactions, which dynamically generate the quartic couplings.  At this juncture, we introduce a slight modification in the DTC sector by assuming that the DTC fermions are doublets of the $\rm SU(2)_R$ symmetry introduced in section \ref{s-pheno}, and transform under the symmetry  $\rm SU(3)_c \times \rm SU(2)_L \times \rm SU(2)_R \times \rm U(1)_Y \times \mathcal{G}$ as,
\begin{eqnarray}
D_L^i&\equiv&  
\begin{pmatrix} C^i \\ S^i \end{pmatrix}_L 
:(1,1, 2, \dfrac{1}{3},1,\rm N_{\rm DTC},1),
\quad
D_R^i \equiv 
\begin{pmatrix} C^i \\ S^i \end{pmatrix}_R
:(1,1, 2, \dfrac{1}{3},1,\rm N_{\rm DTC},1),
\end{eqnarray} 

As an illustration, consider two types of DTC fermions $D_1$ with $Y=1$  corresponding the electric charge $Q=\pm \dfrac{1}{2}$,  and  fermions $D_2$  with $Y=\dfrac{2}{3}$  corresponding the electric charge $Q=\pm \dfrac{1}{3}$.  Moreover, consider the operator
\begin{equation}
\mathcal{L} = \frac{g_{\mathrm{EDTC}}^2}{\Lambda_{\mathrm{EDTC}}^2}
\left(\bar{D}_1 D_1\,\bar{D}_2 D_2-\bar{D}_1 \gamma_5 \tau^i D_1 \,\bar{D}_2\gamma_5 \tau^i D_2
\right),
\end{equation}
where $\tau^i$ are Pauli matrices.

This operator explicitly breaks the separate chiral symmetries of the $D_1$ and $D_2$ fermions. Assuming that this interaction is weakly coupled, the induced mass of the mesons, corresponding to  $\chi_r$ fields in the effective theory of the SHVM,   may be estimated using Dashen's formula~\cite{Dashen:1969eg}:
\begin{equation}
M_\chi^2 =
\frac{1}{F_{\mathrm{DTC}}^2}\,
\langle 0| [Q_5^a,[Q_5^a,\mathcal{H}]] |0 \rangle,
\qquad
\mathcal{H} = -\mathcal{L},
\quad
Q_5^a = \int d^3x\, J_{50}^a(x),
\end{equation}
where $J_{50}^a$ is the axial current associated with the field $\chi$.

Evaluating the commutators yields~\cite{Miransky:1994vk},
\begin{equation}
M_\chi^2 \approx
\frac{1}{F_{\mathrm{DTC}}^2}\,
\frac{g_{\mathrm{EDTC}}^2}{\Lambda_{\mathrm{EDTC}}^2}\,
\langle 0|\,\bar{D}_1 D_1 \,\bar{D}_2 D_2\,|0\rangle
=
\frac{1}{F_{\mathrm{DTC}}^2}\,
\frac{g_{\mathrm{EDTC}}^2}{\Lambda_{\mathrm{EDTC}}^2}\,
\bigl(\Lambda_{\mathrm{DTC}} e^{4k}\bigr)^6 .
\end{equation}

This result generalizes to a condensate with $2 n$ fermion bilinears as
\begin{equation}
M_\chi^2
\approx
\frac{1}{F_{\mathrm{DTC}}^n}\,
\frac{g_{\mathrm{EDTC}}^2}{\Lambda_{\mathrm{EDTC}}^{2(n-1)}}\,
\langle (\bar{\psi}_R\psi_L)^n\rangle
=
\frac{1}{F_{\mathrm{DTC}}^n}\,
\frac{g_{\mathrm{EDTC}}^2}{\Lambda_{\mathrm{EDTC}}^{2(n-1)}}\,
\bigl(\Lambda_{\mathrm{DTC}}\, e^{k_{\mathrm{DTC}}\Delta\chi}\bigr)^{3n}.
\end{equation}

For example, with
$n=4$, $\rm N_{\mathrm{DTC}}=3$, 
$\Lambda_{\mathrm{DTC}}=500~\mathrm{GeV}$,
$\Lambda_{\mathrm{EDTC}}=510~\mathrm{GeV}$,
$k_{\mathrm{DTC}}=0.019$, and  using the Eq. \ref{decay_const} in section \ref{scaling} we obtain $F_{\mathrm{DTC}}=238~\mathrm{GeV}$,  the mass of the eight-fermion scalar  (corresponding to field $\chi_1$) is found to be
\begin{align}
  M_{\chi_1} \simeq 6.5~\mathrm{TeV}.  
\end{align}

\subsection{The FN mechanism based on the $\mathcal{Z}_{\rm N} \times \mathcal{Z}_{\rm M}$ flavor symmetry}
\label{sec3}
The other possible low-energy limit of the DTC paradigm could be the FN mechanism based on the $\mathcal{Z}_{\rm N} \times \mathcal{Z}_{\rm M}$ flavor symmetry \cite{Abbas:2018lga}.  This is extensively discussed in \cite{Abbas:2022zfb,Abbas:2023ion,Abbas:2024dfh}.  To show the FN mechanism based on the $\mathcal{Z}_{\rm N} \times \mathcal{Z}_{\rm M}$ flavor symmetry as a possible low energy limit of the DTC paradigm, we use a model which can provide a unified solution to the flavor problem and dark matter through the emergence of flavonic dark matter, a new class of scalar dark matter \cite{Abbas:2023ion}.

This model is based on the  $\mathcal{Z}_{8} \times \mathcal{Z}_{22}$ flavor symmetry, where the mass of the top quark arises through the dimension-5 operator.   As will be shown in the next subsection, this model can be easily accommodated within the DTC paradigm.  We show the transformation of the SM and the flavon field, $\chi$, under the  $\mathcal{Z}_{8} \times \mathcal{Z}_{22}$ flavor symmetry in table \ref{tab_z8z22}. 

  \begin{table}[H]
 \small
\begin{center}
\begin{tabular}{|c|c|c|c|c|c|c|c|c| c|c|c| c|c|c|}
  \hline
  Fields             &        $\mathcal{Z}_8$                    & $\mathcal{Z}_{22}$ & Fields             &        $\mathcal{Z}_8$                    & $\mathcal{Z}_{22}$   & Fields             &        $\mathcal{Z}_8$                    & $\mathcal{Z}_{22}$    & Fields             &        $\mathcal{Z}_8$                    & $\mathcal{Z}_{22}$     & Fields             &        $\mathcal{Z}_8$                    & $\mathcal{Z}_{22}$        \\
  \hline
  $u_{R}$                 &   $ \omega^2$  &$ \omega^{\prime 2}$        &$c_{R}$                 &   $ \omega^5$  & $ \omega^{\prime 5}$    &$t_{R}$                 &   $ \omega^6$  & $ \omega^{\prime 6}$       & $d_{R}$                 &   $ \omega^3$  &     $\omega^{\prime 3} $           & $s_{R}$                 &   $ \omega^4$  &     $\omega^{\prime 4} $           \\
  $b_{R}$                 &   $ \omega^4$  &     $\omega^{\prime 4} $     &   $\psi_{L,1}^q$                 &    $ \omega^2$  &    $\omega^{\prime 10} $      & $\psi_{L,2}^q$                 &  $ \omega$  &     $\omega^{\prime 9} $       &  $\psi_{L,3}^q$                 &    $\omega^{7} $  &      $\omega^{\prime 7} $ & $\psi_{L,1}^\ell$                 &   $ \omega^3$  &    $\omega^{\prime 3} $          \\
     $\psi_{L,2}^\ell$                  &   $ \omega^2$  &    $\omega^{\prime 2} $    &   $\psi_{L,3}^\ell$                 &   $ \omega^2$  &    $\omega^{\prime 2} $     &  $e_R$                 &   $\omega^{2} $  &     $\omega^{\prime 16} $          & $\mu_R$                 &  $\omega^5 $   &     $\omega^{\prime 19} $      &  $\tau_R $                 &   $ \omega^7$  &     $\omega^{\prime 21} $              \\
            $ \nu_{e_R} $                 &     $\omega^2 $    &     $1 $         & $   \nu_{\mu_R}$                 &     $\omega^5 $    &     $\omega^{\prime 3} $          &  $  \nu_{\tau_R} $                 &     $\omega^6 $    &     $\omega^{\prime 4} $        &   $\chi$                        & $ \omega$  &       $  \omega^\prime$       & $\varphi$              &   1        &     1                  \\          
  \hline
     \end{tabular}
\end{center}
\caption{The charges of the SM as well as the flavon field under the $\mathcal{Z}_8 \times \mathcal{Z}_{22}$  symmetry,  where $\omega$ denotes the 8th,  and $\omega^\prime $ denotes the 22th root of unity respectively. }
 \label{tab_z8z22}
\end{table} 

The mass Lagrangian for the charged fermions originate from the Lagrangian produced by the  $\mathcal{Z}_8 \times \mathcal{Z}_{22}$  flavor symmetry,
\bea
\label{massz11}
-{\mathcal{L}}_{\rm Yukawa} &=&    \left(  \dfrac{ \chi}{\Lambda} \right)^{8}  y_{11}^u \bar{ \psi}_{L_1}^q \tilde{\varphi} u_{R}+  \left(  \dfrac{ \chi}{\Lambda} \right)^{5}  y_{12}^u \bar{ \psi}_{L_1}^q \tilde{\varphi} c_{R} +  \left(  \dfrac{ \chi}{\Lambda} \right)^{4}  y_{13}^u \bar{ \psi}_{L_1}^q \tilde{\varphi}  t_{R} +  \left(  \dfrac{ \chi}{\Lambda} \right)^{7}  y_{21}^u \bar{ \psi}_{L_2}^q \tilde{\varphi} u_{R}\nonumber \\
&+& \left(  \dfrac{ \chi}{\Lambda} \right)^{4}  y_{22}^u \bar{ \psi}_{L_2}^q \tilde{\varphi} c_{R} + \left(  \dfrac{ \chi}{\Lambda} \right)^{3}  y_{23}^u \bar{ \psi}_{L_2}^q \tilde{\varphi} t_{R}+ \left(  \dfrac{ \chi}{\Lambda} \right)^{5}  y_{31}^u \bar{ \psi}_{L_3}^q \tilde{\varphi} u_{R} + \left(  \dfrac{ \chi}{\Lambda} \right)^{2}  y_{32}^u \bar{ \psi}_{L_3}^q \tilde{\varphi} c_{R} \nonumber \\
&+&  \left(  \dfrac{ \chi}{\Lambda} \right) y_{33}^u \bar{ \psi}_{L_3}^q \tilde{\varphi} t_{R} 
+ \left(  \dfrac{ \chi}{\Lambda} \right)^{7} y_{11}^d \bar{ \psi}_{L_1}^q  \varphi d_{R} + \left(  \dfrac{ \chi}{\Lambda} \right)^{6} y_{12}^d \bar{ \psi}_{L_1}^q  \varphi s_{R} + \left(  \dfrac{ \chi}{\Lambda} \right)^{6} y_{13}^d \bar{ \psi}_{L_1}^q  \varphi b_{R} \nonumber \\ &+& \left(  \dfrac{ \chi}{\Lambda} \right)^{6} y_{21}^d \bar{ \psi}_{L_2}^q  \varphi d_{R} 
+ \left(  \dfrac{ \chi}{\Lambda} \right)^{5} y_{22}^d \bar{ \psi}_{L_2}^q  \varphi s_{R} + \left(  \dfrac{ \chi}{\Lambda} \right)^{5} y_{23}^d \bar{ \psi}_{L_2}^q  \varphi b_{R} + \left(  \dfrac{ \chi}{\Lambda} \right)^{4} y_{31}^d \bar{ \psi}_{L_3}^q  \varphi d_{R}\nonumber \\ &+& \left(  \dfrac{ \chi}{\Lambda} \right)^{3} y_{32}^d \bar{ \psi}_{L_3}^q  \varphi s_{R}    
+\left(  \dfrac{ \chi}{\Lambda} \right)^{3} y_{33}^d \bar{ \psi}_{L_3}^q  \varphi b_{R} + 
\left(  \dfrac{ \chi}{\Lambda} \right)^{9} y_{11}^\ell \bar{ \psi}_{L_1}^\ell  \varphi e_{R} + \left(  \dfrac{ \chi}{\Lambda} \right)^{6} y_{12}^\ell \bar{ \psi}_{L_1}^\ell  \varphi \mu_{R} \nonumber \\
&+& \left(  \dfrac{ \chi}{\Lambda} \right)^{4} y_{13}^\ell \bar{ \psi}_{L_1}^\ell  \varphi \tau_{R}   +\left(  \dfrac{ \chi}{\Lambda} \right)^{8} y_{21}^\ell \bar{ \psi}_{L_2}^\ell  \varphi e_{R} + \left(  \dfrac{ \chi}{\Lambda} \right)^{5} y_{22}^\ell \bar{ \psi}_{L_2}^\ell  \varphi \mu_{R} + \left(  \dfrac{ \chi}{\Lambda} \right)^{3} y_{23}^\ell \bar{ \psi}_{L_2}^\ell  \varphi \tau_{R} \nonumber \\
&+&   \left(  \dfrac{ \chi}{\Lambda} \right)^{8} y_{31}^\ell \bar{ \psi}_{L_3}^\ell  \varphi e_{R} + \left(  \dfrac{\chi}{\Lambda} \right)^{5} y_{32}^\ell \bar{ \psi}_{L_3}^\ell  \varphi \mu_{R} \nonumber + \left(  \dfrac{ \chi}{\Lambda} \right)^{3} y_{33}^\ell \bar{ \psi}_{L_3}^\ell  \varphi \tau_{R} 
 + \text{H.c.},
\eea
where $\epsilon = \langle \chi \rangle /\Lambda <1$.

The mass matrices of the charged fermions are given as,
\begin{align}
\label{mass_mat_FN}
\M_u & = \dfrac{v}{\sqrt{2}}
\begin{pmatrix}
y_{11}^u  \epsilon^8 &  y_{12}^u \epsilon^{5}  & y_{13}^u \epsilon^{4}    \\
y_{21}^u \epsilon^7     & y_{22}^u \epsilon^4  &  y_{23}^u \epsilon^{3}  \\
y_{31}^u  \epsilon^{5}    &  y_{32}^u  \epsilon^2     &  y_{33}^u  \epsilon 
\end{pmatrix}, 
\M_d   = \dfrac{v}{\sqrt{2}}
\begin{pmatrix}
y_{11}^d  \epsilon^7 &  y_{12}^d \epsilon^6 & y_{13}^d \epsilon^6   \\
y_{21}^d  \epsilon^6  & y_{22}^d \epsilon^5 &  y_{23}^d \epsilon^5  \\
 y_{31}^d \epsilon^4 &  y_{32}^d \epsilon^3   &  y_{33}^d \epsilon^3
\end{pmatrix}, 
\M_\ell =  \dfrac{v}{\sqrt{2}}
\begin{pmatrix}
y_{11}^\ell  \epsilon^9 &  y_{12}^\ell \epsilon^4  & y_{13}^\ell \epsilon^4   \\
y_{21}^\ell  \epsilon^{10}  & y_{22}^\ell \epsilon^5  &  y_{23}^\ell \epsilon^3  \\
 y_{31}^\ell \epsilon^{8}   &  y_{32}^\ell \epsilon^5   &  y_{33}^\ell \epsilon^3
\end{pmatrix}.
\end{align}

The masses of charged fermions can be written as,
\begin{align}
\label{eqn5}
\{m_t, m_c, m_u\} &\simeq \{|y_{33}^u| \epsilon , ~ \left |y_{22}^u  - \frac {y_{23}^u y_{32}^u} {y_{33}^u  }   \right|  \epsilon^4 ,\\&
~ \left |y_{11}^u- \frac {y_{12}^u y_{21}^u}{y_{22}^u-y_{23}^u y_{32}^u/y_{33}^u}- \frac{y_{13}^u (y_{31}^u y_{22}^u-y_{21}^u y_{32}^u)-y_{31}^u y_{12}^u y_{23}^u}{(y_{22}^u- y_{23}^u y_{32}^u/y_{33}^u) y_{33}^u} \right| \epsilon^8\}v/\sqrt{2}  ,\nonumber \\ 
\{m_b, m_s, m_d\} & \simeq \{|y_{33}^d| \epsilon^3, ~ \left |y_{22}^d- \frac {y_{23}^d y_{32}^d} {y_{33}^d} \right| \epsilon^5,\\ \nonumber 
&  \left |y_{11}^d- \frac {y_{12}^d y_{21}^d}{y_{22}^d-y_{23}^d y_{32}^d/y_{33}^d}- \frac{y_{13}^d (y_{31}^d y_{22}^d-y_{21}^d y_{32}^d)-y_{31}^d y_{12}^d y_{23}^d}{(y_{22}^d- y_{23}^d y_{32}^d/y_{33}^d) y_{33}^d} \right| \epsilon^7\}v/\sqrt{2} ,\\ \nonumber 
\{m_\tau, m_\mu, m_e\} & \simeq \{|y_{33}^l| \epsilon^3, ~ \left|y_{22}^l- \frac {y_{23}^l y_{32}^l} {y_{33}^l} \right| \epsilon^5,\\& ~  \left |y_{11}^l- \frac {y_{12}^l y_{21}^l}{y_{22}^l-y_{23}^l y_{32}^l/y_{33}^l}- \frac{y_{13}^l \left( y_{31}^l y_{22}^l-y_{21}^l y_{32}^l \right) -y_{31}^l y_{12}^l y_{23}^l}{\left(  y_{22}^l- y_{23}^l y_{32}^l/y_{33}^l \right) y_{33}^l} \right| \epsilon^9\}v/\sqrt{2}.
\end{align}
The quark mixing angles are,
\begin{eqnarray}
\sin \theta_{12}  \simeq |V_{us}| &\simeq& \left|{y_{12}^d \over y_{22}^d}  -{y_{12}^u \over y_{22}^u}  \right| \epsilon, ~
\sin \theta_{23}  \simeq |V_{cb}| \simeq  \left|{y_{23}^d \over y_{33}^d}   -{y_{23}^u \over y_{33}^u}   \right| \epsilon^2,~
\sin \theta_{13}  \simeq |V_{ub}| \simeq  \left|{y_{13}^d \over y_{33}^d}    -{y_{12}^u y_{23}^d \over y_{22}^u y_{33}^d}      
- {y_{13}^u \over y_{33}^u}   \right|   \epsilon^3. \qquad
\end{eqnarray}
We obtain Dirac neutrino masses by adding  three right handed  neutrinos $\nu_{eR}$, $\nu_{\mu R}$, $\nu_{\tau R}$  to the SM, and writing the Lagrangian,
\begin{eqnarray}
\label{mass5}
-{\mathcal{L}}_{\rm Yukawa}^{\nu} &=&      y_{ij}^\nu \bar{ \psi}_{L_i}^\ell   \tilde{H}  \nu_{R_{j}} \left[  \dfrac{ \chi}{\Lambda} \right]^{n_{ij}^\nu} +  {\rm H.c.}. 
\end{eqnarray}
The Dirac mass matrix for neutrinos is given as,
\begin{equation}
\label{NM}
\M_{\D} = \dfrac{v}{\sqrt{2}}
\begin{pmatrix}
y_{11}^\nu  \epsilon^{25} &  y_{12}^\nu \epsilon^{22} & y_{13}^\nu   \epsilon^{21}  \\
y_{21}^\nu  \epsilon^{24}  & y_{22}^\nu \epsilon^{21} &  y_{23}^\nu   \epsilon^{20} \\
y_{31}^\nu \epsilon^{24}   &  y_{32}^\nu  \epsilon^{21}   &  y_{33}^\nu \epsilon^{20}
\end{pmatrix}.
\end{equation}
We obtain the masses with normal hierarchy,
\begin{align}
\label{eqn5a}
\{m_3, m_2,  m_1\} & \simeq \{|y_{33}^\nu| \epsilon^{20}, ~ \left|y_{22}^\nu- \frac {y_{23}^\nu y_{32}^\nu} {y_{33}^\nu} \right| \epsilon^{21},\\& ~  \left |y_{11}^\nu- \frac {y_{12}^\nu y_{21}^\nu}{y_{22}^\nu-y_{23}^\nu y_{32}^\nu/y_{33}^\nu}- \frac{y_{13}^\nu \left( y_{31}^\nu y_{22}^\nu-y_{21}^\nu y_{32}^\nu \right) -y_{31}^\nu y_{12}^\nu y_{23}^\nu}{ \left( y_{22}^\nu- y_{23}^\nu y_{32}^\nu/y_{33}^\nu \right) y_{33}^\nu} \right| \epsilon^{25}\}v/\sqrt{2}.\nonumber
\end{align}
 The leptonic mixing angles turn out to be,
\begin{eqnarray}
\label{numixing}
\sin \theta_{12}  &\simeq& \left|  {y_{12}^\ell \over y_{22}^\ell} -{y_{12}^\nu \over y_{22}^\nu}   \right| \epsilon, ~
\sin \theta_{23} \simeq  \left|  {y_{23}^\ell \over y_{33}^\ell} -  {y_{23}^\nu \over y_{33}^\nu}  \right|,~
\sin \theta_{13} \simeq \left| {y_{13}^\ell \over y_{33}^\ell}  - {y_{12}^\nu y_{23}^\ell\over y_{22}^\nu  y_{33}^\ell} -  {y_{13}^\nu \over y_{33}^\nu} \right|  \epsilon. 
\end{eqnarray}

\subsection{The FN mechanism based on the $\mathcal{Z}_{\rm N} \times \mathcal{Z}_{\rm M}$ flavor symmetry within the DTC paradigm}

\begin{figure}[H]
	\centering
 \includegraphics[width=\linewidth]{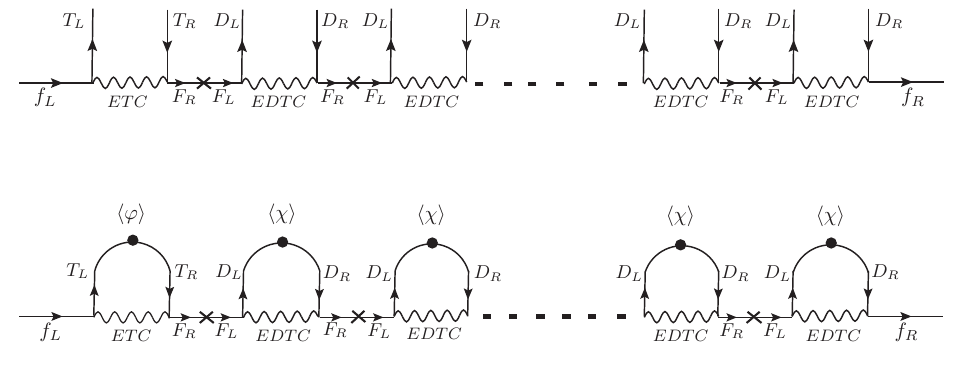}
    \caption{Feynman diagrams for quark and charged lepton masses when the FN mechanism is embedded in the DTC paradigm. 
    Top: generic interactions among SM, TC, DQCD, and DTC fermions. 
    Bottom: formation of condensates (circular blobs) and the resulting SM fermion masses.}
 \label{fig_DTC_FN}	
\end{figure}

The FN mechanism, based on a $\mathcal{Z}_{\rm N} \times \mathcal{Z}_{\rm M}$ flavor symmetry, can be embedded within the DTC paradigm. The interactions responsible for generating the charged fermion mass matrices are shown in the upper part of Fig.~\ref{fig_DTC_FN}. In the lower part of the figure, we illustrate the generic FN mechanism of fermion mass generation, where $\langle \varphi \rangle$ and $\langle \chi \rangle$ correspond to the chiral condensates that play the roles of the Higgs and flavon VEVs, respectively.

The mass matrices corresponding to Eqs.~\ref{mass_mat_FN} and \ref{NM} in the FN framework can be obtained as
\bea
\label{TC_masses}
\M_{f} & = &   |y_{ij}^f| \, \dfrac{\rm N_{\rm TC}}{2 \pi^2 } \, \frac{\Lambda_{\rm TC}^{2}}{\Lambda_{\rm ETC}}  
\left( \dfrac{1}{\Lambda} \, \dfrac{\rm N_{\rm DTC}}{4 \pi^2 } \, \frac{\Lambda_{\rm DTC}^3}{\Lambda_{\rm EDTC}^{2}} \, e^{2 k} \right)^{n_{ij}^f},
\eea
where $f = u,d,\ell,\nu$. The effective expansion parameter $\epsilon$ can be written as
\begin{align}
    \epsilon  \propto  \dfrac{1}{\Lambda} \, \dfrac{\rm N_{\rm DTC}}{4 \pi^2 } \, \frac{\Lambda_{\rm DTC}^3}{\Lambda_{\rm EDTC}^{2}} \, e^{2 k},  
\end{align}
while the analogue of the SM Higgs VEV is identified as
\begin{align}
  \langle \varphi \rangle  \propto  \dfrac{\rm N_{\rm TC}}{2 \pi^2 } \, \frac{\Lambda_{\rm TC}^{2}}{\Lambda_{\rm ETC}}. 
\end{align}

The FN mechanism provides a compelling theoretical framework to address the flavor problem of the SM within effective field theory. However, in the present DTC set-up with a low technicolor scale, $\Lambda_{\rm TC} = 10^3$~GeV, it cannot be consistently realized. The mechanism could still operate if the technicolor scale were as high as $\Lambda_{\rm TC} = 5$~TeV. For this reason, we do not pursue the collider phenomenology of the FN mechanism further in this work.

\section{Why not the minimal form of the DTC paradigm?}
\label{min_dtc}

A minimal realization of the DTC paradigm has been discussed in Ref.~\cite{Abbas:2023dpf}, based on the gauge symmetry 
\[
\mathcal{G} = {\rm SU}(\rm N_{\rm TC}) \times {\rm SU}(\rm N_{\rm D}).
\]  
In this section, we examine this minimal setup in detail and demonstrate that it is not theoretically consistent with reproducing the flavor structure of the SM.  

\subsection{Multi-fermion chiral condensates $\langle \chi_r \rangle$ from TC dynamics}

In the first scenario, the multi-fermion chiral condensates, which play the role of the effective vacuum expectation values (VEVs) $\langle \chi_r \rangle$, are assumed to arise directly from the TC dynamics. To implement this, one introduces TC fermions charged under  
\[
{\rm SU}(3)_c \times {\rm SU}(2)_L \times {\rm U}(1)_Y \times \mathcal{G},
\]  
with the following assignments:
\begin{eqnarray}
T^i  &\equiv& \begin{pmatrix} T \\ B \end{pmatrix}_L : (1,2,0,\rm N_{\rm TC},1), 
\quad 
T_{R}^i : (1,1,1,\rm N_{\rm TC},1), 
\quad 
B_{R}^i : (1,1,-1,\rm N_{\rm TC},1), \\[6pt] \nonumber
D_{L,R} &\equiv& C_{i,L,R} : (1,1,1,\rm N_{\rm TC},1), 
\quad 
S_{i,L,R} : (1,1,-1,\rm N_{\rm TC},1),
\end{eqnarray}
where $i=1,2,3,\ldots$, and the electric charges are $+1/2$ for $(T, C)$ and $-1/2$ for $(B, S)$.  

The ${\rm SU}\rm (N_{D})$ sector contains vector-like fermions transforming as  
\begin{eqnarray}
F_{L,R} &\equiv& U_{L,R}^i : (3,1,4/3,1,\rm N_{\rm D}), 
\quad 
D_{L,R}^i : (3,1,-2/3,1,\rm N_{\rm D}), \\[6pt] \nonumber
N_{L,R}^i &\equiv& (1,1,0,1,\rm N_{\rm D}), 
\quad 
E_{L,R}^i : (1,1,-2,1,\rm N_{\rm D}).
\end{eqnarray}

\begin{figure}[H]
	\centering
 \includegraphics[width=\linewidth]{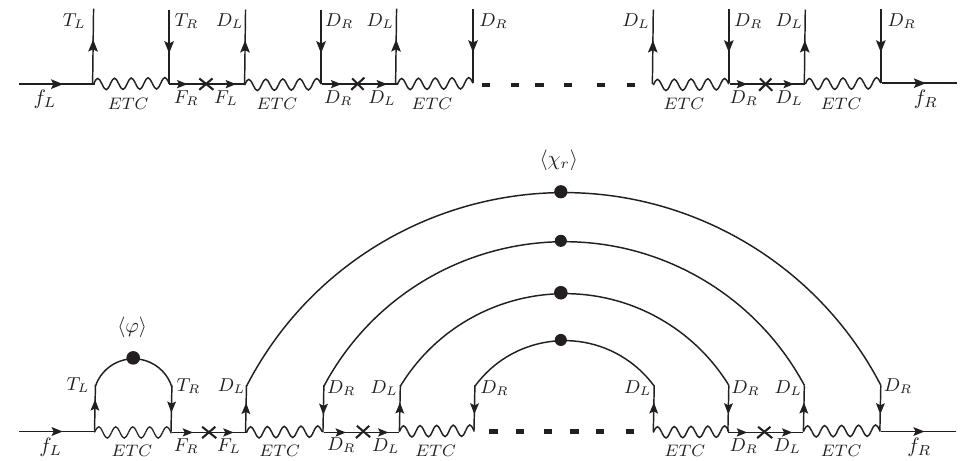}
    \caption{Feynman diagrams for charged fermion mass generation in the minimal version of the DTC paradigm.}
 \label{fig_min_1}	
\end{figure}

We assume that the hierarchical VEVs $\langle \chi_r \rangle$ correspond to TC chiral multi-fermion condensates of the schematic form $\bar{D}_R D_L \cdots \bar{D}_R D_L$. To generate SM fermion masses, all TC fermions ($T$ and $D$) must be embedded into a common ETC symmetry. This setup leads to fermion mass generation via diagrams such as those in Fig.~\ref{fig_min_1}.  

  However, this construction unavoidably induces interactions between right-handed SM fermions and right-handed TC fermions. As a result, the model effectively reduces to the original TC framework based only on ${\rm SU}(\rm N_{\rm TC})$, which is already excluded by experimental constraints.

\subsection{Multi-fermion chiral condensates $\langle \chi_r \rangle$ from DQCD dynamics}

In the second scenario, the multi-fermion chiral condensates $\langle \chi_r \rangle$ are instead assumed to originate from the ${\rm SU}(\rm N_{\rm D})$ (DQCD) dynamics, i.e.  
\[
\langle \chi_r \rangle \sim \bar{F}_R F_L \cdots \bar{F}_R F_L.
\]  
At first sight, this seems more promising from the standpoint of model minimality.  

However, the observed SM flavor structure can only be reproduced if: 
\begin{itemize}
    \item[] 1. left-handed SM fermions, TC fermions, and right-handed $F_R$ fermions are unified in an ETC symmetry, while
    \item[] 2. right-handed SM fermions $f_R$ and left-handed $F_L$ fermions are embedded in a separate extended DTC (EDTC) symmetry. 
\end{itemize}

\begin{figure}[H]
	\centering
 \includegraphics[width=\linewidth]{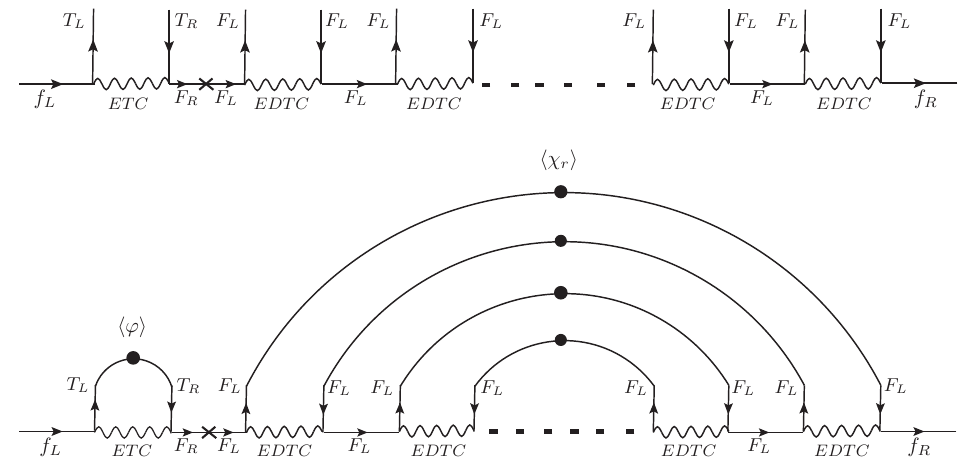}
    \caption{Feynman diagrams for charged fermion mass generation in the minimal version of the DTC paradigm.}
 \label{fig_min_2}	
\end{figure}

This arrangement generates fermion mass terms through diagrams such as those in Fig.~\ref{fig_min_2}.  

A few important issues arise in this scenario:  

\begin{itemize}
    \item The multi-fermion chiral condensates must take the form $\langle \chi_r \rangle \sim \bar{F}_L F_L \cdots \bar{F}_L F_L$, which are chirality-preserving and correspond to zero net chirality. Such states are not the most attractive channels for forming scalar spin-zero bound states.  
    \item 
    Consequently, the expected chiral enhancement is absent, making it impossible to account for the observed SM flavor hierarchies.  
    \item 
    Finally, the construction ultimately requires introduction of  an EDTC symmetry to generate realistic fermion masses. This symmetry is already an essential ingredient of the full DTC paradigm.  
\end{itemize}
Taken together, these considerations show that the minimal DTC framework fails to explain the SM flavor structure. Therefore, the minimal version of the paradigm is theoretically disfavored.  

\section{Scaling relations and mass spectrum}
\label{scaling}

The mass spectrum of the DTC paradigm can be estimated through scaling relations. 
Since the TC, DTC, and DQCD dynamics can be regarded as rescaled replicas of QCD dynamics, 
their properties can be related to those of QCD using the 't~Hooft large-$N$ limit~\cite{tHooft:1973alw,Witten:1979kh}. 
For instance, in the large-$N_{\rm th}$ color limit,
\begin{align}
F_{\rm th} \propto \sqrt{N_{\rm th}} \Lambda_{\rm th}, 
\end{align}
where $\Lambda_{\rm th}$ denotes the confinement scale of a generic non-Abelian strong gauge theory and $N_{\rm th}$ its number of colors. 
This leads to the following scaling relation for decay constants,
\begin{align}
\label{decay_const}
F_{\rm th} = 
\sqrt{\frac{N_{\rm th}}{N_{\rm QCD}}} 
\frac{\Lambda_{\rm th}}{\Lambda_{\rm QCD}} f_\pi,
\end{align}
with $f_\pi = 95.4~\text{MeV}$~\cite{pdg24}  with $m_{\rm dyn} \approx \Lambda_{\rm QCD} \simeq 250~\text{MeV}$~\cite{Bethke:2012jm} . 

A similar scaling relation can be written for the mass of the pseudoscalar $\eta_{\rm th}^\prime$~\cite{Tandean:1995ci},
\begin{equation}
\label{eta_th}
m_{\eta_{\rm th}^\prime} =
\sqrt{\tfrac{2}{3}} \,
\frac{F_{\rm th}}{f_{\pi}}
\frac{N_{\rm QCD}}{N_{\rm th}} \,
m_{\eta_{\rm QCD}^\prime},
\end{equation}
where $m_{\eta_{\rm QCD}^\prime}=957.78$ MeV ~\cite{pdg24}.

The mass of the DTC scalar, using Eq.~\eqref{higgs_mass}, is given by,
\begin{align}
\label{scal_dtc}
m_{H_{\rm DTC}} \approx 2 \Lambda_{\rm  DTC} \exp(k_{\rm DTC} \Delta \chi_{\rm DTC}),
\end{align}
and similarly, for the scalar in the DQCD sector,
\begin{align}
\label{scal_d}
m_{H_{\rm D}} \approx 2  \Lambda_{\rm  D} \exp(k_{\rm D} \Delta \chi_{\rm D}).
\end{align}

\begin{table}[H]
\setlength{\tabcolsep}{6pt} 
\renewcommand{\arraystretch}{1} 
\centering
\begin{tabular}{l|c|c|c}
\toprule
 & $\Lambda = \Lambda_{\rm DTC}$  & $\Lambda > \Lambda_{\rm DTC}$ & $\Lambda < \Lambda_{\rm DTC}$ \\
\midrule
$m_{\eta_{\rm TC}^\prime} $ [GeV] & $2025$ & $2025$ & $2025$   \\
$m_{\eta_{\rm DTC}^\prime} $ [GeV] & $721$ & $721$ & $1442$    \\
$m_{\eta_{\rm D}^\prime}$ [GeV]& $931$ & $1862$  & $931$  \\
\bottomrule
\end{tabular}
\caption{Masses of the pseudoscalar states $\eta_{\rm TC,DTC,D}^\prime$ in the SHVM obtained using Eq.~\eqref{eta_th}.}
\label{tab:eta_mass1}
\end{table}

The masses of the lowest-lying scalar states, determined from Eqs.~\eqref{scal_dtc} and~\eqref{scal_d}, are summarized in Table~\ref{tab:scalar_mass1}.

\begin{table}[H]
\setlength{\tabcolsep}{6pt} 
\renewcommand{\arraystretch}{1} 
\centering
\begin{tabular}{l|c|c|c}
\toprule
 & $\Lambda = \Lambda_{\rm DTC}$  & $\Lambda > \Lambda_{\rm DTC}$ & $\Lambda < \Lambda_{\rm DTC}$ \\
\midrule
$m_{H_{\rm DTC}} $ [GeV] & $1028.4$ & $1038.7$ & $2073.3$   \\
$m_{H_{\rm D}}$ [GeV]& $1002$   & $2119.4$   & $1012$  \\
\bottomrule
\end{tabular}
\caption{Masses of the scalar states in the DTC and DQCD sectors in the SHVM for $\Delta \chi_{\rm DTC,D}= 2$.}
\label{tab:scalar_mass1}
\end{table}

\section{Collider phenomenology of the DTC paradigm}
\label{col_sig}
The mass spectrum derived in Sec.~\ref{scaling} implies that both the DTC and DQCD sectors give rise to 
a rich pseudoscalar spectrum extending from sub-TeV to TeV scales. 
Thus,  all lowest-lying scalars remain accessible to collider experiments. 
This opens up a range of discovery opportunities at the HL-LHC and future hadron colliders. 
The most promising search channels involve diphoton resonances, complemented by $b\bar b$,  $\tau^+\tau^-$ and $\bar{t}t$ final states, 
with production dominated by gluon fusion in analogy to the SM Higgs.

In this section, we investigate the collider signatures of the DTC paradigm at the HL-LHC, the HE-LHC, and a future 100~TeV hadron collider. 
For concreteness, we assume that the TC sector consists of a single fermionic doublet, i.e.~two flavors $T$ and $B$. 
Our primary interest lies in the phenomenology of the lowest-lying scalars and pseudoscalars of the TC, DTC, and DQCD spectrum. 
The focus will be on their production rates and discovery sensitivities at present and future collider facilities.  

We assume that the interactions between the TC and DTC sectors, and between TC and DQCD, are negligible.\footnote{In fact, such mixings are suppressed by factors of $1/\Lambda$ and can be safely ignored.}  
As a result, the mixing between the SM Higgs and the additional $\chi_i$ fields of the SHVM is effectively zero. 
This implies that existing LHC searches for low-mass scalars~\cite{CMS:2018amk} do not constrain this scenario. 
Moreover, decays of composite scalars and pseudoscalars into $WW$ and $ZZ$ occur only at the one-loop level and are therefore strongly suppressed. 
The most relevant experimental constraints instead come from searches for scalar resonances in diboson channels: 
ATLAS excludes masses above $300$~GeV in $WW/ZZ$ final states~\cite{ATLAS:2022eap}, while CMS sets a bound above $200$~GeV~\cite{CMS:2019bnu}. 
In the diphoton channel, ATLAS excludes scalar masses above $200$~GeV~\cite{ATLAS:2021uiz}, while CMS pushes this limit to $500$~GeV~\cite{CMS:2018dqv}.

\subsection{Lagrangian for scalar, pseudoscalar and vector mesons}

\begin{figure}[h]
	\centering
 \includegraphics[width=\linewidth]{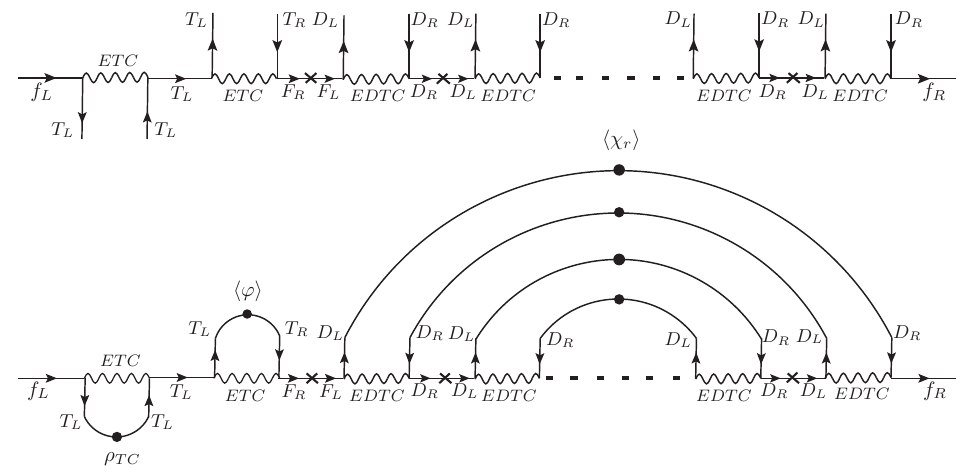}
    \caption{Effective couplings of $\rho_{\rm TC}$ to SM fermions in the DTC paradigm.}
 \label{fig_rho_TC}	
\end{figure}

To investigate the spectrum of the DTC paradigm, we note that the left- and right-handed SM fermions are embedded in different ETC and EDTC symmetries. As a result, the scalar, pseudoscalar, and vector mesons of the DTC spectrum do not couple directly to SM fermions. For instance, the vector meson $\rho_{\rm TC}$ couples to SM fermions only through higher-order interactions, as illustrated in Fig.~\ref{fig_rho_TC}. We assume that the $\rho_{\rm TC}$ is a $\bar{T}_L T_L$ bound state, which corresponds to the most attractive channel within the EMAC hypothesis.

These interactions give rise to the effective Lagrangian
\bea
\label{rho_TC}
{\mathcal{L}} &=& -\sum_{f} \lambda_f \dfrac{1}{\Lambda }
\Bigl[ \bar{\psi}_f \, \varphi \, \psi_f \, \chi _r \Bigr]  
\dfrac{1}{F_{\rm TC}} \gamma_\mu \rho_{\rm TC}^\mu
+  {\rm H.c.}, \\ \nonumber
&=& -\sum_{f} \lambda_f \,\frac{m_f}{F_{\rm TC}}\, 
    \bar{\psi}_f  \gamma_\mu \psi_f \, \rho_{\rm TC}^\mu 
    +  {\rm H.c.},
\eea
where the coefficients $\lambda_f$ encode the effective ETC couplings to SM fermions. Parametrically, one finds
\begin{equation}
 \lambda_f \propto \frac{\Lambda_{\rm TC}^2}{\Lambda_{\rm ETC}^2},
\end{equation}
which is strongly suppressed for $\Lambda_{\rm ETC} \sim 10^7~\text{GeV}$.

Consequently, the direct couplings of $\rho_{\rm TC}$ to SM fermions are highly suppressed, 
conventional Drell--Yan searches are not sensitive, 
making vector-boson fusion (VBF) the more relevant production mechanism. Analogously, the decays of the pseudoscalar $\eta_{\rm TC}^\prime$ into SM fermions are suppressed by the same numerical factor. 

The CMS Collaboration has recently performed a search for heavy vector states in the VBF channel~\cite{CMS:2022pjv}, 
reporting a local excess of $3.6\sigma$ (global significance $2.3\sigma$) around $2.1~\text{TeV}$. While this result is not statistically conclusive, 
it is noteworthy that the excess lies close to the predicted techni-rho mass, 
$m_{\rho_{\rm TC}} \simeq 2~\text{TeV}$. A more definitive assessment of this possible connection requires further experimental data 
and a dedicated collider study, which is beyond the scope of the present work.

\medskip

\begin{figure}[H]
	\centering
 \includegraphics[width=\linewidth]{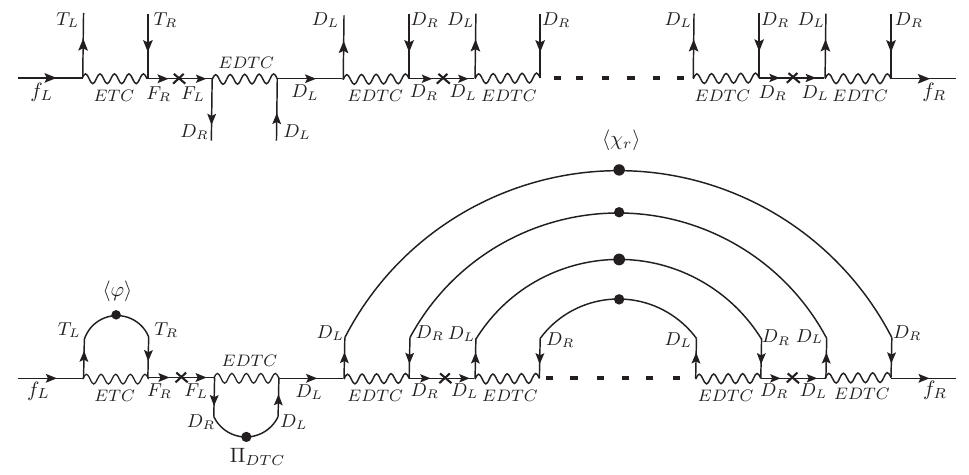}
    \caption{Effective couplings of the DTC pion $\Pi_{\rm DTC}$ to SM fermions.}
 \label{fig_PI_DTC}	
\end{figure}

The interactions of SM fermions with the DTC pion $\Pi_{\rm DTC}$ are shown in Fig.~\ref{fig_PI_DTC}, where we assume that it is a $\bar{D}_L D_R$ bound state, which corresponds to the most attractive channel for spinless mesons.

The resulting Lagrangian is given by
\bea
\label{PI_DTC}
{\mathcal{L}} &=& -\sum_{f} \lambda_f \dfrac{1}{\Lambda }
\Bigl[ \bar{\psi}_f \, \varphi \, \psi_f \, \chi _r \Bigr]  
\dfrac{1}{F_{\rm DTC}} \, \bar{\psi}_f i \gamma_5 \psi_f \, \Pi_{\rm DTC}
+  {\rm H.c.}, \\ \nonumber
&=& -\sum_{f} \lambda_f \,\frac{m_f}{F_{\rm DTC}}\, 
    \bar{\psi}_f  i \gamma_5 \psi_f \, \Pi_{\rm DTC} 
    +  {\rm H.c.},
\eea
where the coefficients $\lambda_f$ parametrize the effective EDTC couplings to SM fermions, with
\begin{equation}
 \lambda_f \propto \frac{\Lambda_{\rm DTC}^2}{\Lambda_{\rm EDTC}^2}.
\end{equation}
From the fit results we note that the scales $\Lambda_{\rm DTC}$ and $\Lambda_{\rm EDTC}$ are numerically close. Therefore, the coefficients $\lambda_f$ are not negligible and can give rise to distinctive DTC collider phenomenology.

\medskip

Finally, the DQCD pion $\Pi_{\rm D}$ arises as a $\bar{F}_L F_R$ bound state, with interactions mediated by GUT bosons, similar to the mechanism in Fig.~\ref{fig_nu}. The corresponding effective Lagrangian is
\bea
\label{PI_DQCD}
{\mathcal{L}} &=& -\sum_{f} \lambda_f \dfrac{1}{\Lambda }
\Bigl[ \bar{\psi}_f \, \varphi \, \psi_f \, \chi _r \Bigr]  
\dfrac{1}{F_{\rm D}} \, \bar{\psi}_f i \gamma_5 \psi_f \, \Pi_{\rm D}
+  {\rm H.c.}, \\ \nonumber
&=& -\sum_{f} \lambda_f \,\frac{m_f}{F_{\rm D}}\, 
    \bar{\psi}_f  i \gamma_5 \psi_f \, \Pi_{\rm D} 
    +  {\rm H.c.},
\eea
where the coefficients $\lambda_f$ encode the effective GUT couplings to SM fermions. Parametrically,
\begin{equation}
 \lambda_f \propto \frac{\Lambda^2}{\Lambda_{\rm GUT}^2},
\end{equation}
which are strongly suppressed for $\Lambda_{\rm GUT} \sim 10^8 - 10^{16}~\text{GeV}$. Therefore, collider signatures of $\Pi_{\rm D}$ to a fermionic pair are expected to be negligible and will not be explored further in this work.

In summary, the technicolor mesons $\rho_{\rm TC}$ and $\eta_{\rm TC}^\prime$ couple  very weakly to SM fermions due to the large ETC scale, suppressing their direct collider signatures. By contrast, the DTC pion $\Pi_{\rm DTC}$ can couple appreciably to SM fermions since $\Lambda_{\rm DTC}$ and $\Lambda_{\rm EDTC}$ are of comparable magnitude, making it the most promising state for collider phenomenology within this sector. Finally, the DQCD pion $\Pi_{\rm D}$ is essentially inert at collider scales, as its interactions are suppressed by the ultra-high GUT scale.  For clarity, we summarize these results in Table~\ref{tab:meson_summary}.  The table concludes that only relevant collider physics in the fermionic final states  comes from only DTC scalars and pseudoscalars.

\begin{table}[h]
\centering
\begin{tabular}{|c|c|c|c|}
\hline
\textbf{State} & \textbf{Constituents} & \textbf{Suppression Scale} & \textbf{Collider relevance } \\ 

&                                      &                            &         \textbf{for fermionic final states}                                        \\
\hline
$\rho_{\rm TC}$  & $\bar{T}_L T_L$ & $\Lambda_{\rm ETC} \sim 10^7~\text{GeV}$ & Negligible \\
\hline
$\eta_{\rm TC}^\prime$  & $\bar{T} T$ & $\Lambda_{\rm ETC} \sim 10^7~\text{GeV}$ & Negligible \\
\hline
$\Pi_{\rm DTC}$, $H_{\rm DTC}$  & $\bar{D}_L D_R$ & $\Lambda_{\rm DTC} \sim \Lambda_{\rm EDTC}$ & Promising \\
\hline
$\Pi_{\rm D}$, $H_D$ & $\bar{F}_L F_R$ & $\Lambda_{\rm GUT} \sim 10^8 - 10^{16}~\text{GeV}$ & Negligible \\
\hline
\end{tabular}
\caption{Comparison of suppression scales and collider relevance for technicolor and related bound states in the DTC paradigm.}
\label{tab:meson_summary}
\end{table}

The main collider signatures of the DTC sector therefore arise from inclusive production of pions, etas, and scalars, followed by decays into SM fermions or photons:
\begin{align}
pp \;\to\; \Pi_{\rm DTC}/\eta_{\rm DTC}^\prime/H_{\rm DTC} \;\to\; f_i f_j \,, \;\gamma \gamma \,,
\end{align}
where $f_i$ denotes a generic SM fermion.  

To connect with experimental searches, we adopt benchmark scalar and pseudoscalar mass values of $500$~GeV and $1000$~GeV in our collider analysis. 
These benchmarks are motivated by current ATLAS and CMS searches for new resonances, which probe precisely this mass range. 
They also serve as representative points to study the reach of future facilities. 
A dedicated discussion of the projected sensitivities at the HL-LHC, HE-LHC, and a 100~TeV collider is presented below. 
This strategy follows the methodology of several recent collider studies~\cite{Abbas:2024jut,Abbas:2024dfh}.

The production cross section for a scalar or pseudoscalar resonance $\phi=\phi_S (\mbox{scalar}),\phi_P (\mbox{pseudoscalar})$ of mass $M$ and decaying to a final state  X is given by \cite{Arcadi:2023smv},
\begin{equation}
\sigma(p p \rightarrow \phi \rightarrow X) =  \frac{1}{M  s}  C_{gg}\Gamma(\phi\rightarrow gg) BR(\phi\rightarrow X),
\end{equation}

where $C_{gg}$ is the weight factor accounting for the PDFs of the proton and the color factors, and $s$ denotes the squared center of mass energy. The values of the $C_{gg}$ are determined from the PDFs as follows \cite{Arcadi:2023smv},
\begin{equation}
C_{gg} = \frac{\pi^2}{8} \int_{M^2/s}^1 \frac{dx}{x} g(x)g\left(\frac{M^2}{sx}\right).
\end{equation}
We use the {\tt MSTW2008} PDF \cite{Martin:2009iq} to generate the production cross-sections of these particles through various modes. 

The partial decay widths to $f\bar{f}$ are given by \cite{Arcadi:2023smv},
\begin{eqnarray}
\Gamma(\phi_S\rightarrow ff) &=& \frac{N_C^f g_{sff}^2 m_f^2(M) M}{8\pi v^2}\left(1-\frac{4m_f^2}{M^2}\right)^{3/2},\\
\Gamma(\phi_P\rightarrow ff) &=& \frac{N_C^f g_{pff}^2 m_f^2(M) M}{8\pi v^2}\left(1-\frac{4m_f^2}{M^2}\right)^{1/2},
\end{eqnarray}
where $g_{sff},g_{pff}$ are the ratios between the quark coupling to the spin-0 particle and the SM Yukawa couplings and the color factor $N_C^f=3$ for quarks and 1 for leptons.

The partial decay widths to $gg$ and $\gamma\gamma$ are expressed as \cite{Arcadi:2023smv},
\begin{eqnarray}
\Gamma(\phi_S\rightarrow gg) &=& \frac{ \alpha_s^2 M^3}{32\pi^3v^2} |\sum_f g_{sff} F_S\left(\frac{M^2}{4m_f^2}\right)|^2,\\
\Gamma(\phi_P\rightarrow gg) &=& \frac{ \alpha_s^2 M^3}{32\pi^3 v^2} |\sum_f g_{pff} F_P\left(\frac{M^2}{4m_f^2}\right)|^2,\\
\Gamma(\phi_S\rightarrow \gamma\gamma) &=& \frac{ \alpha^2 M^3}{256\pi^3v^2} |\sum_f 2N_C^f Q_f^2 g_{sff} F_S\left(\frac{M^2}{4m_f^2}\right)|^2,\\
\Gamma(\phi_P\rightarrow \gamma\gamma) &=& \frac{ \alpha^2 M^3}{256\pi^3 v^2} |\sum_f 2N_C^f Q_f^2 g_{pff} F_P\left(\frac{M^2}{4m_f^2}\right)|^2.
\end{eqnarray}
The form factors $F_S(x)$ and $F_P(x)$ can be written as,
\begin{eqnarray}
F_S(x) &=& x^{-1} (1+(1-x^{-1})f(x)),\\
F_P(x) &=&x^{-1}f(x).
\end{eqnarray}
where,
\[
f(x) =
\begin{cases}
\arcsin^2(\sqrt{x}), & x \leq 1 \\
-\dfrac{1}{4} \left( \log\left( \dfrac{\sqrt{x} + \sqrt{x - 1}}{\sqrt{x} - \sqrt{x - 1}} \right) - i\pi \right)^2, & x > 1
\end{cases}
\]
and $x= \frac{M^2}{4m_f^2}$.


The $\chi_r$ fields couple to the SM fermions through the Lagrangian  in equation (41).  Therefore, their coupling to a pair of fermion is of the order $\dfrac{y_{ij}^f m_f}{\Lambda}$.  On the other side, the DTC pions  couple to the SM fermions through the Lagrangian  in equation (109), and their coupling to a pair of fermion is of the order $\dfrac{\lambda^f m_f}{F_{\Pi_{\rm DTC}}}$.  Therefore, fields $\chi_r$ fields do not affect the production and decays of DTC pions, DTC eta and DTC higgs at leading order.

\subsection{Current and future sensitivities}
The sensitivities of the production cross-sections of  a heavy  pseudoscalar in different modes for the  HL-LHC, the HE-LHC, and a 100 TeV hadron collider are estimated in reference \cite{Abbas:2024dfh}, and are given in table \ref{tab:futurelimits11} 

\begin{table}[H]
\setlength{\tabcolsep}{6pt} 
\renewcommand{\arraystretch}{1} 
\centering
\begin{tabular}{l|cc|cc|cc}
\toprule
 & \multicolumn{2}{c|}{HL-LHC [14 TeV, $3~\iab$] } & \multicolumn{2}{c|}{HE-LHC [27 TeV, $15~\iab$]} & \multicolumn{2}{c}{100 TeV, $30~\iab$} \\
$m$~[GeV] &  500 &  1000 &  500 &  1000 &  500 &  1000 \\
\midrule
jet-jet [pb] &   & $4 \e{-2}$ & & $3 \e{-2}$  &     & $4\e{-2}$\\
$\tau \tau$ [pb]  & $7\e{-3}$ & $1\e{-3}$ & $4\e{-3}$ & $7\e{-4}$ & $5\e{-3}$ & $8\e{-4}$   \\
$e e$, $\mu \mu$ [pb] & $2\e{-4}$ & $4\e{-5}$  & $1\e{-4}$ & $3\e{-5}$  & $1\e{-4}$ & $3\e{-5}$  \\
$\gamma \gamma$ [pb] & $1\e{-4}$  & $2\e{-5}$  & $6\e{-5}$ & $1\e{-5}$  & $7\e{-5}$ & $1\e{-5}$  \\
$b  \bar{b}$ [pb]    & & $9\e{-3}$    &  & $5\e{-3}$    &  & $7\e{-3}$  \\
$t \bar{t}$ [pb]     & 4   & $5\e{-2}$      & 3      & $4\e{-2}$    & $8$ & $0.1$  \\
\bottomrule
\end{tabular}
\caption{Projected reach  $\sigma \times \text{BR}$ for high-mass scalar or pseudoscalar resonance searches through inclusive production channels at the HL-LHC, the HE-LHC, and a future 100 TeV collider.}
\label{tab:futurelimits11}
\end{table}

In addition to the above sensitivities of masses, the mass spectrum of the DTC sector contains other different masses of scalars and pseudoscalars.  The sensitivities of these masses are not given in reference   \cite{Abbas:2024dfh}. To estimate the sensitivities of these masses in the HL-LHC, the HE-LHC, and a future 100 TeV collider, we use the prescription discussed in  reference  \cite{Abbas:2024dfh}.  For this purpose, we use 
square root scaling of the luminosity of the LHC by,
\begin{equation}
 \mathcal{S} \simeq \frac{S}{\sqrt{B}} \simeq    \sqrt{\mathcal{L}} \frac{\sigma_s}{\sqrt{\sigma_B}},
\end{equation} 
where $S$ denotes the number of signal events, $B$ is the background events, $\sigma_s$ stands for the signal cross-section, and 
$\sigma_B$ shows the background cross-section.

As discussed in reference \cite{Abbas:2024dfh}, a conservative estimate of the sensitivities of the HL-LHC, the HE-LHC and 100 TeV collider can be made with the following assumptions:
\begin{enumerate}
    \item   The significance $\mathcal{S} \simeq \frac{S}{\sqrt{B}}$ does not change among colliders.
    \item The reconstruction efficiencies and  background rejection remain constant among  colliders.
\end{enumerate}
These assumptions are also used in the ``Collider Reach" tool, which is capable of providing an estimate of the mass of a BSM physics at the LHC and a future collider \cite{CR}.

Thus, the sensitivity of a  signal of scalar mass  at a future collider (FC) is given by 
\begin{align}
  \sigma_s^{\rm FC}  = \sqrt{\dfrac{\mathcal{L}_{\rm LHC}}{\mathcal{L}_{\rm FC}}} \sqrt{\dfrac{\sigma_{ B}^{ FC}}{\sigma_{ B}^{ LHC}}} \sigma_{s}^{\rm LHC},
\end{align}
where FC= HL-LHC, HE-LHC, and  a 100 TeV collider.  As observed in reference \cite{Abbas:2024dfh},  $\sigma_{ B}^{ FC}$ and $\sigma_{ B}^{ LHC}$  turn out to be $\sigma_{ B}^{ HE-LHC} \leq 2 \sigma_{ B}^{ LHC}$ and $\sigma_{ B}^{ 100 TeV} \leq 10 \sigma_{ B}^{ LHC}$,   and  $ \sigma_{ s}^{ LHC}$ represents the current limits  given in tables \ref{tab:limits_LHC1}.

\begin{table}[H]
\setlength{\tabcolsep}{5.2pt}
\renewcommand{\arraystretch}{1}
\centering

\begin{tabular}{l|cc|ccc|ccc}
\toprule
& \multicolumn{2}{c|}{$\mathcal{L}[fb^{-1}]$ [References]} 
& \multicolumn{3}{c|}{ATLAS 13 TeV} 
& \multicolumn{3}{c}{CMS 13 TeV} \\
$m$~[GeV] 
& \myalign{c}{ATLAS} & \myalign{c|}{CMS} 
& \myalign{c}{721} & \myalign{c}{1442} & \myalign{c|}{2025} 
& \myalign{c}{721} & \myalign{c}{1442} &  \myalign{c}{2025} \\
\midrule
$\tau \tau$~[pb]  
& $36.1$ \cite{ATLAS:2017eiz} 
& $35.9$ \cite{CMS:2018rmh}  
&$2\e{-2}$  & $6\e{-3}$  & $6\e{-3}$
& $2\e{-2}$  & $7\e{-3}$ & $4\e{-3}$ \\
$e e$, $\mu \mu$~[pb]   
& $139$ \cite{ATLAS:2019erb}
& $140$ \cite{CMS:2021ctt}  
& $4\e{-4}$ & $1\e{-4}$ & $6\e{-5}$ 
&  $6\e{-4}$ & $2\e{-4}$ & $8\e{-5}$\\
$\gamma \gamma$~[pb]
& $139$ \cite{ATLAS:2021uiz}  
& $35.9$ \cite{CMS:2018dqv}  
& $2\e{-4}$  & $8\e{-5}$ & $6\e{-5}$  
&   $9\e{-4}$ & $4\e{-4}$ & $1\e{-4}$\\
$b  \bar{b}$~[pb]
& $36.1$ \cite{ATLAS:2019npw,ATLAS:2020lks}  
& $35.9$ \cite{CMS:2018rkg} 
&  & $4 \e{-2}$ & $1\e{-2}$ 
&  &  & $2\e{-2}$ \\
$t  \bar{t}$~[pb]
& $36.1$ \cite{ATLAS:2019npw,ATLAS:2020lks}  
& $35.9$ \cite{CMS:2018rkg} 
& $3.7$ & $1.3$ & $4\e{-2}$  
&   $0.94$ & $0.1$ & $2\e{-2}$\\
\bottomrule
\end{tabular}
\caption{Current limits on production cross-section times branching ratio ($\sigma \times BR$) at 13 TeV LHC from ATLAS and CMS  resonance searches for scalars or pseudoscalars.}
\label{tab:limits_LHC1}
\end{table}

Our estimate of the sensitivities of the masses  given in  tables \ref{tab:limits_LHC1} at the HL-LHC, the HE-LHC, and a future 100 TeV collider are given in table \ref{tab:estimated_HL1a}.

\begin{table}[H]
\setlength{\tabcolsep}{6pt} 
\renewcommand{\arraystretch}{1} 
\centering
\begin{tabular}{l|ccc|ccc|ccc}
\toprule
 & \multicolumn{3}{c|}{HL-LHC [14 TeV, $3~\iab$] } & \multicolumn{3}{c|}{HE-LHC [27 TeV, $15~\iab$]} & \multicolumn{3}{c}{100 TeV, $30~\iab$} \\
$m$~[GeV] &  721 &  1442 & 2025 & 721 &  1442 & 2025 &  721 &  1442 & 2025\\
\midrule

$\tau \tau$ [pb]  & $2\e{-3}$ & $6\e{-4}$ & $4\e{-4}$ 
& $1\e{-3}$ & $4\e{-4}$ & $3\e{-4}$
& $2\e{-3}$ & $7\e{-4}$  & $4\e{-4}$ \\
$e e$, $\mu \mu$ [pb] & $8\e{-5}$ & $2\e{-5}$  & $1\e{-5}$
& $5\e{-5}$ & $1\e{-5}$  & $8\e{-6}$
& $9\e{-4}$ & $2\e{-5}$ & $1\e{-5}$ \\
$\gamma \gamma$ [pb] & $4\e{-5}$  & $2\e{-5}$ & $9\e{-6}$
& $3\e{-5}$ & $1\e{-5}$  & $7\e{-6}$
& $4\e{-5}$ & $2\e{-5}$  & $1\e{-5}$\\
$b  \bar{b}$ [pb]    & & $8\e{-3}$   & $2\e{-3}$
&  & $5\e{-3}$  & $1\e{-3}$ 
&  & $8\e{-3}$  & $2\e{-3}$\\
         
$t \bar{t}$ [pb]     & $0.1$   & $1\e{-2}$ & $2\e{-3}$
& $6\e{-2}$      & $7\e{-3}$ & $2\e{-3}$
& $0.1$ & $1\e{-2}$  & $2\e{-3}$\\
\bottomrule
\end{tabular}
\caption{Projected reach  $\sigma \times \text{BR}$ for high-mass scalar or pseudoscalar resonance searches through inclusive production channels at the HL-LHC, the HE-LHC, and a future 100 TeV collider.}
\label{tab:estimated_HL1a}
\end{table}

\subsection{Signatures of the DTC-sector}
We apply the fit results from sub-section \ref{SHVM_UV} in the case when the SHVM is accommodated within  the DTC paradigm.   These are obtained for three scenarios given, as $\Lambda_{\rm DTC} = \Lambda$, $\Lambda_{\rm DTC} < \Lambda$, and $\Lambda_{\rm DTC} > \Lambda$.The number of colors $\rm N_{\rm DTC}$  is identical in all three scenarios.  We investigate the collider signatures of the $\Pi_{\rm DTC}$, $\eta_{\rm DTC}^\prime$ and $H_{\rm DTC}$ states at the 14 TeV HL-LHC, the 27 TeV HE-LHC, and a 100 TeV collider. As discussed earlier, collider signatures of $\rho_{\rm TC}$, $\eta_{\rm TC}^\prime$, and spectrum of the DQCD are highly suppressed in the fermionic final sates.  Therefore, we do not discuss them in this work.

\begin{figure}[H]
	\centering
	\begin{subfigure}[]{0.4\linewidth}
    \includegraphics[width=\linewidth]{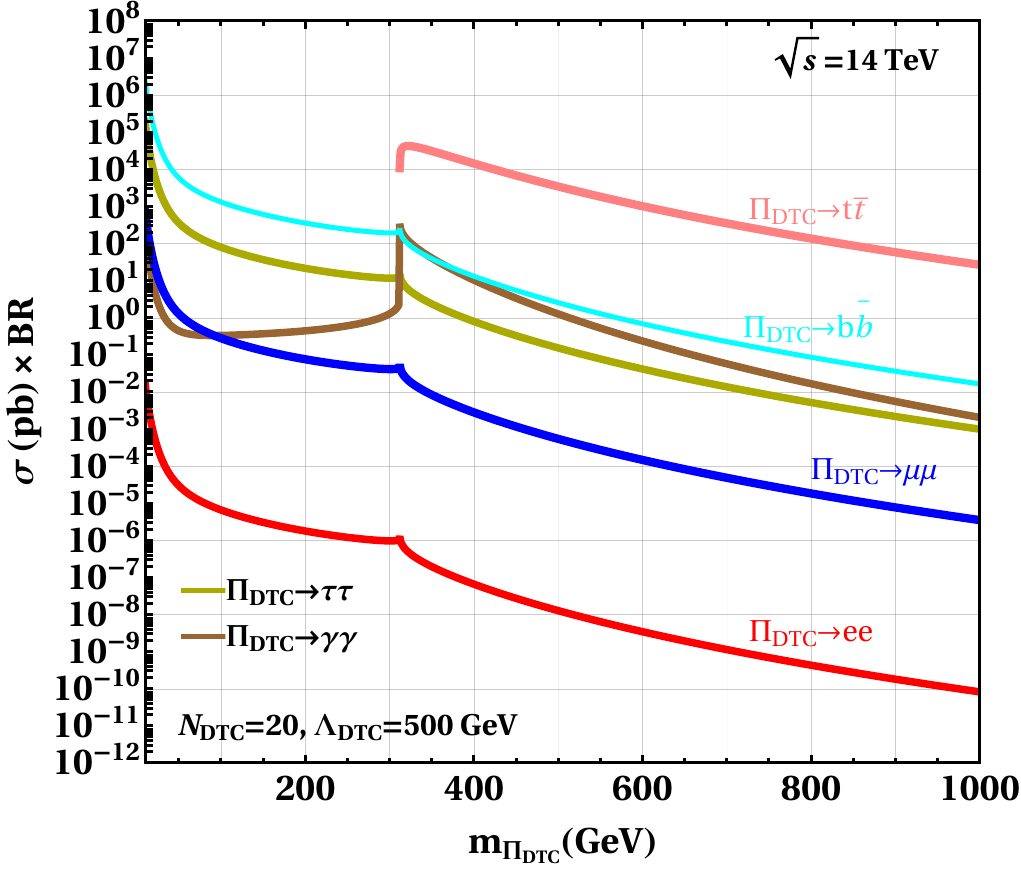}
    \caption{}
         \label{fig_8a}	
\end{subfigure}
\begin{subfigure}[]{0.4\linewidth}
    \includegraphics[width=\linewidth]{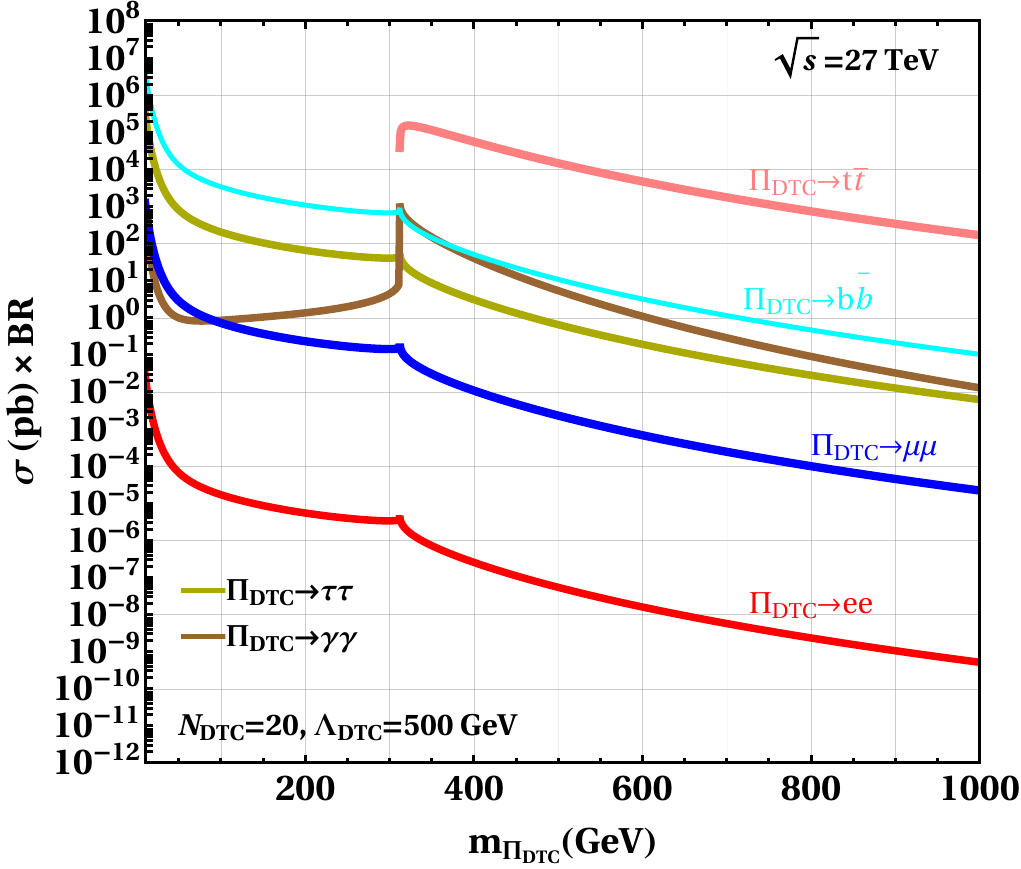}
    \caption{}
         \label{fig_8b}	
\end{subfigure}
 \begin{subfigure}[]{0.4\linewidth}
 \includegraphics[width=\linewidth]{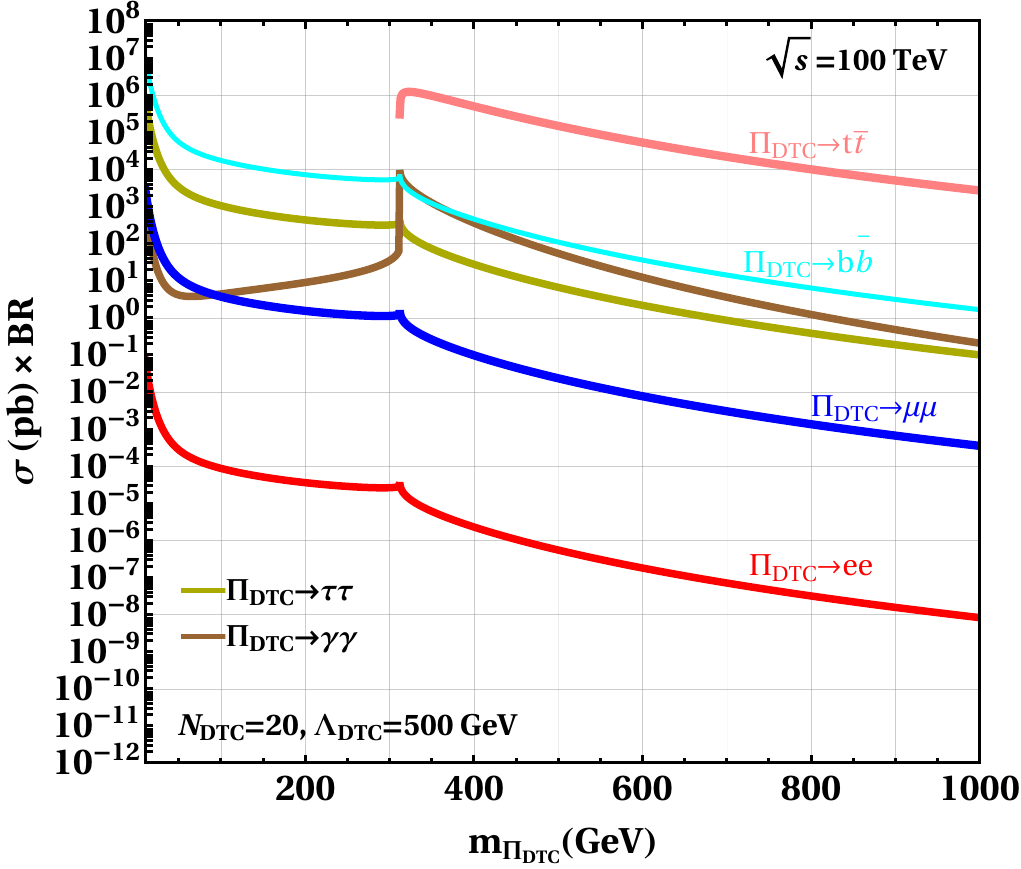}
 \caption{}
         \label{fig_8c}
 \end{subfigure}
 
 \caption{ $\sigma$ $\times$ BR of the various possible decay modes of the $\Pi_{\rm DTC}$ into photons, quark pairs and lepton pairs in the SHVM mechanism for $\rm N_{\rm DTC}=20$,  $\Lambda_{\rm DTC}=500$ GeV at  (\ref{fig_8a}) the 14 TeV HL-LHC (\ref{fig_8b})  the 27 TeV HE-LHC  and (\ref{fig_8c}) a 100 TeV future collider . }
  \label{fig_pidtc_decays_shvm}
	\end{figure}

The variations of  cross-sections of the DTC-pions $\Pi_{\rm DTC}$  is  shown  in figure \ref{fig_pidtc_decays_shvm} .  The benchmark predictions for the production of DTC-pion  for $\Lambda_{\rm DTC} = 500$ GeV  are recorded  in table \ref{tab:bench_pi_dtc_500GeV}  for heavy  masses  at the HL-LHC, the HE-LHC and a 100 TeV collider.


\label{shvm_col1}
\graphicspath{{plots/}}

    
    \begin{table}[H]
\setlength{\tabcolsep}{6pt} 
\renewcommand{\arraystretch}{1} 
\centering
\begin{tabular}{l|cc|cc|cc}
\toprule
 & \multicolumn{2}{c|}{HL-LHC [14 TeV, $3~\iab$] } & \multicolumn{2}{c|}{HE-LHC [27 TeV, $15~\iab$]} & \multicolumn{2}{c}{100 TeV, $30~\iab$} \\

$m_{\pi_{\rm DTC}}$~[GeV] &  500 & 1000&  500 &  1000 &  500 &  1000 \\
\midrule
$\tau \tau$ [pb]    & \fbox{$0.2$} & $9.9\e{-4}$ & \fbox{$0.7$} & \fbox{$6.8\e{-3}$}  & \fbox{$7.4$} & \fbox{$0.1$}   \\
 $\mu \mu$ [pb] & \fbox{$5.4\e{-4}$} & $3.5\e{-6}$  & \fbox{$2.6\e{-3}$} & $2.4\e{-5}$   & \fbox{$2.6\e{-2}$} & \fbox{$3.9\e{-4}$} \\
$e e$ [pb]    & $1.3\e{-8}$ &  $8.2\e{-11}$  & $6.1\e{-8}$ &  $5.6\e{-10}$  & $6.2\e{-7}$ & $9.0\e{-9}$  \\
$\gamma \gamma$ [pb] & \fbox{$1.3$} & \fbox{$2.1\e{-3}$ }  & \fbox{$3.0$} & \fbox{$7.0\e{-3}$ }  & \fbox{$30$} & \fbox{$0.1$}  \\
$b  \bar{b}$ [pb]    & $2.5$ & \fbox{$1.6\e{-2}$}     & $12$ & \fbox{$0.1$}    & $124$ & $1.8$  \\
$t \bar{t}$ [pb]        & \fbox{$3456$}  & \fbox{$29$}    & \fbox{$16561$}  & \fbox{$185$}   & \fbox{$1.7 \e{5}$} & \fbox{$2952$}  \\
\bottomrule
\end{tabular}
\caption{Benchmark points for $\Pi_{\rm DTC}$ production channels for high mass $m_{\Pi_{\rm DTC}}$  at the 14 TeV HL-LHC, 27 TeV HE-LHC and a 100 TeV collider.}
\label{tab:bench_pi_dtc_500GeV}
\end{table}

The benchmark values of the signatures of production of  DTC-eta at the HL-LHC, the HE-LHC and a 100 TeV collider is shown in table \ref{tab:DTC_eta1} for $\Lambda_{\rm DTC}=500$ GeV.  We notice that the mass of the  DTC-eta  is exactly predicted by the scaling relations.  Therefore, we do not show the variation of corresponding cross-sections for the  DTC-eta .
\begin{table}[H]
\setlength{\tabcolsep}{6pt} 
\renewcommand{\arraystretch}{1} 
\centering
\begin{tabular}{l|c|c|c}
\toprule
 & HL-LHC [14 TeV, $3~\iab$]  & HE-LHC [27 TeV, $15~\iab$] & 100 TeV, $30~\iab$ \\
\midrule
$\tau \tau$ [pb]  & \fbox{$1.1\e{-2}$} & \fbox{$5.6\e{-2}$} & \fbox{$0.7$}\\
 $\mu \mu$ [pb]  & $3.9\e{-5}$  & \fbox{$2.0\e{-4}$} & \fbox{$2.5\e{-3}$}  \\
$e e$ [pb]     & $9.1\e{-10}$  & $4.7\e{-9}$ & $5.9\e{-8}$  \\
$\gamma \gamma$ [pb]  & \fbox{$4.3\e{-2}$}  & \fbox{$0.2$} & \fbox{$2.8$}  \\
$b  \bar{b}$ [pb]   & $0.2$    & $0.9$ & $11.8$  \\
$t \bar{t}$ [pb]      & \fbox{$285$}  & \fbox{$1454$} & \fbox{$18367$}  \\
\bottomrule
\end{tabular}
\caption{Benchmark points for $\eta_{\rm DTC}^\prime$ production channels at the 14~TeV HL-LHC, 27~TeV HE-LHC, and a 100~TeV collider  where $m_{\eta_{\rm DTC}^\prime} = 721~\text{GeV}$.
.
}
\label{tab:DTC_eta1}
\end{table}

Our benchmark prediction for the scalar of the DTC paradigm are presented in table \ref{tab:higgs_DTC1} at the HL-LHC, HE-LHC, and a 100 TeV future collider for $\Lambda_{\rm DTC}=500$ GeV. 

\begin{table}[H]
\setlength{\tabcolsep}{6pt} 
\renewcommand{\arraystretch}{1} 
\centering
\begin{tabular}{l|c|c|c}
\toprule
 & HL-LHC [14 TeV, $3~\iab$]  & HE-LHC [27 TeV, $15~\iab$] & 100 TeV, $30~\iab$ \\
\midrule
$\tau \tau$ [pb]  & $4.6\e{-5}$ & \fbox{$2.9\e{-3}$} & \fbox{$4.8\e{-2}$}\\
 $\mu \mu$ [pb]  & $1.6\e{-6}$  & $1.0\e{-5}$ & $1.7\e{-4}$  \\
$e e$ [pb]     & $3.8\e{-11}$  & $2.4\e{-10}$ & $3.9\e{-9}$  \\
$\gamma \gamma$ [pb]  & \fbox{$3.9\e{-4}$}  & \fbox{$2.5\e{-3}$} & \fbox{$4.1\e{-2}$}  \\
$b  \bar{b}$ [pb]   & $7.6\e{-3}$    & \fbox{$4.8\e{-2}$} & \fbox{$0.8$ } \\
$t \bar{t}$ [pb]      & \fbox{$11.3$ }   & \fbox{$73$} & \fbox{$1190$}  \\
\bottomrule
\end{tabular}
\caption{Benchmark points for $H_{\rm DTC}$ production channels  at the 14 TeV HL-LHC, 27 TeV HE-LHC and a 100 TeV collider  with $m_{H_{\rm DTC}}=1028$ GeV.}
\label{tab:higgs_DTC1}
\end{table}
 \begin{table}[H]
\setlength{\tabcolsep}{6pt} 
\renewcommand{\arraystretch}{1} 
\centering
\begin{tabular}{l|cc|cc|cc}
\toprule
 & \multicolumn{2}{c|}{HL-LHC [14 TeV, $3~\iab$] } & \multicolumn{2}{c|}{HE-LHC [27 TeV, $15~\iab$]} & \multicolumn{2}{c}{100 TeV, $30~\iab$} \\

$m_{\pi_{\rm DTC}}$~[GeV] &  500 & 1000&  500 &  1000 &  500 &  1000 \\
\midrule
$\tau \tau$ [pb]  & \fbox{$2.8 \e{-2}$} & $1.7\e{-4}$ & \fbox{$0.1$} & \fbox{$1.1\e{-3}$} & \fbox{$1.2$} & \fbox{$1.8\e{-2}$}   \\
 $\mu \mu$ [pb] & $9.8\e{-5}$ &  $6.2\e{-7}$  & $4.2\e{-5}$ & $3.9\e{-6}$  & \fbox{$4.2\e{-3}$} & $6.2\e{-5}$  \\
$e e$ [pb]    & $2.3\e{-9}$ &  $1.4\e{-11}$ & $9.8\e{-9}$ & $9.1\e{-11}$  & $9.9\e{-8}$ & $1.5\e{-9}$  \\
$\gamma \gamma$ [pb] & \fbox{$0.1$} &   \fbox{$1.8\e{-4}$}  & \fbox{$0.5$} & \fbox{$1.1\e{-3}$}  & \fbox{$4.8$} & \fbox{$8.3\e{-2}$}  \\
$b  \bar{b}$ [pb]   & $0.4$ & $2.9\e{-3}$    & $2.0$ & \fbox{$1.8\e{-2}$ }   & $20$ & \fbox{$0.3$}  \\
$t \bar{t}$ [pb]    &    \fbox{$627$} & \fbox{$4.7$}      & \fbox{$2662$}      & \fbox{$0.2$ }   & \fbox{$26938$} & \fbox{$474$}  \\
\bottomrule
\end{tabular}
\caption{Benchmark points for $\Pi_{\rm DTC}$ production channels for high mass $m_{\Pi_{\rm DTC}}$  at the 14 TeV HL-LHC, 27 TeV HE-LHC and a 100 TeV collider.}
\label{tab:bench_SHVM2_pi_dtca}
\end{table}

\begin{figure}[H]
	\centering
	\begin{subfigure}[]{0.4\linewidth}
    \includegraphics[width=\linewidth]{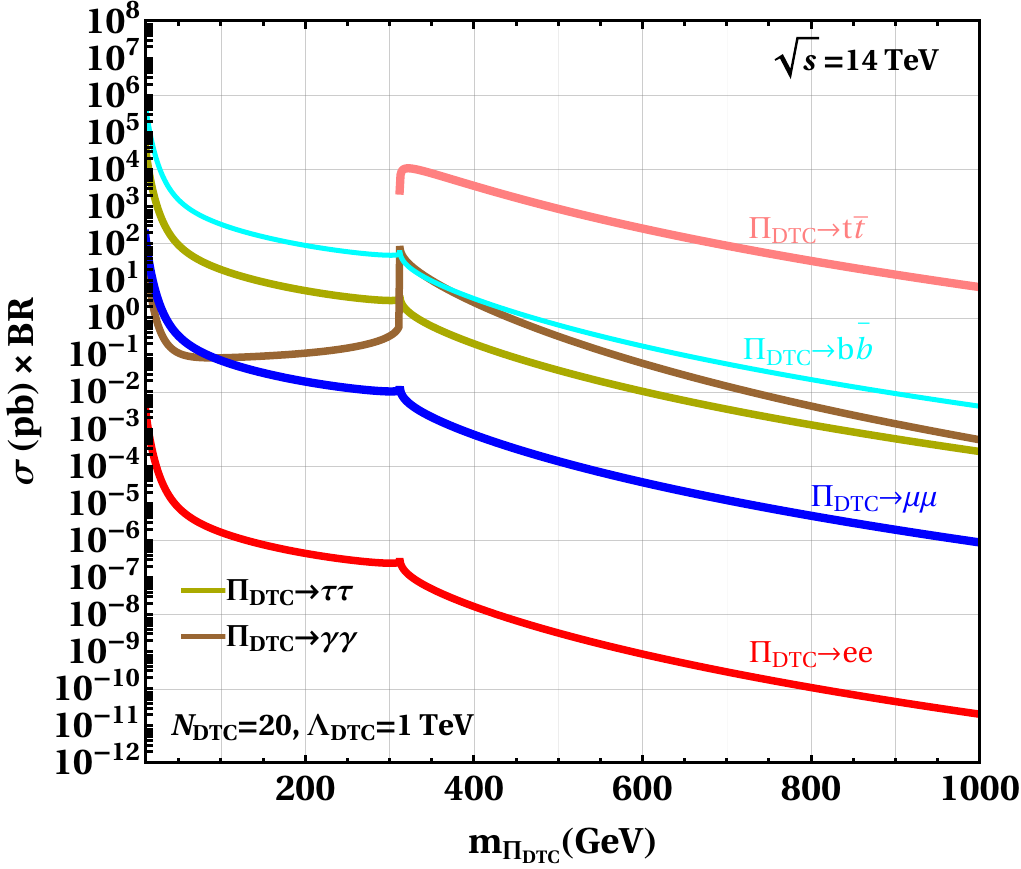}
    \caption{}
         \label{fig11_a}	
\end{subfigure}
   \begin{subfigure}[]{0.4\linewidth}
    \includegraphics[width=\linewidth]{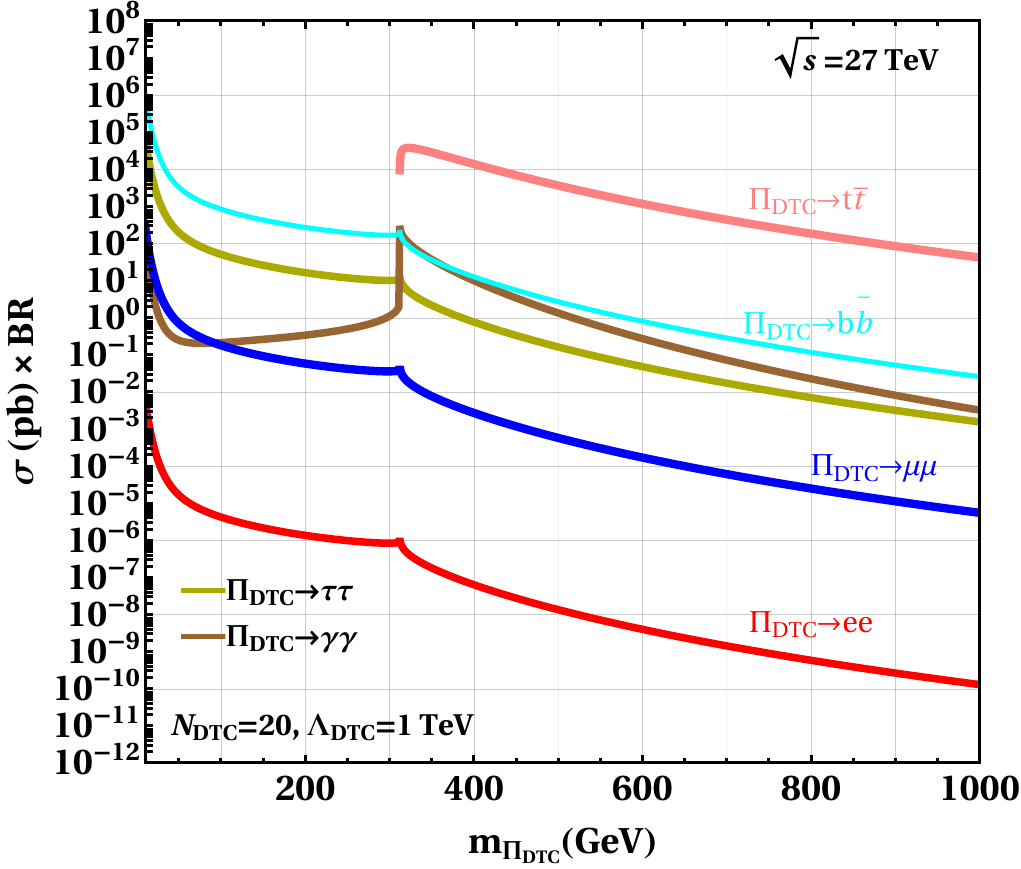}
    \caption{}
         \label{fig11_b}	
\end{subfigure}
 \begin{subfigure}[]{0.4\linewidth}
 \includegraphics[width=\linewidth]{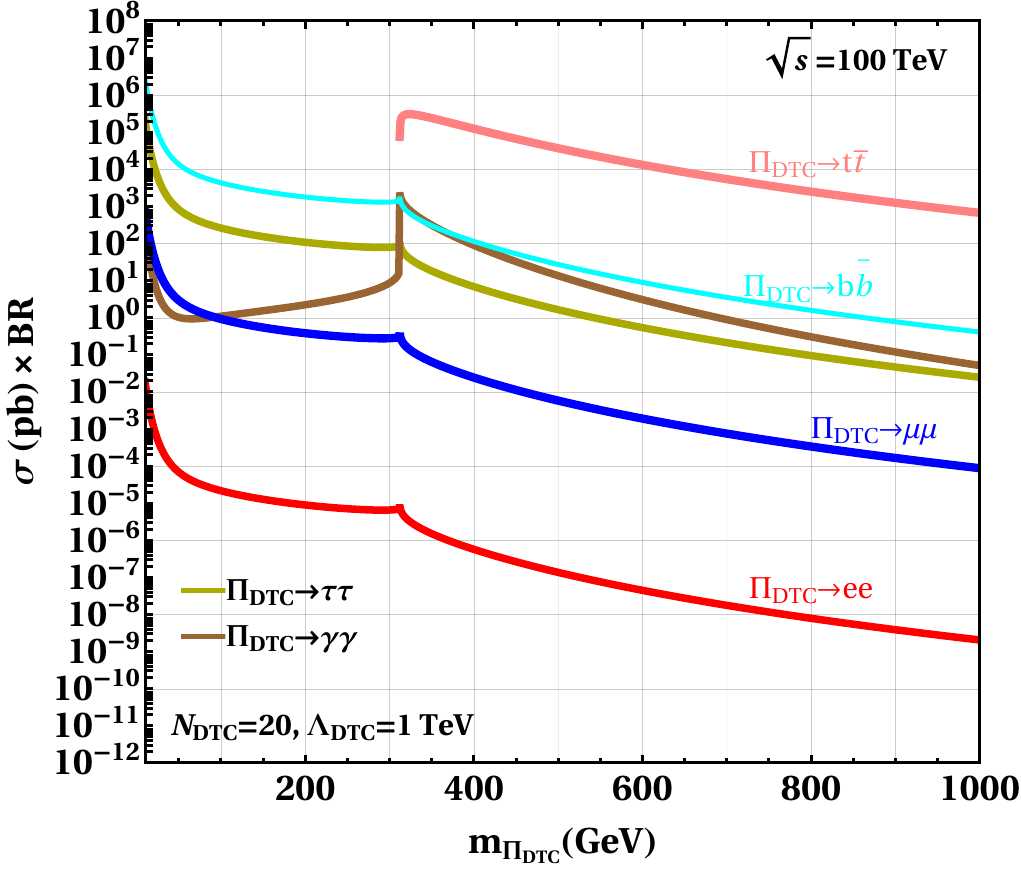}
 \caption{}
         \label{fig11_c}
 \end{subfigure} 
 \caption{ $\sigma$ $\times$ BR of the various possible decay modes of the $\Pi_{\rm DTC}$ into photons, quark pairs and lepton pairs for $\rm N_{\rm DTC}=20, \Lambda_{\rm DTC}= 1$ TeV at the (\ref{fig11_a}) 14 TeV HL-LHC, (\ref{fig11_b}) 27 TeV HE-LHC, and (\ref{fig11_c}) a 100 TeV collider. }
  \label{fig_piDTC_decays_SHVM2}
	\end{figure}

  The production cross-sections of DTC-pion for $\Lambda_{\rm DTC} =10^3$ GeV for    various channels is shown in figure \ref{fig_piDTC_decays_SHVM2} at the HL-LHC, HE-LHC, and a 100 TeV future collider.  Benchmark signatures  for $\Pi_{\rm DTC}$,  $\eta_{\rm DTC}^\prime$ and $H_{\rm DTC}$ are given in tables \ref{tab:bench_SHVM2_pi_dtca}-\ref{tab:higgs_DTC3a} for $\Lambda_{\rm DTC}=10^3$ GeV.

\begin{table}[H]
\setlength{\tabcolsep}{6pt} 
\renewcommand{\arraystretch}{1} 
\centering
\begin{tabular}{l|c|c|c}
\toprule
 & HL-LHC [14 TeV, $3~\iab$]  & HE-LHC [27 TeV, $15~\iab$] & 100 TeV, $30~\iab$ \\
\midrule
$\tau \tau$ [pb]  & $1.4\e{-5}$ & $1.2\e{-4}$ & \fbox{$9.1\e{-4}$}\\
 $\mu \mu$ [pb]  & $5.1\e{-8}$  & $4.3\e{-7}$ & $3.2\e{-6}$  \\
$e e$ [pb]     & $1.2\e{-12}$  & $1.0\e{-11}$ & $7.5\e{-11}$  \\
$\gamma \gamma$ [pb]  & $1.5\e{-5}$  & \fbox{$1.2\e{-4}$} & \fbox{$5.8\e{-4}$}  \\
$b  \bar{b}$ [pb]   & $2.4\e{-4}$    & $2.0\e{-3}$ & $1.5\e{-2}$  \\
$t \bar{t}$ [pb]      & \fbox{$0.4$}  & \fbox{$3.4$} & \fbox{$25$}  \\
\bottomrule
\end{tabular}
\caption{Benchmark points for $\eta_{\rm DTC}^\prime$ production channels at the 14~TeV HL-LHC, 27~TeV HE-LHC, and a 100~TeV collider  where $m_{\eta_{\rm DTC}^\prime} = 1442~\text{GeV}$.
.
}
\label{tab:DTC_eta2a}
\end{table}

\begin{table}[H]
\setlength{\tabcolsep}{6pt} 
\renewcommand{\arraystretch}{1} 
\centering
\begin{tabular}{l|c|c|c}
\toprule
 & HL-LHC [14 TeV, $3~\iab$]  & HE-LHC [27 TeV, $15~\iab$] & 100 TeV, $30~\iab$ \\
\midrule
$\tau \tau$ [pb]  & $7.0\e{-7}$ & $8.9\e{-6}$ & $2.9\e{-4}$\\
 $\mu \mu$ [pb]  & $2.5\e{-9}$  & $3.2\e{-8}$ & $1.0\e{-6}$  \\
$e e$ [pb]     & $5.8\e{-14}$  & $7.4\e{-13}$ & $2.4\e{-11}$  \\
$\gamma \gamma$ [pb]  & $3.7\e{-7}$  & $4.6\e{-6}$ & \fbox{$1.5\e{-5}$}  \\
$b  \bar{b}$ [pb]   & $1.2\e{-5}$    & $1.5\e{-4}$ & \fbox{$4.8\e{-3}$ } \\
$t \bar{t}$ [pb]      & \fbox{$1.9\e{-2}$ }   & \fbox{$0.2$} & \fbox{$8.0$}  \\
\bottomrule
\end{tabular}
\caption{Benchmark points for $H_{\rm DTC}$ production channels  at the 14 TeV HL-LHC, 27 TeV HE-LHC and a 100 TeV collider with $m_{H_{\rm DTC}}=2073$ GeV.}
\label{tab:higgs_DTC3a}
\end{table}

\section{Summary}
\label{sum}

QCD-like TC models were originally proposed to provide an elegant and natural mechanism for electroweak symmetry breaking, with mass generation arising dynamically through chiral symmetry breaking. However, these models encountered severe challenges in reproducing the observed fermion mass spectrum of the SM. The primary difficulty stems from FCNC interactions, which push the required scale of ETC to around $10^6$~GeV. This, in turn, suppresses fermion masses to phenomenologically unrealistic values.  

Moreover, QCD-like TC models typically predict a Higgs boson much heavier than the observed 125~GeV and are generally incompatible with electroweak precision observables. Even alternative strong-dynamics approaches, such as walking technicolor, face difficulties in generating realistic fermion masses. These scenarios often rely on hierarchical breaking of large non-Abelian flavor symmetries, producing distinct scales for SM fermion masses. Yet, the practical implementation remains complex, and a fully consistent description of both fermion mass hierarchies and mixing patterns is still lacking.  

In this work, we have presented a DTC based framework, which offers a novel fermionic mass mechanism for TC type theories by avoiding issues faced by  the conventional TC or walking type theories.  The DTC paradigm rests on the following key principles:
\begin{enumerate}
\item  
The underlying gauge group is defined as  
\be
\mathcal{G} \equiv \mathrm{SU}(\rm N_{\rm TC}) \times \mathrm{SU}(\rm N_{\rm DTC}) \times \mathrm{SU}(\rm N_{\rm D}),
\ee  
representing a set of QCD-like gauge sectors, each asymptotically free and confining at low energies.  

\item
Fermion masses and mixing, including those of neutrinos, are generated dynamically through multi-fermion condensates. At low energies, these condensates manifest as hierarchical VEVs, effectively reducing to the SHVM and thereby providing a dynamical solution to the flavor problem.  
\end{enumerate}

Within this framework, both the SHVM and, in principle, FN mechanisms can be naturally embedded. However, we find that a simple realization of the FN mechanism is not viable within the DTC paradigm.  Thus, the SHVM embedded in the DTC framework solves the problem of flavor of the SM.  This is an important development for technicolor type theories.

We emphasize that the mass-generation mechanism explored in this work represents a generic framework. For example, the underlying TC dynamics could be substituted with walking dynamics, leading to richer and potentially distinctive phenomenology. Alternatively, the gauge symmetry $\mathrm{SU}(\rm N_{\rm TC})$ could be replaced by other groups, as realized in composite Higgs models~\cite{Agashe:2004rs,Ma:2015gra}. For further discussion of such possibilities, see Refs.~\cite{Cacciapaglia:2014uja,Bellazzini:2014yua,Panico:2015jxa}.

From a phenomenological perspective, we investigate the collider signatures of the DTC model in the scenario where SHVM dynamics are realized. Inclusive decay channels such as $\bar{b}b$, $\tau^+\tau^-$, $t\bar{t}$, and $\gamma\gamma$ are studied in detail. Several of these signatures lie within the sensitivity reach of the HL-LHC across a wide mass range. We further extend our analysis to the HE-LHC and a future 100~TeV collider such as the FCC-hh.  

A key feature is that couplings of TC bound states, such as $\rho_{\rm TC}$, $\eta_{\rm TC}^\prime$, and DQCD mesons, to fermionic final states are highly suppressed, leading to vanishing direct collider signatures in those channels. This suppression motivates to search for these particles in alternative channels such as VBF.

\section*{Acknowledgement} 
This work is supported by the Council of Science and Technology, Government of Uttar Pradesh, India through the project ``A new paradigm for flavour problem'' (Project No.~CST/D-1301), and by the Anusandhan National Research Foundation (SERB), Department of Science and Technology, Government of India through the project ``Higgs Physics within and beyond the Standard Model'' (Project No.~CRG/2022/003237).

\section*{Appendix}

\subsection*{Outline of a possible extended technicolor and extended dark-technicolor}
In this appendix, we present an outline of a possible extended  and dark-extended technicolur scenario  as discussed in reference \cite{Abbas:2023ivi}. For ETC model,  the TC fermions, left-handed SM fermions, and $F_R$ fermions are accommodated in an $\rm SU(\rm N_{\rm TC} + 12 ) $  symmetry in the following way:
\begin{eqnarray}
\psi^{\rm ETC}_L  &\equiv&   \begin{pmatrix}
T ~B \\
T ~B \\
T ~B  \\
\cdots \\
\cdots \\
 T ~B  \\
u ~d \\
u ~d \\
u ~d \\
\nu_e~ e\\
c ~s \\
c ~s \\
c ~s \\
\nu_\mu~ \mu\\
t ~b \\
t ~b \\
t ~b \\
\nu_\tau~ \tau
\end{pmatrix}_L,     
\psi^{\rm ETC}_R  \equiv   \begin{pmatrix}
T \\
T  \\
T  \\
\cdots \\
\cdots \\
T  \\
U^1 \\
U^1  \\
U^1  \\
N^1\\
U^2 \\
U^2  \\
U^2  \\
N^2\\
U^3 \\
U^3  \\
U^3  \\
N^3\\
\end{pmatrix}_R,   
\psi^{\rm ETC}_R  \equiv   \begin{pmatrix}
B \\
B  \\
B  \\
\cdots \\
\cdots \\
B  \\
D^1 \\
D^1  \\
D^1  \\
E^1\\
D^2 \\
D^2  \\
D^2  \\
E^2\\
D^3 \\
D^3  \\
D^3  \\
E^3\\
\end{pmatrix}_R.   
\end{eqnarray}

The  group $SU(\rm N_{\rm DTC} +1)$ defines the EDTC where the first family qaurk-multiplet is given by:
\begin{eqnarray}
\psi^{\rm EDTC,q}_{L,i}  &\equiv&   \begin{pmatrix}
c_i ~c_i~c_i~c_i \cdots U^i \\
s_i ~s_i~s_i~s_i \cdots D^i \\
\end{pmatrix}_L,     
\psi^{\rm EDTC, q}_R  \equiv  \begin{pmatrix}
c_i ~c_i~c_i~c_i \cdots f_u \\
s_i ~s_i~s_i~s_i \cdots f_d \\
\end{pmatrix}_R,   
\end{eqnarray}
where $i=1,2,3 \cdots$ show the number of generations, and $f_u= u, c, t$ and $f_d= d, s, b$ stand for  the right-handed quark SM fields.

The leptonic multiplet can be defined in a similar manner as,
\begin{eqnarray}
\psi^{\rm EDTC,\ell}_{L,i}  &\equiv&   \begin{pmatrix}
e_i ~e_i~e_i~e_i \cdots N^i \\
n_i ~n_i~n_i~n_i \cdots E^i \\
\end{pmatrix}_L,     
\psi^{\rm EDTC,\ell}_{R,i}  \equiv   \begin{pmatrix}
e_i ~e_i~e_i~e_i \cdots f_\nu \\
n_i ~n_i~n_i~n_i \cdots f_e \\
\end{pmatrix}_R,      
\end{eqnarray}
where $f_\nu= \nu_e,  \nu_\mu, \nu_\tau$ and $f_e= e, \mu, \tau$ denote the right-handed leptonic SM fields.

We assume that ETC and DETC are further accommodated in GUT symmetry $\mathcal{G}_{\rm GUT}$, which is broken as,
\be 
\mathcal{G}_{\rm GUT} \rightarrow \rm SU(\rm N_{\rm TC} + 12 ) \times SU(\rm N_{\rm D}) \times SU(\rm N_{\rm DTC} + 1 ) \rightarrow SU(3)_c \times SU(\rm N_{\rm TC} )  \times SU(\rm N_{\rm D}) \times SU(\rm N_{\rm DTC}).
\ee 




\end{document}